\documentclass{article}
\usepackage{graphicx}
\usepackage{xcolor}
\usepackage{amsfonts}
\usepackage{amssymb}
\def\ligne#1{\hbox to\hsize{#1}}
\def\leurre{\noindent\leftskip0pt\small\baselineskip 10pt}
\newtheorem{thm}{\textbf{Theorem}}
\newtheorem{fig}{\textbf{Figure}}
\newtheorem{tab}{\textbf{Table}}
\newtheorem{lem}{\textbf{Lemma}}

\newtheorem{prop}{\textbf{Proposition}}
\newtheorem{requ}{\textbf{Requirement}}

\author{Maurice {\sc Margenstern}}
\title{A strongly universal cellular automaton in the dodecagrid with five states}
\begin{document}
\maketitle

\begin{abstract}
In this paper, we prove that there is a strongly universal cellular automaton in the 
dodecagrid, the tessellation $\{5,3,4\}$ of the hyperbolic $3D$ space, with five states 
which is rotation invariant. This improves a previous paper of the author where the 
automaton required ten states.
\end{abstract}

\section{Introduction}~\label{intro}

    In many papers, the author studied the possibility to construct universal cellular
automata in tilings of the hyperbolic plane, a few ones in the hyperbolic $3D$ space. 
Most often, the constructed cellular automaton was weakly universal. By {\it weakly 
universal}, we mean that the automaton is able to simulate a universal device starting
from an infinite initial configuration. However, the initial configuration should not be
arbitrary. It was the case as far as it was periodic outside a large enough circle, in 
fact it was periodic outside such a circle in two different directions:
the simulated device was a two-registered machine, which is enough for that purpose
from a theorem established by Coke-Minsky, \cite{minsky}. 

\newcount\compterel\compterel=1
\def\numerrel{\the\compterel\global \advance\compterel by 1}
   In the present paper, we consider the tiling of the hyperbolic $3D$-space which we 
call the {\bf dodecagrid} whose signature is, by definition, \hbox{$\{5,3,4\}$}. 
In that signature, 5 is the number of sides of a face, 3 is the number of edges which 
meet at a vertex, 4 is the number of dodecahedrons around an edge. In the hyperbolic 
$3D$-space, there is another tessellation based on another dodecahedron whose signature is
\hbox{$\{5,3,5\}$} which means that around an edge, there should be five dodecahedrons.
That latter dodecahedron, $\mathcal D$, is different from the one we consider in the 
present paper. Our dodecahedron, called also Poincar\'e's dodecahedron has right-angles
between contiguous sides of its faces and the dihedral angle between adjacent faces
is also a right angle. It is not the case for $\mathcal D$ whose dihedral angle
is smaller, as far as it is \hbox{$\displaystyle{{2\pi}\over5}$} and, consequently, it
is bigger than Poincar\'e's dodecahedron. From now on, we consider the dodecagrid only 
and its dodecahedrons are always copies of Poincar\'e's dodecahedron, which we denote 
by~$\Delta$.

Below, Figure~\ref{fdodecs}
provides us with a representation of~$\Delta$ according to Schlegel representation
of the solid. That representation could be used to represent the dodecagrid too.
However, here, we shall use another representation illustrated by Figure~\ref{stab_fix0} 
which we shall again meet later in Section~\ref{scenario}. That figure makes use of a 
property we shall see with the explanations about  Figure~\ref{fdodecs}. Before turning 
to that argumentation, we presently deal with the Schlegel representation.

   Figure~\ref{fdodecs} is obtained by the projection of the vertices and the edges of
$\Delta$ on the plane of one of its faces. Note that such a projection can be performed
both in the Euclidean $3D$-space and in the hyperbolic one. 
\vskip 10pt
\vtop{
\ligne{\hfill
\includegraphics[scale=0.5]{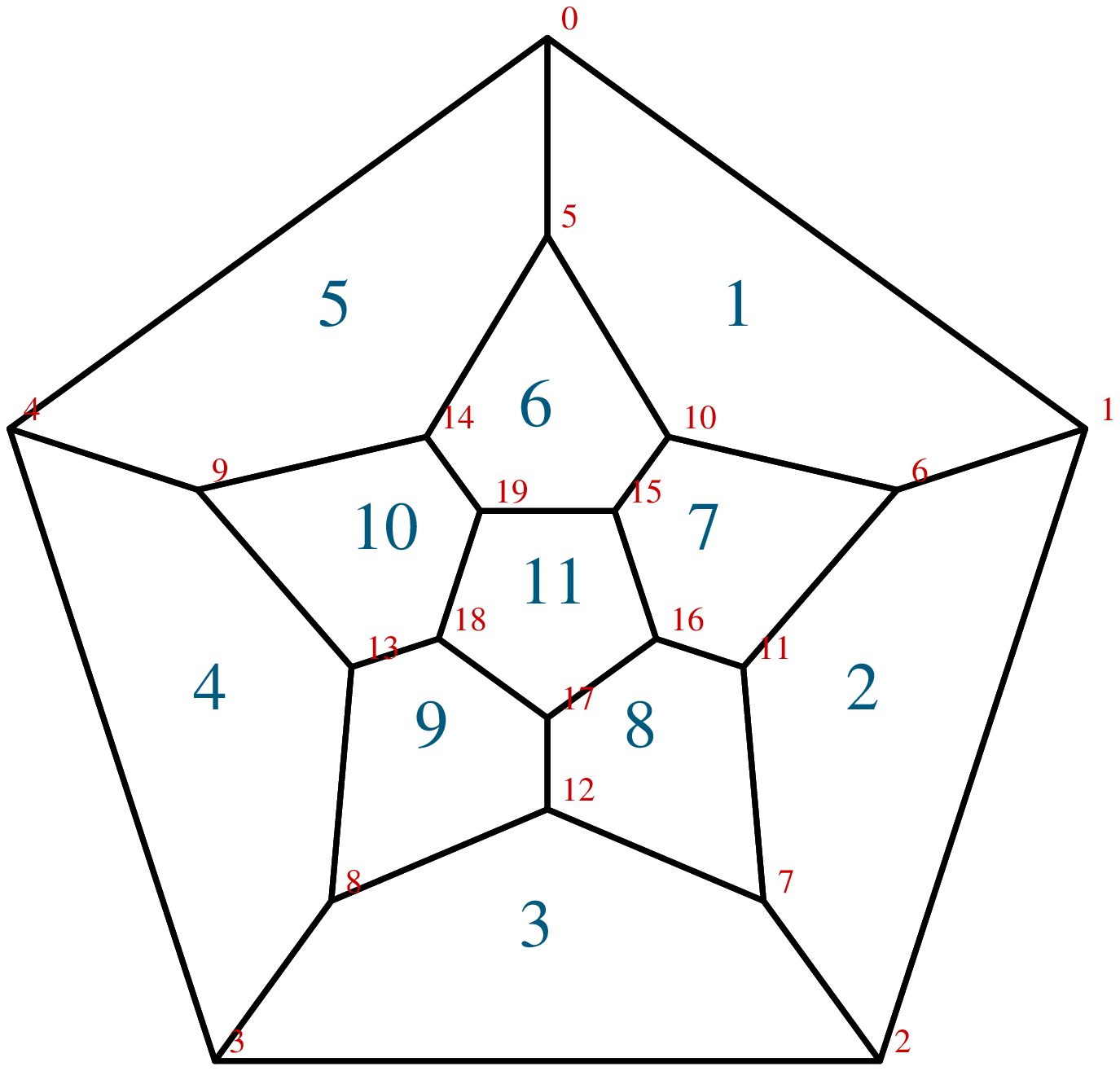}
\hfill}
\vspace{-5pt}
\begin{fig}\label{fdodecs}
\leurre
The Schlegel representation of the dodecahedron~$\Delta$.
\end{fig}
}

As shown on the figure, we number the faces of~$\Delta$. 
Face~0 is the face whose plane is the one on which the projection is 
performed. We number the faces around face~0 by clockwise turning around the face when 
we look at the plane from a point which stands over~$\Delta$. 'Over' means that $\Delta$
and the centre of the projection are on the same half-space defined by the plane on which
the projection is performed. If we consider the face which is opposed to face~0, we
also number the remaining faces of~$\Delta$, they are also clockwise numbered from~6
to~10 with face~6 sharing a side with face~5 and another one with face~1. The not yet
numbered face which is opposite to face~0 is now numbered~11. That numbering will be
the basic numbering from which we derive a numbering of the vertices as shown on the
figure: we first number the vertices of face~0, then the other end of the edge
abutting the already numbered vertices. We repeat the process as far as three edges meet
at a vertex, until all vertices are numbered. The numbering which we performed in that
way is also clockwise performed. However, later on, when we shall have to point at a 
vertex or an edge, we shall indicate the faces which share that element: two faces for 
an edge, three faces for a vertex.

Later, we shall consider the neighbours of a dodecahedron in the dodecagrid. Consider
such a tile, say $\Delta$ again, and assume a numbering of its faces obtained as in 
Figure~\ref{fdodecs}. Let $F_i$ be the face numbered by~$i$, with \hbox{$i\in\{0..11\}$},
we say later on {\bf face~$i$}. We say that $F_i$ and $F_j$, with \hbox{$i\not=j$} are 
{\bf contiguous} if and only if they share an edge. There is a single dodecahedron of 
the dodecagrid which shares $F_i$ with~$\Delta$, we denote that dodecahedron 
by~$\Delta_i$. We say that $\Delta_i$ is a {\bf neighbour} of~$\Delta$. We shall often 
say that $\Delta$ and~$\Delta_i$ {\bf can see} each other through face~$i$ and other
expressions connected with vision. Of course, in face~$i$, $i$ is the number
of $F_i$ in~$\Delta$. Now the face may receive another number in $\Delta_i$. Often,
we decide that face~$i$ in~$\Delta$ is face~0{} in~$\Delta_i$ and that a face 
of~$\Delta_i$ which is in the same plane as a face~$j$ of~$\Delta$ receives the 
number~$j$ in~$\Delta_i$ too. We shall do that in what follows if not otherwise specified.
The neighbours of~$\Delta$ share an important property: 

\begin{prop}\label{pneighs}
if $F_i$ and $F_j$ are contiguous
faces, $\Delta_i$ and~$\Delta_j$ cannot see each other, but 
\hbox{$\Delta_{i_j}=\Delta_{j_i}$}.
\end{prop}

The proposition is an easy corollary of the following assertion:

\begin{lem}\label{lneighs}
Two tiles~$T_1$ and~$T_2$ of the dodecagrid which share and edge~$s$ can see each other 
if and only if there is a plane~$\Pi$ containing~$s$ such that $T_1$ and~$T_2$ are 
completely contained in the same closed half-space defined by~$\Pi$.
\end{lem}

\noindent
Proof of the proposition and of the lemma.\vskip 0pt
By construction, let~$\Pi_i$ be the plane containing~$F_i$. That plane defines a 
half-space $\mathcal S$ which contains~$\Delta$. Clearly, $\Delta_i$ and~$\Delta$
are not both in~$\mathcal S$: otherwise we would have \hbox{$\Delta=\Delta_i$}.
From a similar argument,  $\Delta_j$ and~$\Delta$ are not on the same side with respect
to~$\Pi_j$.By construction, as $F_i$ and $F_j$ are contiguous, \hbox{$\Pi_i\perp\Pi_j$}.
Now, if \hbox{$\ell=\Pi_i\cap\Pi_j$}, $\Pi_i$ and $\Pi_j$ define four dihedral right 
angles around~$\ell$. By construction, $\Delta$ lies inside one angle, $\Delta_i$ lies
in a second one which is in the same side as $\Delta$ with respect to~$\Pi_j$ and, 
symmetrically, $\Delta_j$ lies in a third angle which is in the same side as $\Delta$
with respect to~$\Pi_i$. The fourth angle may contain a dodecahedron which shares the 
same edge with $\Delta$, $\Delta_i$ and~$\Delta_j$: it is a neighbour of~$\Delta_i$ 
seen from its face~$j$ but also, for the same reason, a neighbour of~$\Delta_j$ seen 
from its face~$i$. Now, by construction of the dodecagrid, there can exactly be four 
dodecahedrons around an edge, so that the neighbour~$j$ 
of~$\Delta_i$ is the neighbour~$i$ of~$\Delta_j$ which can be written
\hbox{$\Delta_{i_j} = \Delta_{j_i}$} as in Proposition~\ref{pneighs}. \hfill$\Box$

In particular, if two faces of tiles~$T_1$ and~$T_2$ sharing and edge~$s$ are in the same
plane~$\Pi$, and if $T_1$ and~$T_2$ are not in the same half-space defined by~$\Pi$
the tiles cannot see each other. {\it A fortiori}, if the tiles have no common element,
they cannot see each other.

\def\HH{$\mathcal H$}
   The cellular automaton we construct in Section~\ref{scenario} evolves in the
hyperbolic $3D$ space but a large part of the construction deals with a single plane
which we shall call the {\bf horizontal plane} denoted by \HH. In fact, both sides
of~\HH{} will be used by the construction and most tiles of our construction have a face
on~\HH.

   We take advantage of that circumstance to define a particular representation which we
find more appropriate to our purpose. 

   The trace of the dodecagrid on~\HH{} is a tiling of the hyperbolic plane, namely the
tiling $\{5,4\}$ we call the {\bf pentagrid}. The left-hand side part of 
Figure~\ref{penta} illustrates the tiling and its right-hand side part illustrates a way 
to locate the cells of the pentagrid. Let us look closer at the figure whose pictures 
live in Poincar\'e's disc, a popular representation of the hyperbolic plane, 
see~\cite{mmbook1}.

\vskip 10pt
\vtop{
\ligne{\hfill
\includegraphics[scale=0.75]{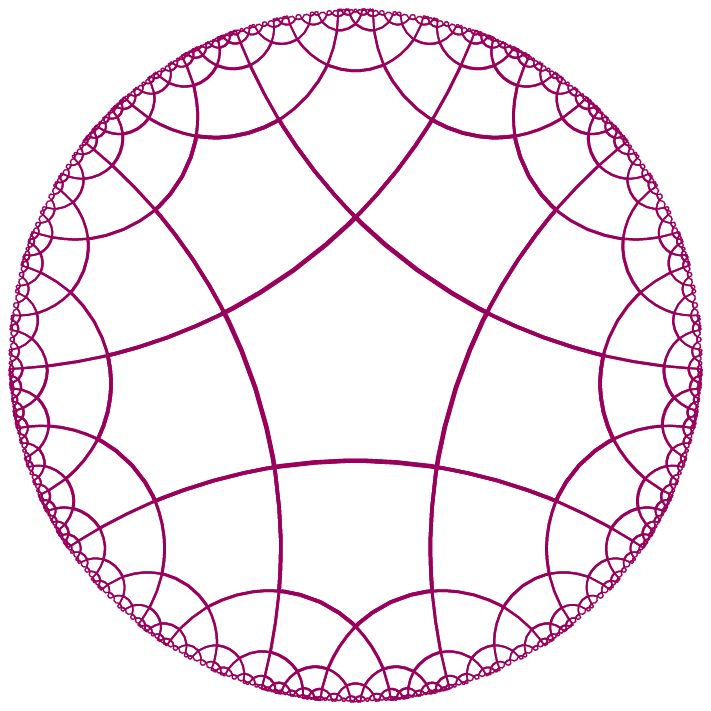}
\raise-20pt\hbox{\includegraphics[scale=0.475]{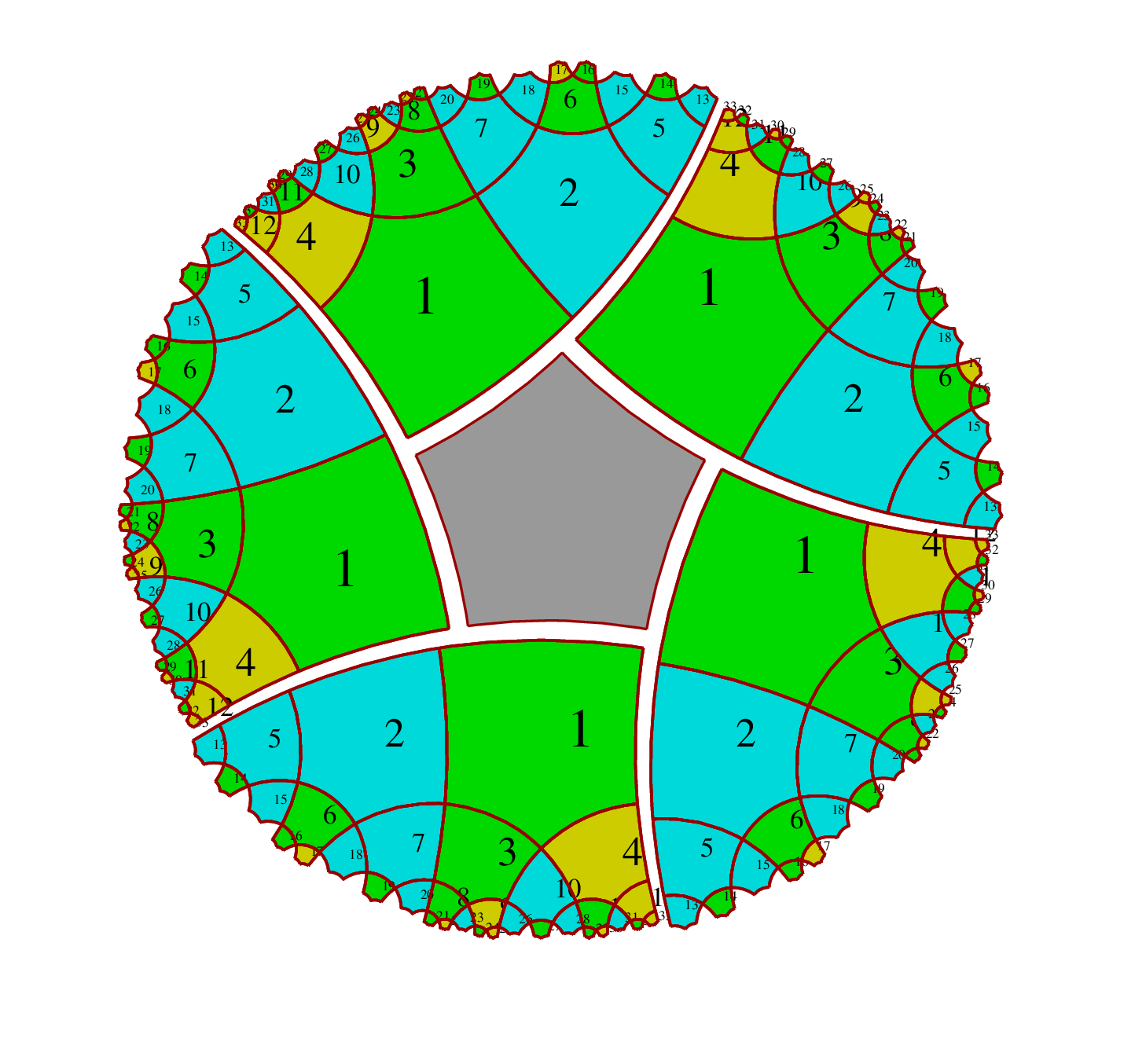}}
\hfill}
\vspace{-5pt}
\begin{fig}\label{penta}
\leurre
To left: a representation of the pentagrid in Poincar\'es disc model of the hyperbolic 
plane. To right: a splitting of the hyperbolic plane into five sectors around a once
and for all fixed central tile.
\end{fig}
}
   In the right-hand side picture, we can see five tiles which are counter-clockwise
numbered from~1 up to~5, those tiles being the neighbours of a tile which we call
the {\bf central tile} for convenience. Indeed, there is no 
central tile in the pentagrid
as there is no central point in the hyperbolic plane. We can see the disc model as a
window over the hyperbolic plane, as if we were flying over that plane in an abstract
spacecraft. The centre of the circle is the point on which are attention is focused while
the circle itself is our horizon. Accordingly, the central tile is the tile which is 
central with respect to the area under our consideration. It is also the reason to
number the central tile by~0.

The right-hand side picture shows us five blocs of tiles we call {\bf sectors}. 
Each sector is defined by a unique tile which shares and edge with the central one.
We number that tile by~1. It is a green tile on the picture. The sector is delimited
by two rays~$u$ and~$v$ issued from a vertex of tile~1: they continue two consecutive 
sides of tile~0.  Those rays are supported by straight lines in the hyperbolic plane 
and they define a right angle. Tile~1 is called the {\bf head} of the sector it defines :
the sector is the set of tiles contained in the angle defined by~$u$ and~$v$.
We also number the sectors from~1 to~5 by counter-clockwise turning around tile~0.
We also say that two tiles are {\bf neighbouring} or that they are {\bf neighbours} of 
each other if and only if they have a common side.

\vskip 10pt
\vtop{
\ligne{\hfill
\includegraphics[scale=1.5]{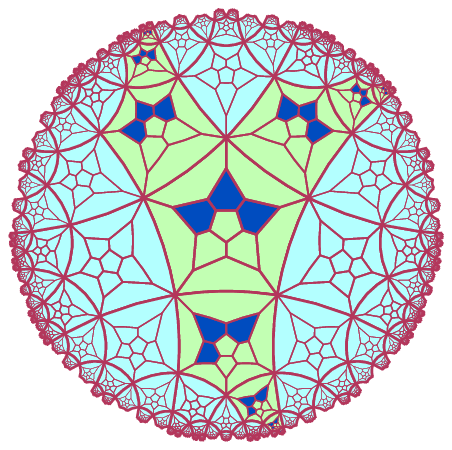}
\hfill}
\vspace{-5pt}
\begin{fig}\label{stab_fix0}
\leurre
A configuration we shall later meet in Section~{\rm\ref{scenario}}. The central tile
is, in some sense, the centre of that configuration. A dark blue face means that a
blue dodecahedron is put on that face.
\end{fig}
}

   Consider the configuration illustrated by Figure~\ref{stab_fix0}. We can see the 
Schlegel projection of a dodecahedron on each tile of the picture. A few dodecahedrons
have another light colour and on each of them, three faces have a dark blue face. We call
them the elements of a {\bf track}.  Tile~0 is the central tile of the picture. 
Around tile~0, we can see three elements of track exactly, the other neighbouring tiles
having a uniform light colour: we call them {\bf blank} tiles. We number the elements of 
tracks by the number of their sectors, here 1, 3 and~5 by counter-clockwise turning around 
tile~0. On the picture, tile~3 is just below tile~0. Let us look closer at tiles~0 and~3
of~\HH. Let $\Delta_0$ and~$\Delta_3$ be the tiles of the dodecagrid which are put on 
tiles~0 and~3 of~\HH{} respectively.  Assume that the face~1 of $\Delta_0$ and 
of~$\Delta_3$ share the side which is shared by the tiles~0 and~3 of~\HH. By construction
of the figure, tiles~0 and~3 are on the same side with respect to~\HH. As a corollary, 
the faces~1 of $\Delta_0$ and~$\Delta_3$ coincide so that,
according to Lemma~\ref{lneighs} and to Proposition~\ref{pneighs}, 
\hbox{$(\Delta_0)_1 = \Delta_3$} and \hbox{$(\Delta_3)_1=\Delta_0$}. 
From the same proposition, we can see that $(\Delta_0)_6$
and $(\Delta_0)_7$ cannot see each other but $(\Delta_0)_6$ and $(\Delta_3)_7$
can see each other and, for a similar reason, $(\Delta_0)_7$ and $(\Delta_3)_6$ can also
see each other.

Accordingly, we have to pay attention to dodecahedrons which are put 
on the faces of a dodecahedron in the representation as defined on Figure~\ref{stab_fix0}:
we call that representation the {\bf \HH-representation}. The rule is simple: a face~$F$ 
of a dodecahedron~$\Delta$ takes the colour of the neighbour that $\Delta$ can see 
through~$F$. The rule also applies to the faces of two neighbouring dodecahedrons whose
face~0 is on~\HH{} through which they see each other. Each face take the colour of its 
neighbour.

By {\it abus de langage}, we also call \HH{} the restriction of the 
dodecagrid to those which sit on that plane. When it will be needed to clarify, we denote 
by \HH$_u$, \HH$_b$ the set of dodecahedrons which are placed upon, below~\HH{} 
respectively.

   Now that the global setting is given, we shall proceed as follows: 
Section~\ref{scenario} indicates the main lines of the implementation which is precisely
described in Subsection~\ref{newrailway}. At last, Section~\ref{srules} gives us the rules
followed by the automaton. That section also contain a few figures which illustrate the 
application of the rules. Those figures were established from pieces of figures drawn by 
a computer program which applied the rules of the automaton to an appropriate window in 
each of the configurations described in Subsection~\ref{newrailway}. The computer 
program also checked that the set of rules is coherent and that rules are pairwise 
independent with respect to rotation invariance. As far as that latter property is
much stronger as one could think we start our study by a short section on the rotations of
the dodecahedron, see Section~\ref{rotododec}.

   That allowed us to prove the following property:

\begin{thm}\label{letheo}
There is a strongly universal cellular automaton in the dodecagrid
which is rotation invariant, truly spatial and which has five states.
\end{thm}

\section{Rotations of the dodecahedron}\label{rotododec}

We give here an extended version of the study we have produced in~\cite{mmarXivh3D3st}.
The section deals with the rotations of the dodecahedron. As indicated in the quoted 
paper, it is classically known that there are sixty rotations which leave the dodecahedron
globally invariant: the image can exactly be put on the same place of the space as the 
initial dodecahedron. Using the Schlegel representation, it means that a rotation
which leaves $\Delta$ globally invariant performs a permutation on the number of the 
faces. Below, Figures~\ref{rotadodec1}, \ref{rotadodec2} and~\ref{rotadodec3} illustrate 
those rotations. As can be checked on the figures there are sixty such rotations.
They are characterized by the axis around which the rotation is performed. The change in 
the faces define the angle of the rotation. In order to see how a rotation operate,
we reproduce on the pictures of the dodecahedrons the numbering of the faces.
Figure~\ref{rotadodec1} illustrates the rotations around an axis which joins the 
mid-points of opposite faces, where opposite, here and further, means image of each other
under the symmetry in the centre of~$\Delta$. That defines six axes and, 
for each of them there are four possible positive angles of rotation: each one performs 
a circular permutation  on the numbers
of the faces which are neighbouring a face crossed by the axis. The figure shows us
six rows: each one indicates the faces crossed by the axis. Below that indication, we
have the initial dodecahedron $\Delta$ with the numbering of its faces as 
\ligne{\hfill} 

\vskip 10pt
\vtop{
\ligne{\hfill
\includegraphics[scale=0.725]{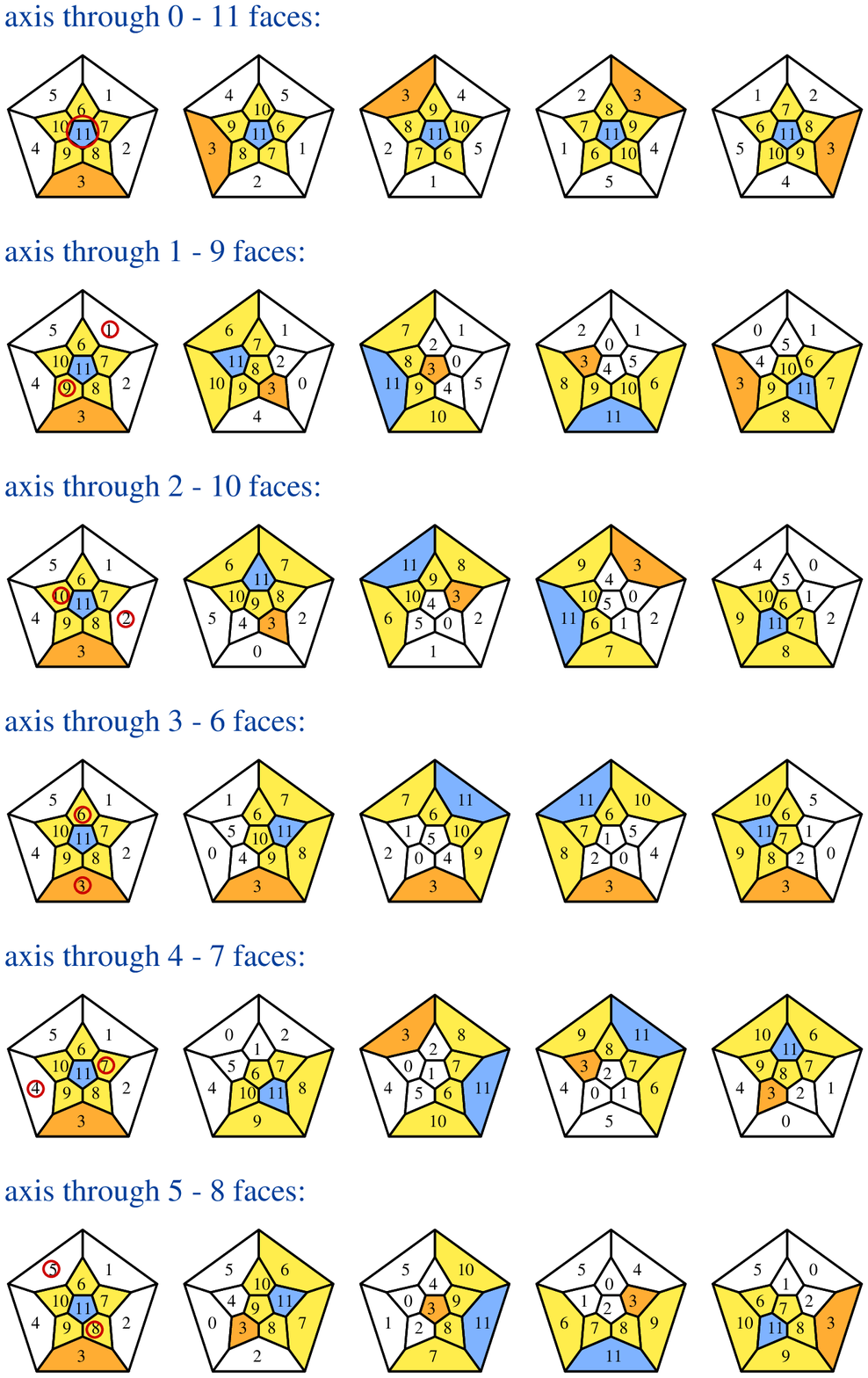}
\hfill}
\vspace{-5pt}
\begin{fig}\label{rotadodec1}
\leurre
The rotations which leave a dodecahedron globally invariant. Here,  the rotations around
the axes joining the mid-points of two opposite faces.
\end{fig}
}
\vskip 5pt
\noindent
defined by 
Figure~\ref{fdodecs}. Two red circles indicate the opposite faces under consideration.
To the right-hand side of that image of~$\Delta$, four dodecahedrons illustrate
the rotations around the indicated axes and, on each image, the number of the faces 
indicate how the transformation operates on $\Delta$.
Accordingly, Figure~\ref{rotadodec1} shows us the 24 rotations around an axis crossing
two opposite faces through their mid-point.

\vskip 10pt
\vtop{
\ligne{\hfill
\includegraphics[scale=0.586]{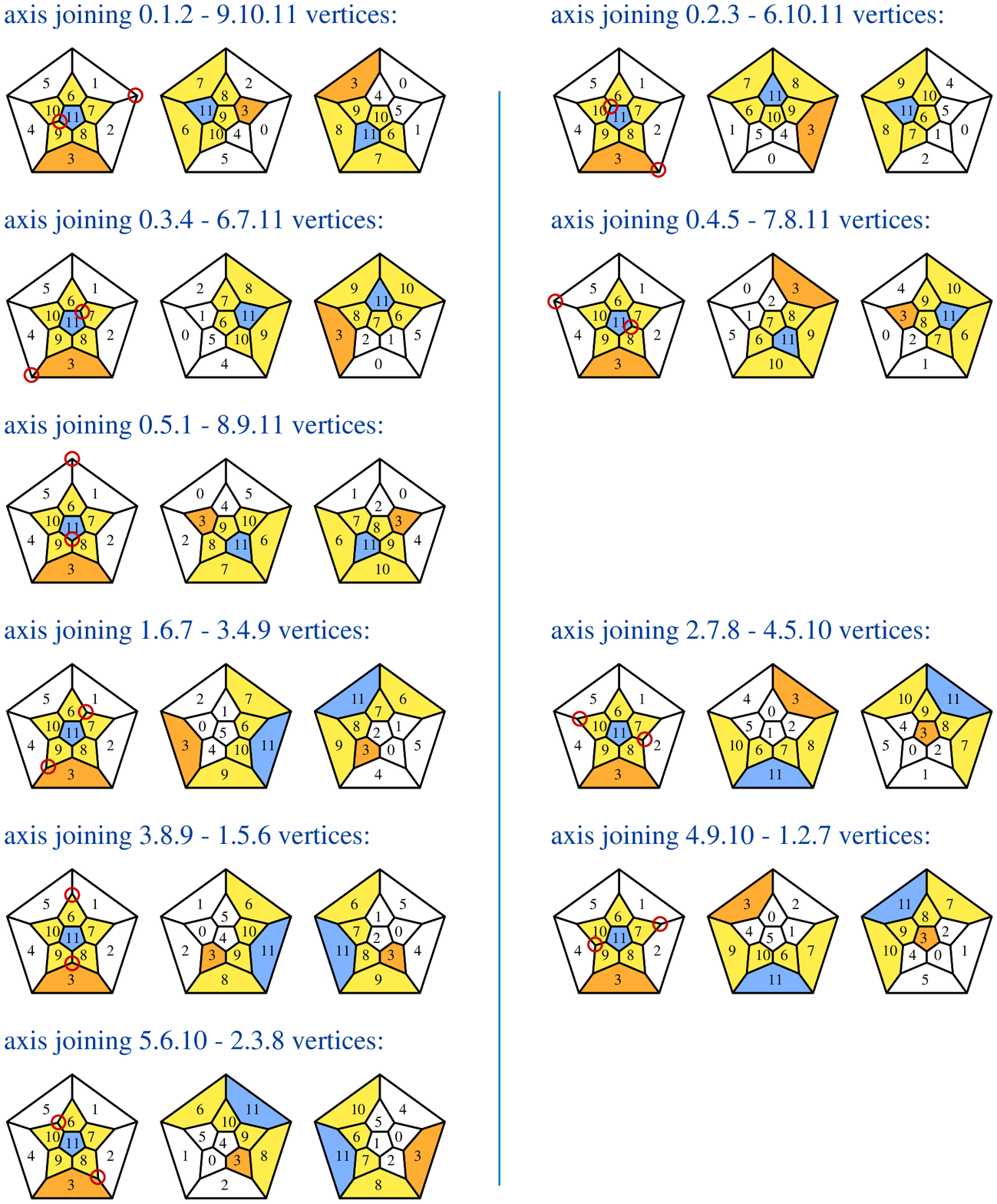}
\hfill}
\vspace{-5pt}
\begin{fig}\label{rotadodec2}
\leurre
The rotations which leave a dodecahedron globally invariant. Here,  the rotations around
the axes joining two opposite vertices.
\end{fig}
}
\vskip 5pt
Figure~\ref{rotadodec2} does the same with the axes joining two opposite vertices 
of~$\Delta$. We can check on Figure~\ref{fdodecs} that $\Delta$ possesses twenty vertices,
so that Figure~\ref{rotadodec2} is structured upon the ten axes defined in that way.
The upper half of the figure illustrate the rotations through the axes joining 
a vertex of face~0 to the opposite vertex of face~11.

Call {\bf lower crown}, {\bf upper crown}, the set of tiles which share an edge with
face~0, face~11 respectively. Each face of the lower crown is an image of a unique one
in the upper crown with respect to the symmetry in the centre of~$\Delta$.
The lower part of Figure~\ref{rotadodec2} concerns the axes joining a vertex of a face
neighbouring face~0 to the opposite vertex of face~11. That correspondence defines five
axes. The five others are defined by opposite vertices which belong to both crowns.
Red circles indicate the vertices through which the axis passes. Around each axis there 
can be two rotations outside the identity. That
provides us with twenty additional rotations. In the figure, the vertices are indicated
by the numbers of the three faces to which they belong.
\vskip 10pt
\vtop{
\ligne{\hfill
\includegraphics[scale=0.55]{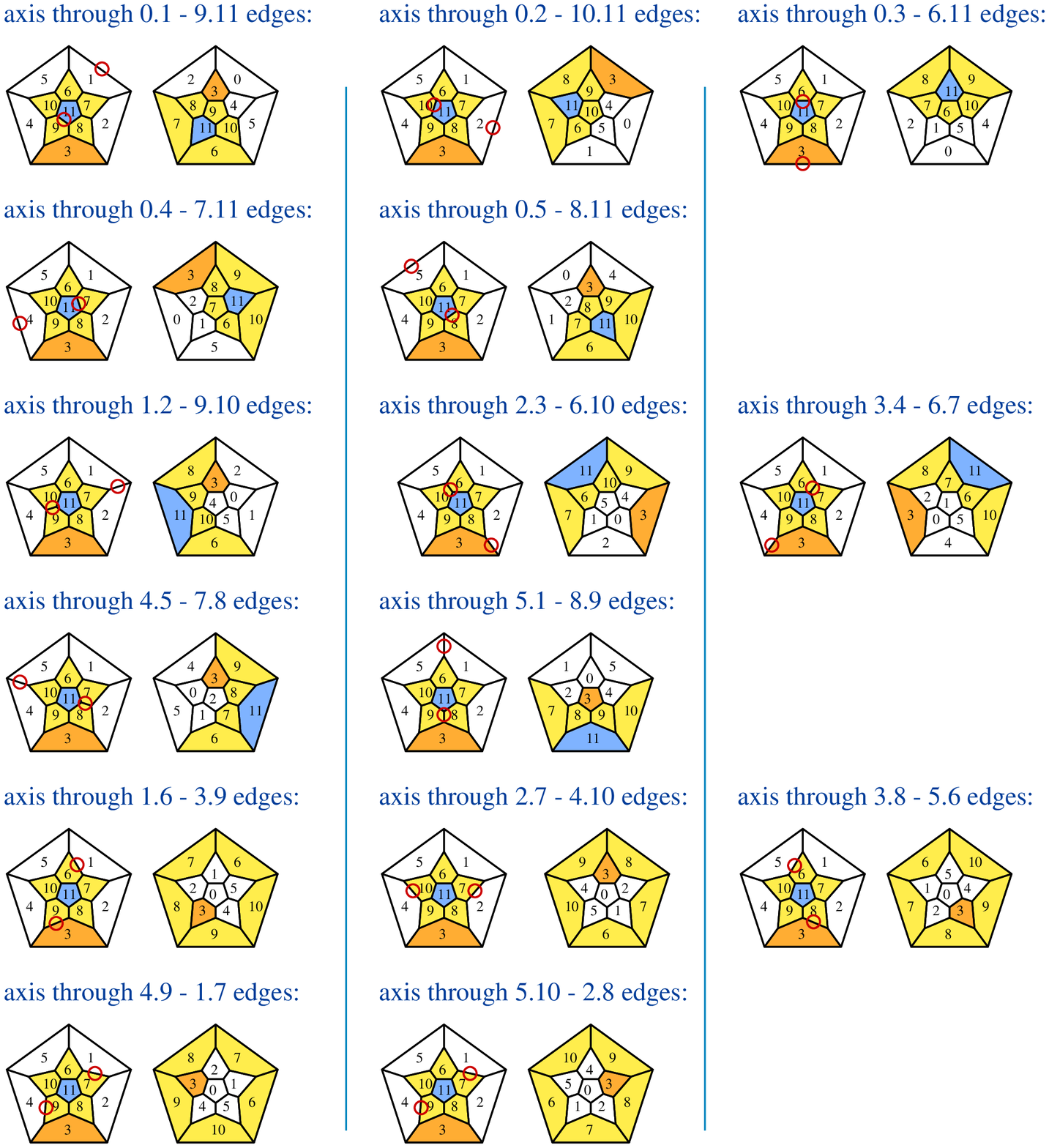}
\hfill}
\vspace{-5pt}
\begin{fig}\label{rotadodec3}
\leurre
The rotations which leave a dodecahedron globally invariant. Here,  the rotations around
the axes joining the mid-points of two opposite edges.
\end{fig}
}
\vskip 5pt
At last and not the least, Figure~\ref{rotadodec3} shows us the rotations performed 
around an axis which joins the mid-points of opposite edges. As can be checked on
Figure~\ref{fdodecs}, $\Delta$ possesses thirty edges which define fifteen non-trivial
rotations. The red circles on the image of $\Delta$ indicate the mid-points through which
the considered axis passes. In the figure, the edges are indicated by two numbers, those
of the faces to which they belong. Such a numbering will be used from now on.

   Accordingly, the figures give us 24, 20 and 15 rotations which, together with the
identity constitute the sixty rotations leaving $\Delta$ globally invariant.

In order to correctly draw the figures we need to make an easy correspondence
between a rotation and its image, starting from the image. We do that in 
Table~\ref{trelate}.

\def\ffs#1{\footnotesize{#1 }}
\newdimen\wrot \wrot=24pt
\newdimen\wlarge \wlarge=160pt
\def\lignerota #1 #2 #3 #4 #5 {%
\ligne{
\hbox to \wrot{\ffs{#1 }\hfill}\hskip 5pt
\hbox to \wrot{\ffs{#2 }\hfill}\hskip 5pt
\hbox to \wrot{\ffs{#3 }\hfill}\hskip 5pt
\hbox to \wrot{\ffs{#4 }\hfill}\hskip 5pt
\hbox to \wrot{\ffs{#5 }\hfill}\hskip 5pt
\hfill}
}
\def\lignerotb #1 #2 #3 #4 #5 {%
\ligne{
\hbox to \wrot{\bf{\ffs{#1 }}\hfill}\hskip 5pt
\hbox to \wrot{\bf{\ffs{#2 }}\hfill}\hskip 5pt
\hbox to \wrot{\bf{\ffs{#3 }}\hfill}\hskip 5pt
\hbox to \wrot{\bf{\ffs{#4 }}\hfill}\hskip 5pt
\hbox to \wrot{\bf{\ffs{#5 }}\hfill}
\hfill}
}
\vtop{
\begin{tab}\label{trelate}
\leurre
Finding the rotation from the image identified by its faces which play the roles of
face~$0$ and~$1$. When the image faces are known, the rotation is on the just below row
in the same column. The first number to identify the rotation is $1$, $2$ and~$3$ 
for Figures~$4$, $5$ and~$6$ respectively. In each figure, a rotation is identified by 
the rank of the axis in the figure, counting the rank line after line, from left to
right on each line. At last, when there is a third number, it indicates which rotation
for that axis. As an example, it can be checked that the image $(9\ 4)$ is obtained 
by the rotation of Figure~$5$, attached to the tenth axis as its second rotation.
\end{tab}
\vskip-7pt
\ligne{\hfill
\vtop{\leftskip 0pt\parindent 0pt\hsize=\wlarge
\lignerota {0 1} {0 2} {0 3} {0 4} {0 5} 
\lignerotb {id} {1 1 4} {1 1 3} {1 1 2} {1 1 1} 
\vskip 2pt
\lignerota {1 0} {1 2} {1 7} {1 6} {1 5} 
\lignerotb {3 1} {2 1 1} {1 3 1} {1 6 1} {2 5 1} 
\vskip 2pt
\lignerota {2 0} {2 3} {2 8} {2 7} {2 1} 
\lignerotb {2 1 2} {3 2} {2 2 1} {1 4 1} {1 2 1} 
\vskip 2pt
\lignerota {3 0} {3 4} {3 9} {3 8} {3 2} 
\lignerotb {1 3 4} {2 2 2} {3 3} {2 3 1} {1 5 1} 
\vskip 2pt
\lignerota {4 0} {4 5} {4 10} {4 9} {4 3} 
\lignerotb {1 6 4} {1 4 4} {2 3 3} {3 4} {1 4 1}
\vskip 2pt
\lignerota {5 0} {5 1} {5 6} {5 10} {5 4} 
\lignerotb {2 5 2} {1 2 1} {1 5 4} {2 4 2} {3 5}
}
\hfill
\vtop{\leftskip 0pt\parindent 0pt\hsize=\wlarge
\lignerota {6 1} {6 7} {6 11} {6 10} {6 5} 
\lignerotb {1 2 2} {2 9 2} {2 7 2} {1 6 2} {3 10} 
\vskip 2pt
\lignerota {7 1} {7 2} {7 8} {7 11} {7 6} 
\lignerotb {1 2 3} {3 6} {1 3 2} {2 10 1} {2 8 1} 
\vskip 2pt
\lignerota {8 2} {8 3} {8 9} {8 11} {8 7} 
\lignerotb {2 9 1} {1 3 3} {3 7} {1 4 2} {2 6 1} 
\vskip 2pt
\lignerota {9 3} {9 4} {9 10} {9 11} {9 8} 
\lignerotb {2 7 1} {2 10 2} {1 4 3} {3 8} {1 5 2} 
\vskip 2pt
\lignerota {10 4} {10 5} {10 6} {10 11} {10 9} 
\lignerotb {1 6 3} {2 8 2} {2 6 2} {1 5 3} {3 9}
\vskip 2pt
\lignerota {11 6} {11 7} {11 8} {11 9} {11 10} 
\lignerotb {3 11} {3 14} {3 12} {3 15} {3 13}
}
\hfill}
}
\vskip 10pt
It is not difficult to see with the help of those figures that the six rotations around
a face and the opposite one generate all the rotations. Figure~\ref{fgenrot} illustrates
how to proceed if we fix a face, an edge, a vertex. The colours help us to identify the
elements which would remain fixed in the rotation: one colour for a face, two colours 
for an edge, three of them for a vertex. Two generators allow to obtain the expected
rotations, perhaps using a power of one of them. From that figure, by an appropriate
rotation we can see how to generate all rotations of Figures~\ref{rotadodec2}
and~\ref{rotadodec3} by those of Figure~\ref{rotadodec1}. 
Note that in the second line of Figure~\ref{fgenrot}, the second rotation is twice the
angle of the first one while, in the last line, both rotations make use of an angle of
$\displaystyle{{2\pi}\over5}$. 
  
  In order to prepare the construction of the rules explained in Section~\ref{srules}, 
we need to define a simple criterion in order to see the neighbourhood of a tile.

\vskip 10pt
\vtop{
\ligne{\hfill
\includegraphics[scale=0.6]{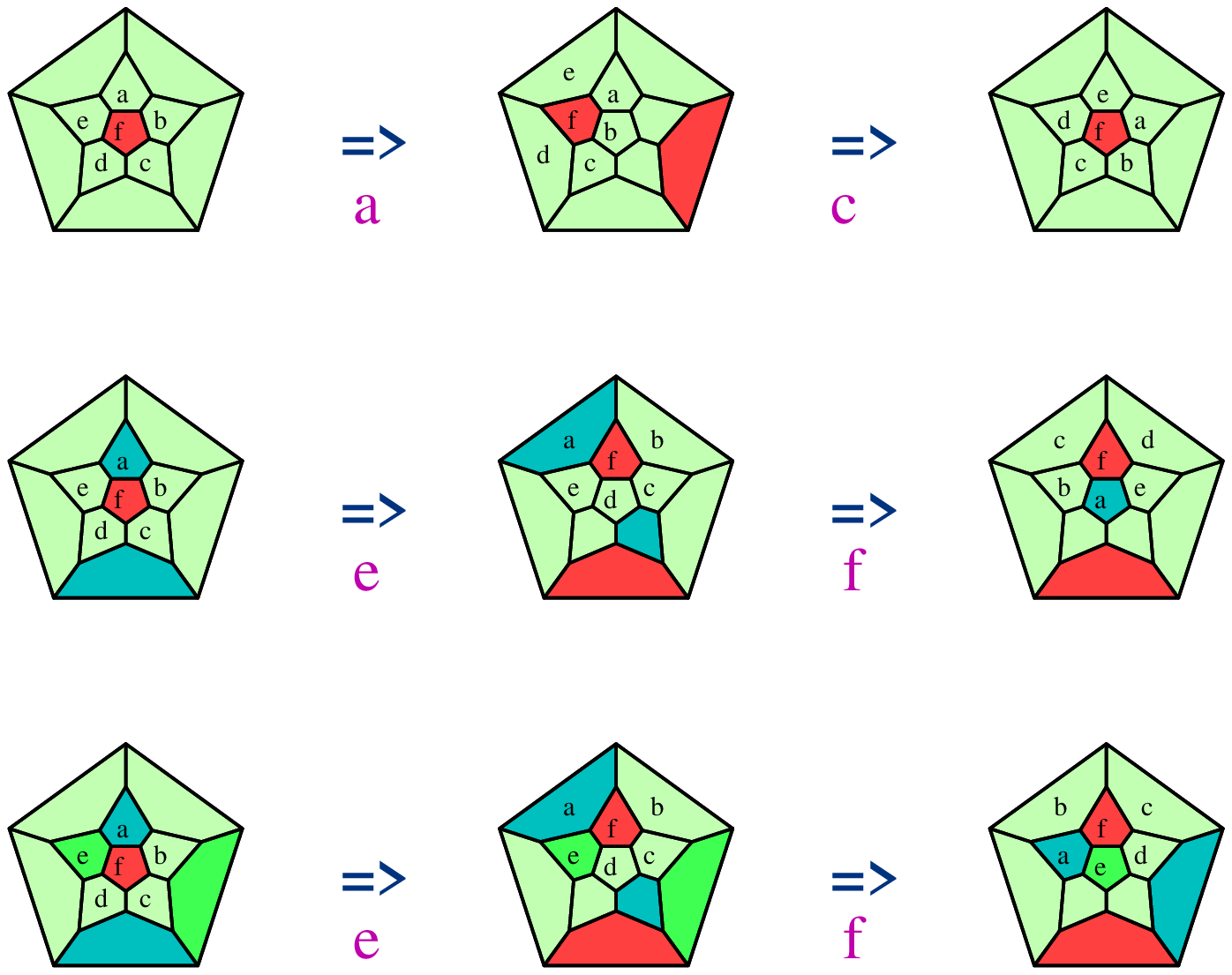}
\hfill}
\vspace{-5pt}
\begin{fig}\label{fgenrot}
\leurre
How to generate the three types of rotations thanks to six fixed rotations around a face.
The letter under an arrow indicates the face around which the rotation is performed.
\end{fig}
}
\vskip 10pt
Figure~\ref{fles4} helps us to establish simple criterion. In the right-hand side part 
of the figure, we have the representation of eight tiles which share a common vertex.
The figure shows us four of them only, those which belong to \HH$_u$. Those which belong
to \HH$_b$ are not visible.
\vskip 10pt
\vtop{
\ligne{\hfill
\includegraphics[scale=1]{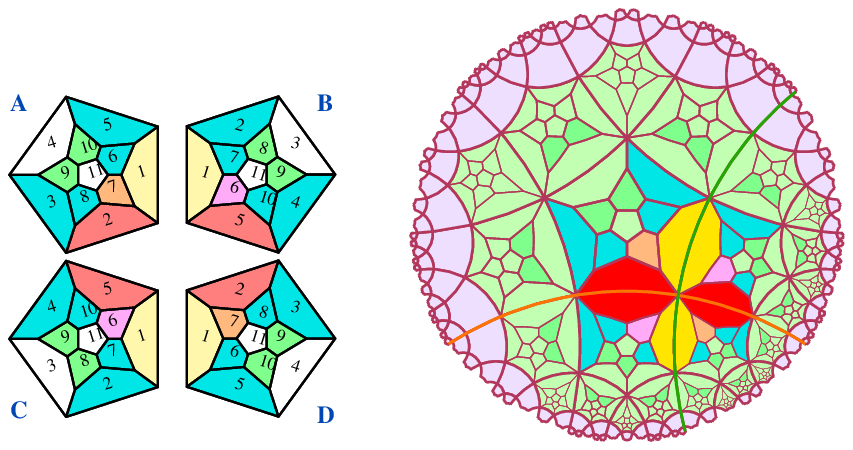}
\hfill}
\vspace{-5pt}
\begin{fig}\label{fles4}
\leurre
Correspondence between neighbours of neighbouring tiles.
\end{fig}
}
\vskip 5pt
\def\VV{\hbox{$\mathcal V$}}
Note that two lines~$\ell$ and~$m$, in green and orange respectively on the right-hand 
side part of the figure, bear sides of those tiles. We fix a numbering of the sides of 
the tiles which is indicated in the 
left-hand side part of the figure. Our four tiles have their face~0 on \HH{} and in all
of them, face~0 and face~1 have their common side on~$\ell$. As we already know, faces~1
of the four tiles lie on the same plane~\VV{} which is orthogonal to~\HH, cutting it 
along~$\ell$. Say that those tiles belong to generation~0 and denote them by {\bf A},
{\bf B}, {\bf C} and {\bf D} as indicated on the left-hand side part of the figure. 
The neighbours of those tiles belong to generation~1. We denote them as {\bf L}$_i$
where {\bf L} is one of the just mentioned letters and $i$ is in \hbox{$\{0..11\}$},
indicating a face on which is lying the neighbour. As far as the face of {\bf A}$_6$, 
{\bf B}$_7$ on~\VV{} is orthogonal to the face~1 of {\bf A}, {\bf B} respectively,
{\bf A}$_6$ and {\bf B}$_7$ share a face on that plane so that they can see each other
according to Proposition~\ref{pneighs}.
That property is illustrated on the left-hand side part of Figure~\ref{fles4} by the fact
that the face~6 of~{\bf A} and the face~7 of~{\bf B} are in blue. Similarly,
{\bf A}$_7$ and {\bf B}$_8$ can see each other. Denote that relationship by
\hbox{\bf A$_7$ $\Bumpeq$ B$_6$}. Here are the similar relationships which can be 
established :
\vskip 5pt
\ligne{\hfill
$\vcenter{\hbox{\vtop{\leftskip 0pt\parindent 0pt\hsize=80pt
\ligne{{\bf A$_6$ $\Bumpeq$ B$_7$}\hfill}
\ligne{{\bf A$_7$ $\Bumpeq$ B$_6$}\hfill}
\ligne{{\bf C$_6$ $\Bumpeq$ D$_7$}\hfill}
\ligne{{\bf C$_7$ $\Bumpeq$ D$_6$}\hfill}
}
\hskip 30pt
\vtop{\leftskip 0pt\parindent 0pt\hsize=80pt
\ligne{{\bf A$_7$ $\Bumpeq$ C$_6$}\hfill}
\ligne{{\bf A$_8$ $\Bumpeq$ C$_{10}$}\hfill}
\ligne{{\bf B$_6$ $\Bumpeq$ D$_7$}\hfill}
\ligne{{\bf B$_{10}$ $\Bumpeq$ D$_8$}\hfill}
}}}$
\hfill(\numerrel)\hskip 20pt}
\vskip 5pt
We can see that in (1) several neighbours can see two neighbours. Indeed, {\bf A} can see
both {\bf B} and {\bf C} but it cannot see {\bf D} as far as {\bf C} and {\bf D} are not
on the same half-space with respect to~\VV. From (1), we can see that {\bf A}$_7$ can see
both {\bf B}$_6$ and {\bf C}$_6$. Let~$s$ be the side which is shared by the faces~1 of
{\bf A}, {\bf B} {\bf C} and {\bf D}. Those tiles are the four tiles around~$s$. Now, if 
$n$ is the line which bears~$s$, we can see that {\bf A}$_7$, {\bf B}$_6$, {\bf C}$_6$
and {\bf D}$_7$ also share a common side which lies on~$n$. 

We defined {\bf L}$_i$ where {\bf L} stands for {\bf A}, {\bf B}, {\bf C} or {\bf D} as
the neighbours of generation~1 for those tiles. There is room for another generation:
indeed, as far as, already noticed, {\bf L}$_i$ and {\bf L}$_j$ do not see each other 
when $i\not=j$. But if {\bf L}$_i$ and {\bf L}$_j$ with $i\not=j$ share an edge~$\sigma$, 
they have a common neighbour~{\bf N}$_{i,j}$: around ~$\sigma$ there must be four tiles:
{\bf L}, {\bf L}$_i$, {\bf L}$_j$ and {\bf N}$_{i,j}$. Consider, for instance, 
{\bf L}$_{11}$, {\bf L}$_6$ and {\bf L}$_7$. We can define {\bf N}$_{11,6}$, 
{\bf N}$_{11,7}$ and {\bf N}$_6,7$. Consider the planes $\Pi_6$, $\Pi_7$ and $\Pi_{11}$
containing the faces~6, 7 and~11 of {\bf L} respectively. For each one, say $\Pi_i$,
{\bf L}$_i$ is in one half-space defined by $\Pi_i$ while {\bf L}$_j$ and {\bf L}$_k$
with \hbox{$\{i,j,k\}=\{6,7,1\}$} are in the other half-space. The intersections of those 
half-spaces define eight regions of the space which we call {\bf octants}. Each octant 
contains a tile touching {\bf L} as far as that situation is the same as the one we can 
see with planes giving rise to $\ell$ and $m$ together with the intersections of the 
half-spaces that they define. We already know seven tiles for seven octants: {\bf L},
{\bf L}$_i$ with $i$ in $\{6,7,11\}$, {\bf N}$_{6,7}$, {\bf N}$_{6,11}$ and 
{\bf N}$_{7,11}$. There is an eighth tile: a common neighbour to those last three tiles,
{\bf N}$_{6,7,11}$. That last tile can see all {\bf N}$_i$'s we have defined, but it
cannot see neither the {\bf L}$_i$'s nor {\bf L} itself. We say that {\bf N}$_{i,j,k}$
with, here, \hbox{$\{i,j,k\}=\{6,7,11\}$} belongs to generation~3.

Now, there is a simple criterion to distinguish those generations: the tiles of
generation~1 have a common face with {\bf L}; those of generation~2 share just a single
edge with {\bf L} while those of generation~3 share just a vertex with~{\bf L}. We have
seen that criterion for the upper crown of~{\bf L}. For the lower crown, things are a bit
different: generation~1 is still defined by the tiles sharing a face with~{\bf L} and
generation~2 by the tiles which share with~{\bf L} one edge only. Taking into account
the tiles which hang below~\HH, we can see that we have all the possible neighbours 
of~{\bf L}. Indeed, consider the vertex~$v$ belonging to two faces of the lower crown and
to a third one of the upper crown. Let $e$ be the edge joining~$v$ to a vertex of 
the face~0 of~{\bf L}. Then the three other tiles sharing with {\bf L} the edge $e$ only
share also~$v$. But, as can be shown on the figure, on each tile, a neighbour of a face
of the upper crown also shares~$v$. So that there is no tile of generation~3 at such a 
vertex. At last, consider a vertex~$w$ belonging to one face~$a$ of the lower crown and to
two faces of the upper crown. Then {\bf L} with its neighbours {\bf L}$_i$ of generation~1
also sharing~$w$ defines three tiles sharing~$w$. The neighbour~{\bf N} of generation~2
defined by those {\bf L}$_i$'s is the fourth tile around the edge~$f$ shared by the two
faces of the upper crown defining~$w$. A similar argument can be formulated for the 
tile~{\bf K} sharing the face~$a$ with~{\bf L} so that we get the eight tiles around~$w$.
So for such a vertex too, there is no tile of generation~3. Accordingly we proved:

\begin{prop}\label{pneighgen}
The neighbours of a tile~{\bf L} whose face~$0$ is on~\HH{} belong to generation~$1$.
The neighbours of the neighbours which share an edge only with~{\bf L} belong
to generation~$2$. There are neighbours of those of generation~$2$ which shares a vertex
only with~{\bf L}: they are the neighbours of generation~$3$. The tiles of generation~$3$
around~{\bf L} touch vertices of~{\bf L} which belong to the upper crown but which do not
belong to the lower crown.
\end{prop}

At last but not the least, we have to see the connection of neighbours of a tile with
the distance from a tile to another one. Say that a sequence of 
tiles~$\{T_i\}_{i\in\{0..n\}}$ is a {\bf path} if $T_i$ and $T_{i+1}$ share a face for 
$0\leq i<n$. In that situation, we say that $n$ is the {\bf length} of the path. We also
say that the path {\bf joins}~$T_0$ to $T_n$ and conversely. The {\bf distance}
between two tiles~$U$ and $V$ is the length of the shortest path joining~$U$ to~$V$.
A {\bf ball} around $T$ of radius~$k$ in the space is the set of tiles whose distance 
to~$T$ is at most~$k$. Those which are at distance~$k$ exactly constitute the 
{\bf sphere} around~$T$ of radius~$k$. Note that two consecutive tiles on a path are
neighbour of generation~1 for one another according to the definition of a path. 

\begin{lem}\label{ldistneigh}
Let $T$ be a tile at distance~$k$ from a tile~$A$. Let $U$ be the tile on a shortest
path from $A$ to~$T$ which is at the distance~$k$$-$1 from~$A$. Define a numbering 
of~$T$ under which the face~0 of~$T$ is the face it shares with~$U$. Then all neighbours
$T_i$ of~$T$ with $i>0$ are at the distance~$k$+1 from~$A$. Let $\Pi_i$ be the plane 
supporting a face~$i$ of~$T$ belonging to its upper crown. Then a shortest path from~$A$
to~$T$ is completely contained in the half-space defined by~$\Pi_i$ which 
contains~$T$.
\end{lem}

\noindent
Proof of the lemma. We proceed by a complete induction on the length of the path.
As the dodecahedron is a convex set, the lemma is true for~$k=0$. Assume that it is
true for~$k$. Let $T$ be a tile at the distance~$k$+1 from~$A$ and let $U$
be a tile on a shortest path from~$A$ to~$T$ which is at distance~$k$. Define the face
$F_0$ shared by~$T$ and~$U$ as the face~0 of~$T$. Let $\Pi_0$ the plane supporting $F_0$.
That half-space defined by $\Pi_0$ which contains~$U$ also contains the path from $A$
to~$U$ by the induction hypothesis. Let $F_j$ be a face of the upper crown of~$T$ and let
$\Pi_j$ be the plane supporting it. If $F_j$ is opposed to~$F_0$, the line joining the
centre of~$F_0$ to that of $F_j$ is a common perpendicular to $\Pi_0$ and to~$\Pi_j$.
Accordingly, those planes are not secant, so that the half-space of~$\Pi_j$ containing~$T$
also contains the half-space of~$\Pi_0$ containing~$U$. Consequently, that half-space
also completely contains the path from~$T_0$ to~$U$. If $F_j$ is not opposed to~$F_0$
there is a unique edge~$s$ joining $F_0$ to~$F_j$. As $s$ is the intersection of two faces
of the lower crown which are, by construction, mutually perpendicular and both 
perpendicular to~$F_0$, $s$ is the common perpendicular line to both~$F_0$ and $F_j$
so that those planes are non-secant. Accordingly me may repeat the previous argument
which allows us to conclude that the half-space defined by $\Pi_j$ which contains~$T$
also contains the whole path from~$A$ to~$T$.

Let~$V$ be a neighbour $T_i$ or~$T$ with $i>0$. Assume that $V$ is at distance~$h$ 
from~$A$ with $h<k$. Consider a shortest path from~$A$ to~$T_i$. As that path and the 
shortest path we considered from~$A$ to~$T$ start both of them from~$A$, there is a 
tile~$B$ on both paths such that the part of those paths from~$A$ to~$B$ is the longest 
common part of the paths from~$A$ to~$T$ and from~$A$ to~$V$. Let~$X_i$, $Y_i$ constitute 
the path from $B$ to~$T$, to~$V$ respectively with $X_0$ and $Y_0$ being neighbours
of~$B$. We know that $X_0$ and~$Y_0$ cannot see each other. Consider $X_1$ and~$Y_1$.
We claim that if $X_1\not=Y_1$ those tiles cannot see each other which is easy to check
on Figure~\ref{fles4}. Accordingly, unless there are few tiles $X_a$, $X_{a+1}$, $X_{a+2}$and $Y_a$, $Y_{a+1}$, $Y_{a+1}$ such that $X_a=Y_a$, $X_{a+1}\not=Y_{a+1}$
and $X_{a+2}=Y_{a+2}$ and as far as $T\not=V$, the conclusion is that $T$ and $V$ 
cannot see each other which is a contradiction with our assumption. That proves the 
lemma. \hfill$\Box$

Note that from the proof of the lemma, we can see that the shortest path from a tile
to another one may not be unique.

The lemma allows us to better see the relation between spheres $\mathcal S$$_k$ of 
radius~$k$ around the same tile. We can see that $\mathcal S$$_{k+2}$ contains 
neighbours of generation~2 of tiles belonging to $\mathcal S$$_k$ and that
$\mathcal S$$_{k+3}$ contains neighbours of generation~3 of tiles belonging 
to~$\mathcal S$$_k$. Accordingly, only spheres $\mathcal S$$_{k+h}$ with $h>3$ have
no contact with $\mathcal S$$_k$.

\section{Main lines of the computation}\label{scenario}

   The first paper about a universal cellular automaton in the pentagrid, the 
tessellation $\{5,4\}$ of the hyperbolic plane, was \cite{fhmmTCS}. That cellular 
automaton was also rotation invariant and, at each step of the computation, the set of 
non quiescent states had infinitely many cycles: we shall say that it is a truly planar 
cellular automaton. That automaton had 22~states. That result was improved by a cellular 
automaton with 9~states in~\cite{mmysPPL}. Recently, it was improved with 5~states, 
see~\cite{mmpenta5st}. A bit later, I proved that in the heptagrid, the tessellation 
$\{7,3\}$ of the hyperbolic plane, there is a weakly universal cellular automaton with 
three states which is rotation invariant and which is truly planar, \cite{mmhepta3st}. 
Later, I improved the result down to two states but the rules are no more rotation 
invariant, see~\cite{mmpaper2st}. Paper \cite{JAC2010} constructs three cellular
automata which are strongly universal and rotation invariant: one in the pentagrid, one 
in the heptagrid, one in the tessellation \hbox{$\{5,3,4\}$} of the hyperbolic 
$3D$-space. By strongly universal we mean that the initial configuration is finite, 
{\it i.e.} it lies within a large enough circle. Recently, I succeeded to implement a 
strongly universal cellular automata in the heptagrid, the tessellation $\{7,3\}$ of
the hyperbolic plane, which improves that latter result: the automaton is rotation 
invariant and it has seven states, see~\cite{mmarXiv21,mmJAC2021}.

    In the present paper, we mainly follow the construction of those papers. However,
it is not true to see this paper as a simple transposition of those previous ones in
the dodecagrid. Here, we take advantage of the third dimension to simplify a few points
of the simulation. However, the third dimension also entails its constraints which make 
it difficult to strongly reduce the number of states: the rotation invariance of the 
rules is a heavy constraint. There are sixty rotations which leave a dodecahedron 
globally invariant. As indicated later, we cannot replace that constraint by the 
invariance under a set of generators.

We simulate a register machine, not necessarily using
the property that two registers are enough to get the strong universality, a result
proved by Coke-Minsky in the sixties, see~\cite{minsky}. 

    The simulation is based on the railway model devised in~\cite{stewart} revisited
by the implementations given in the author's papers, see for instance~\cite{mmpaper2st}.
Sub-section~\ref{railway} describes the main structures of the model. We mainly borrow
its content from previous papers for the sake of the reader. In 
Sub-section~\ref{newrailway} we indicate the new features used in the present simulation.

\subsection{The railway model}\label{railway}

   The railway model of~\cite{stewart} lives in the Euclidean plane. It consists of
{\bf tracks} and {\bf switches} and the configuration of all switches at time~$t$
defines the configuration of the computation at that time. There are three kinds of
switches, illustrated by Figure~\ref{switches}. The changes of the switch configurations
are performed by a locomotive which runs over the circuit defined by the tracks and their
connections organised by the switches.

A switch gathers three tracks $a$, $b$ and~$c$ at a point. In an active crossing,
the locomotive goes from~$a$ to either~$b$ or~$c$. In a passive crossing, it goes
either from~$b$ or~$c$ to~$a$. 

\vskip 10pt
\vtop{
\ligne{\hfill
\includegraphics[scale=0.8]{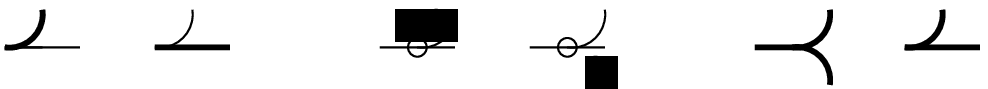}
\hfill}
\begin{fig}\label{switches}
\leurre
The switches used in the railway circuit of the model. To left, the fixed switch, in the
middle, the flip-flop switch, to right the memory switch. In the flip-flop switch, the 
bullet indicates which track has to be taken.
\end{fig}
}

In the fixed switch, the locomotive goes from~$a$ 
to always the same track: either~$b$ or~$c$. The passive crossing of the fixed switch is
possible. The flip-flop switch is always crossed actively only. If the locomotive
is sent from~$a$ to~$b$, $c$ by the switch, it will be sent to~$c$, $b$ respectively at 
the next passage. The memory switch can be crossed actively or passively. Now, the track 
taken by the locomotive in an active passage is the track taken by the locomotive in the 
last passive crossing. Of course, at the initial time of the computation, for the 
flip-flop switch and for the memory one, the track which will be followed by the 
locomotive is defined by the implementation.

   As an example, we give here the circuit which stores a one-bit unit of information,
see Figure~\ref{basicelem}. The locomotive may enter the circuit either through the 
gate~$R$ or through the gate~$W$.

\vtop{
\ligne{\hfill
\includegraphics[scale=0.6]{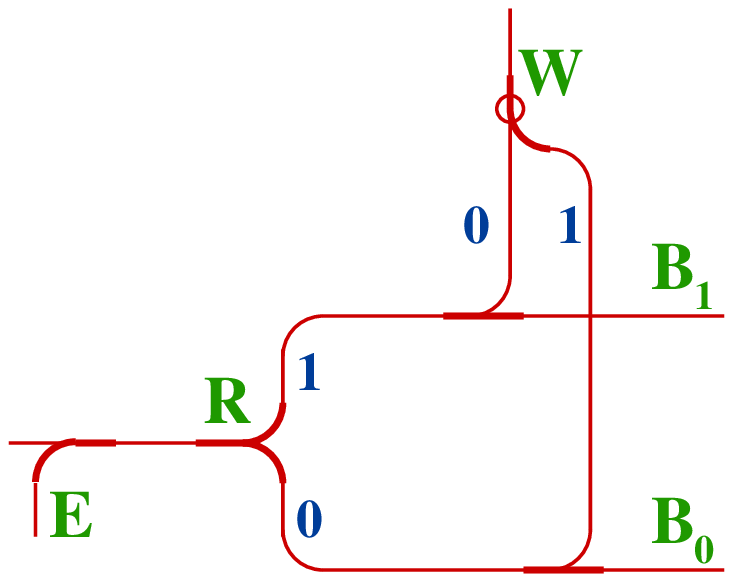}
\hfill}
\begin{fig}\label{basicelem}
\leurre
The basic element containing one bit of information.
\end{fig}
}

  If it enters through the gate~$R$ where a memory switch sits, it goes either through
the track marked with~1 or through the track marked with~0. When it crossed the switch
through track~1, 0, it leaves the unit through the gate~$B_1$, $B_0$ respectively.
Note that on both ways, there are fixed switch sending the locomotive to the appropriate
gate~$B_i$. If the locomotive enters the unit through the gate~$W$, it is sent to the 
gate~$R$, either through track~0 or track~1 from~$W$. Accordingly, the locomotive
arrives to~$R$ where it crosses the switch passively, leaving the unit through the 
gate~$E$ thanks to a fixed switch leading to that latter gate. When the locomotive 
took track~0, 1 from~$W$, the switch after that indicates track~1, 0 respectively and the 
locomotive arrives at~$R$ through track~1, 0 of~$R$. The track are numbered according to 
the value stored in the unit. By definition, the unit is~0, 1 when both tracks from~$W$ 
and from~$R$ are~0, 1 respectively. So that, as seen from that study, the entry 
through~$R$ performs a reading of the unit while the entry through~$W$, changes the unit
from~0 to~1 or from~1 to~0: the entry through~$W$ should be used when it is needed to
change the content of the unit and only in that case. The structure works like a memory
which can be read or rewritten. It is the reason to call it the {\bf one-bit memory}.

   We shall see how to combine one-bit memories in the next sub-section as far as we 
introduce several changes to the original setting for the reasons we indicate there.

\subsection{Tuning the railway model}\label{newrailway}

   We start our presentation with a look on the global aspect of the simulation 
illustrated by Figure~\ref{fglobconfig}. On a part of~\HH, we have the program: a green
quadrangle. On another part, we have two boxes and, attached to each one of them a
long segment of line which represents a register. As two registers are enough to simulate
a Turing machine, see~\cite{minsky}, our illustration contains two registers only.
At last, and not at all the least, segments of line go in between the program and the 
registers. We can imagine that some of them go in that direction and that the others 
go from the registers to the program. Those segments represent {\bf tracks} we study
in the next subsection. They play a key role as far as they convey an information to
the register and go back to the program in order to give a feed back which will determine
the next information to bring to the registers.

\vskip 10pt
\vtop{
\ligne{\hfill
\includegraphics[scale=0.4]{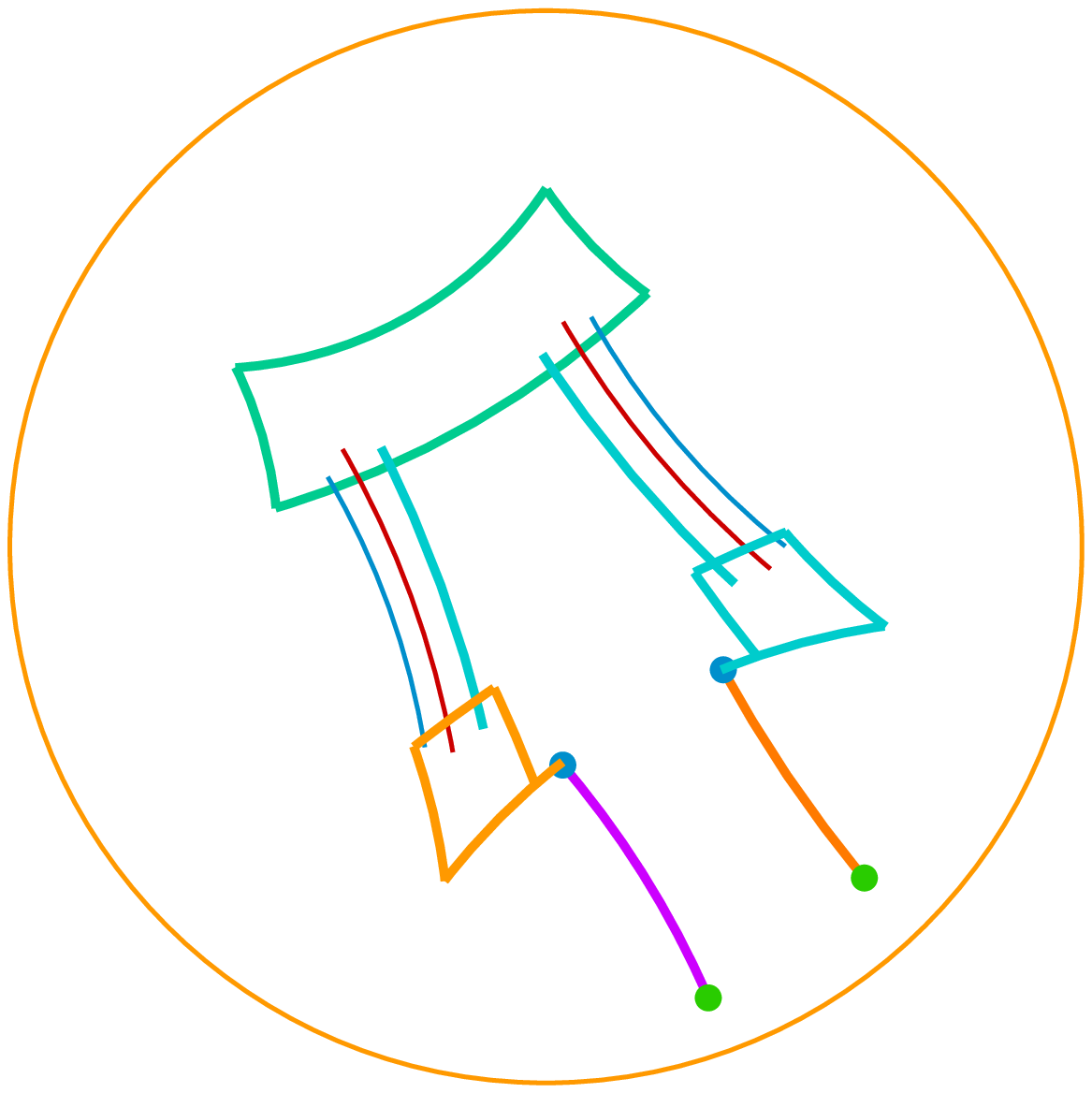}
\hfill}
\begin{fig}\label{fglobconfig}
\leurre
A global view on the simulation.
\end{fig}
}

   We first look at the implementation of the tracks in Sub-subsection~\ref{sbbtracks}
and how it is possible to define the crossing of two tracks.
In Sub-subsection~\ref{sbbswitch} we see how the switches are implemented.
Then, in 
Sub-subsection~\ref{sbbunit}, we see how the one-bit memory is implemented in the new 
context and then, in Sub-section~\ref{sbbregdisp}, how we use it in various places. 
At last but not the least, we shall indicate how registers are implemented
in Sub-subsection~\ref{sbbreg}.

\subsubsection{The tracks}\label{sbbtracks}

    We mentioned the key role played by the tracks in the computation. Without them,
the locomotive could not perform any operation on the registers.
Without them any computation is impossible.  Moreover, as can be seen in many papers of 
the author, that one included, it is not an obvious issue which must always be addressed.
  
    It is not useful to list the similarities and the distinctions between the present 
implementation and those of my previous papers. The best is to explain the implementation
used by this paper. If the reader is interested by the comparison with previous 
implementations the references already indicated give him/her access to the corresponding
papers. We just mentioned that the presentation of the present paper significantly
simplifies what is written in~\cite{mmarXiv21}.

   The tracks are one-way. It is useful to reduce the number of states but it raises
a problem as far as a two-way circulation is required in some portions of the circuit.
It is a point where the third dimension comes to help us. In many portions, the
circuit can be implemented on a fixed plane we call \HH, already mentioned in the 
introduction. Roughly speaking, the traffic in one direction will occur over
\HH{} while the reverse running will be performed below \HH. Occasionally and locally, 
we shall use a plane \VV{} which is orthogonal to~\HH. Note that if \HH{} is fixed
in all the paper, \VV{} may change according to the context where it is mentioned.

The present organisation of the one-way tracks is very different from that 
of~\cite{mmpaper2st}. Contrarily to that paper, we go back to mark the elements of a 
path by milestones. The milestones are blue, in contrast with the blank colour of most 
of the cells: all of them outside a large enough disk containing the initial 
configuration. The blank is the quiescent state of our cellular automaton: if a cell is
blank and if all its neighbours are blank too, the cell remains blank. 
We say {\bf tracks} for portions of a path which follow a line~$\ell$
defined by a side of the tiles in~\HH. Combining several tracks gives rise to paths.
On the path, a blue locomotive is running which consists of a single blue cell which
moves in between blue milestones. Figure~\ref{ftracks} illustrates the constitution of 
the tracks. On the figure, the lines are followed by the side~1 of the elements of tracks.

\vskip 10pt
\vtop{
\ligne{\hfill
\raise 20pt\hbox{\includegraphics[scale=0.4]{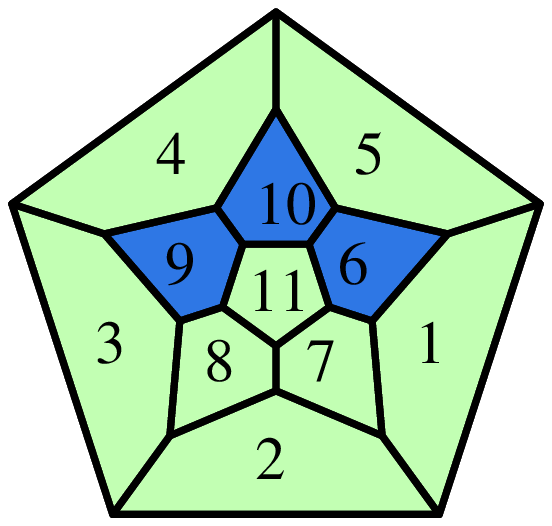}}
\includegraphics[scale=0.8]{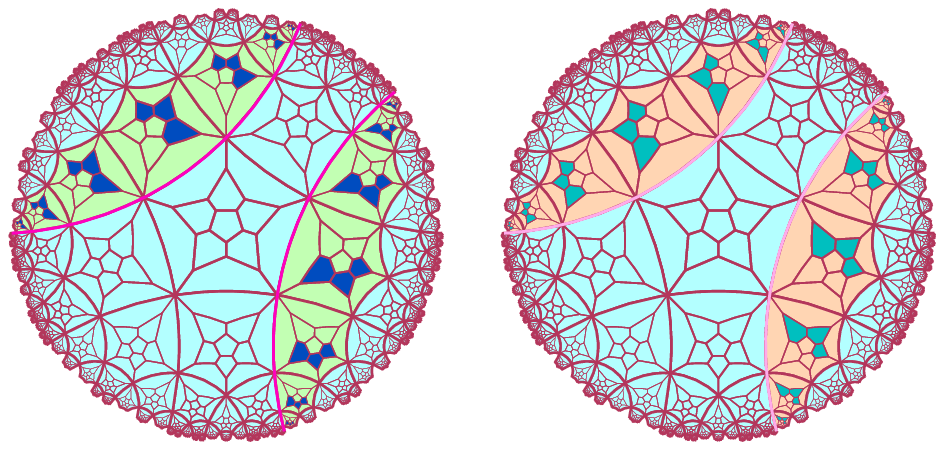}
\hfill}
\begin{fig}\label{ftracks}
\leurre
Leftmost picture: an element of a track. Middle picture: tracks over \HH.
Rightmost picture, return tracks of the previous ones below \HH. Note the lines on both
picture, in the middle and to right.
\end{fig}
}

The leftmost picture of the figure shows us that an element of a track consists
of a white dodecahedron $\Delta$ whose face~0 lie on~\HH{} and on the faces~6, 9 and~10
of~$\Delta$ we have three dodecahedrons $\Delta_6$, $\Delta_9$ and $\Delta_{10}$ which
are blue. Those dodecahedrons close to the face opposite to face~11, opposite to face~0, 
are called the {\bf decoration} of~$\Delta$. In general, the 
locomotive enters an element of the track through its face~5 or~4 and it leaves the 
element through its face~2. In order to allow
variations which we further describe, for instance in Figure~\ref{fpaths}, the 
locomotive may also enter through face~1, the exit face being always face~2.

The middle picture of Figure~\ref{ftracks} shows us two tracks which lie on \HH. As 
clearly seen, the left-hand side track follows a line of~\HH, in mauve in the pictures, 
which contains an edge of the face~1 of all the elements constituting that track. 
A similar remark can be formulated for the right-hand side track. According to the 
convention we indicated on the way a locomotive may cross an element of a track, the 
left-hand side track is going down while the right-hand side one is going up. The 
rightmost picture of the figure shows us tracks which are below~\HH, their face~0 also 
belonging to~\HH. Imagine that the tracks of the middle picture are removed and that we 
see those tracks below~\HH{} from above~\HH{} as if that plane were translucent. We 
can see that the image we need is an image of~$\Delta$ under a symmetry in the plane 
orthogonal to~\HH{} which
contains the opposite edges 1.5 and 8.9 of~$\Delta$. The numbering of the faces is thus
increasing  while counter-clockwise turning around face~0 while the numbering is
clockwise on Figure~\ref{fdodecs}. Note that below~\HH, the direction of the tracks is
opposite to that of the tracks upon~\HH{} which follow the same lines. It is an important
feature which allows us to define a two-way path: one direction is performed upon~\HH{}
while the opposite one occurs below~\HH.

\vskip 10pt
\vtop{
\ligne{\hfill
\includegraphics[scale=0.35]{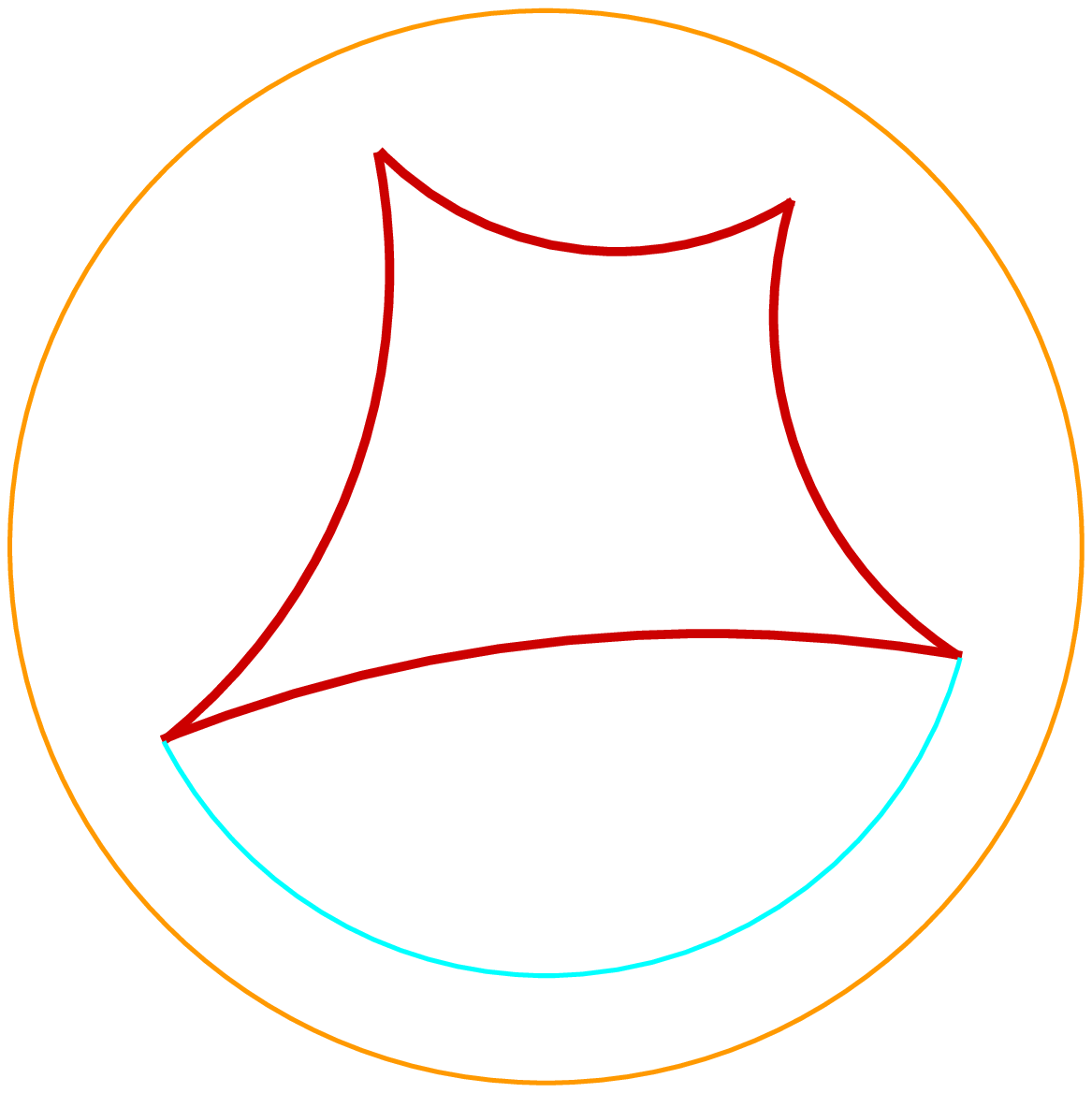}
\hfill}
\vskip-20pt
\ligne{\hfill
\includegraphics[scale=1]{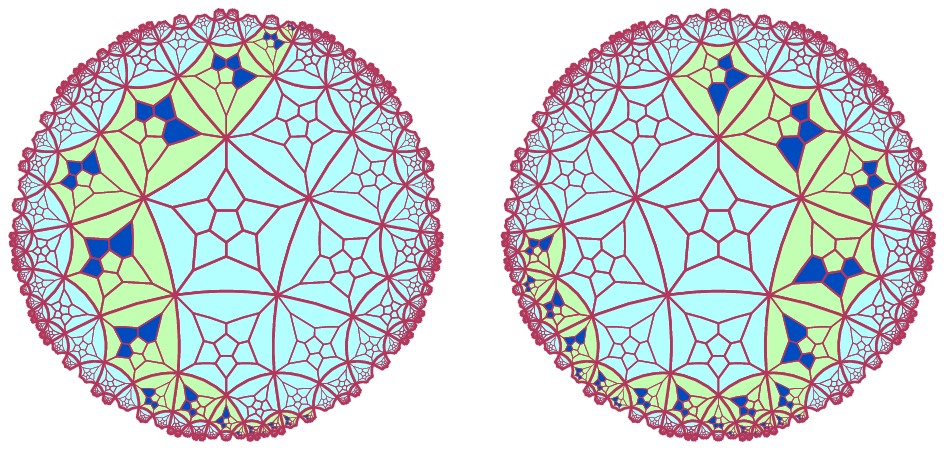}
\hfill}
\vskip-20pt
\begin{fig}\label{fpaths}
\leurre
Basic patterns for the paths. Above, a scheme of a path constituting a quadrangle.
Below: to left, left-hand side top corner, to right, right-hand side bottom corner.
\end{fig}
}

Moreover, in most parts of the circuit we shall 
describe a single locomotive is running on the circuit. In particular, in the case
when a two-way track occurs in some part of the circuit, there can never be a locomotive
in a track~$\tau$ over~\HH{} and another one, at the same time, in the track below $\tau$.

Figure~\ref{fpaths} illustrates how pieces of tracks indicated in Figure~\ref{ftracks}
can be organised into paths. As long as it will be possible, paths for the locomotive
will follow a straight line of the tiling which is the trace of the dodecagrid on~\HH.
However, as the locomotive goes from the program to the register and back, such a circuit
is organised along a kind of quadrangle $Q$ in the hyperbolic plane as illustrated by 
the top picture of Figure~\ref{fpaths}.

On that picture, we can see two paths for drawing the bottom side of~$Q$.
The red one is a segment of a straight line in the hyperbolic plane. The blue one is an
arc of a hyperbolic circle. On the model, the arc is longer than the segment. In fact,
the arc is much longer: its length is proportional to an exponential function of the
radius of the circle supporting the arc. The bottom pictures of the figure show us 
a zoom on two corners of the quadrangle: the left-hand side picture illustrates the
right-hand side bottom corner. By rotating those pictures we can see how to connect
the sides of~$Q$. Note that the right-hand side picture connects a segment of a straight 
line with an arc of a circle. The elements with an entry through side~1 play an important
role in the realisation of an arc of circle and on the connection between segments of 
different directions.

   We can see that those constructions are flexible enough, so that we have a relative
freedom in the construction of the circuit.

\subsubsection{The switches}\label{sbbswitch}

The section follows the implementation described in~\cite{mmarXiv21,mmJAC2021}. We 
reproduce it here for the reader's convenience. The illustration show us what we
call an {\bf idle configuration}. The view given by such pictures is a window focusing
on what we call the centre of the switch and we can see the tiles on~\HH{} within a
circle of radius~3 from the central tile, what we call the {\bf window}. An idle 
configuration is a configuration where there is no locomotive within the just defined 
window. The left-hand side of Figure~\ref{fstab_fxfk} shows us such an idle 
configuration for the passive fixed switch. From what we said in 
Section~\ref{newrailway}, we know that there is no active fixed switch, so that 
illustration of such a switch concerns the passive one only.

We can see that the central tile is an ordinary element of the track. As a locomotive
may enter through its face~5 or through its face~4, the constitution of the switch is
easy: one branch of the switch arrives at face~5{} in the central tile while the other
branch arrives through face~4. We have a single locomotive in the window focused on the
centre of the fixed switch.

Two important structures are required for implementing the remaining swit-ches :
the fork and the controller. We start with the fork and we examine the case of the
controller after the implementation of the flip-flop switch.

The right-hand side picture of
   Figure~\ref{fstab_fxfk} illustrates the implementation of the {\bf fork} in its idle
configuration. That structure receives one locomotive and it yields two ones which leave 
the structure in opposite directions. As can be seen on the figure, the central cell 
differs from an element of the track: its decoration consists of five dodecahedrons
which are placed on faces~6, 7, 8, 9 and~11. Moreover, the locomotive enters the
tile through its face~2 and the two new locomotives exit through faces~4 and~5.

The flip-flop switch and both parts of the memory switch require a much more involved
situation. The global view of an idle configuration is illustrated by 
Figure~\ref{fschflfl}.

\vskip 10pt
\vtop{
\ligne{\hfill
\includegraphics[scale=1]{hyp3d_fix.ps}
\includegraphics[scale=1]{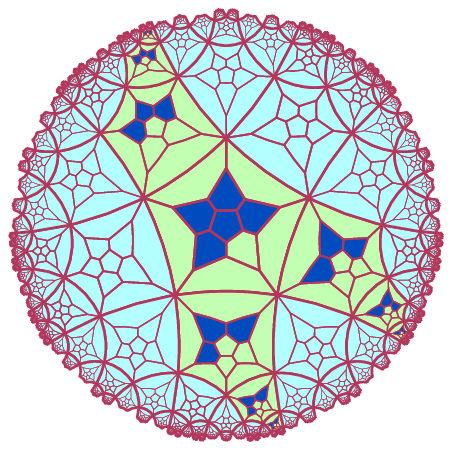}
\hfill}
\vskip-20pt
\begin{fig}\label{fstab_fxfk}
\leurre
Idle configurations, to left of a passive fixed switch, to right of a fork.
\end{fig}
}

\vskip 10pt
\vtop{
\ligne{\hfill
\includegraphics[scale=0.45]{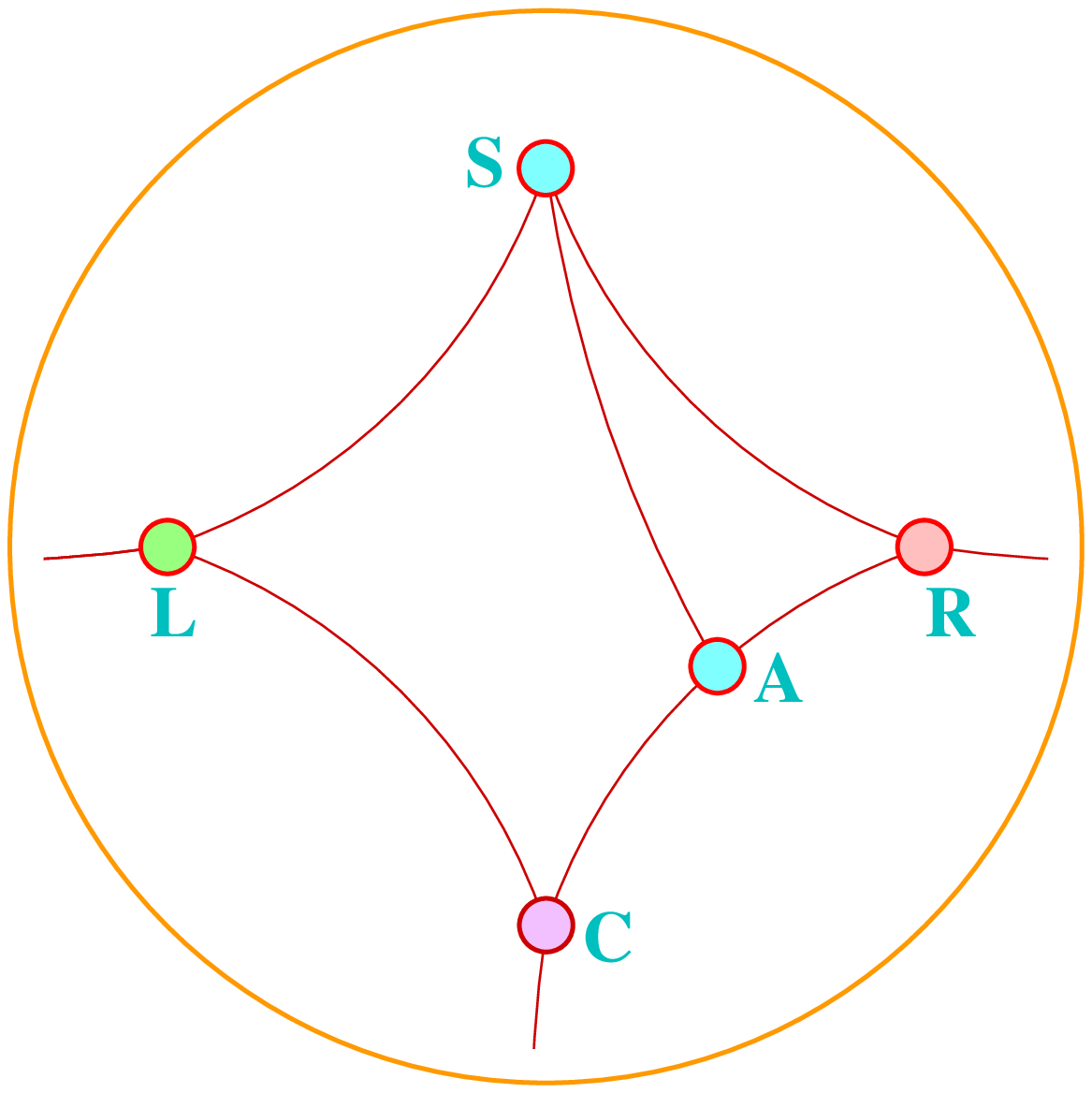}
\raise 50pt\hbox{\includegraphics[scale=0.25]{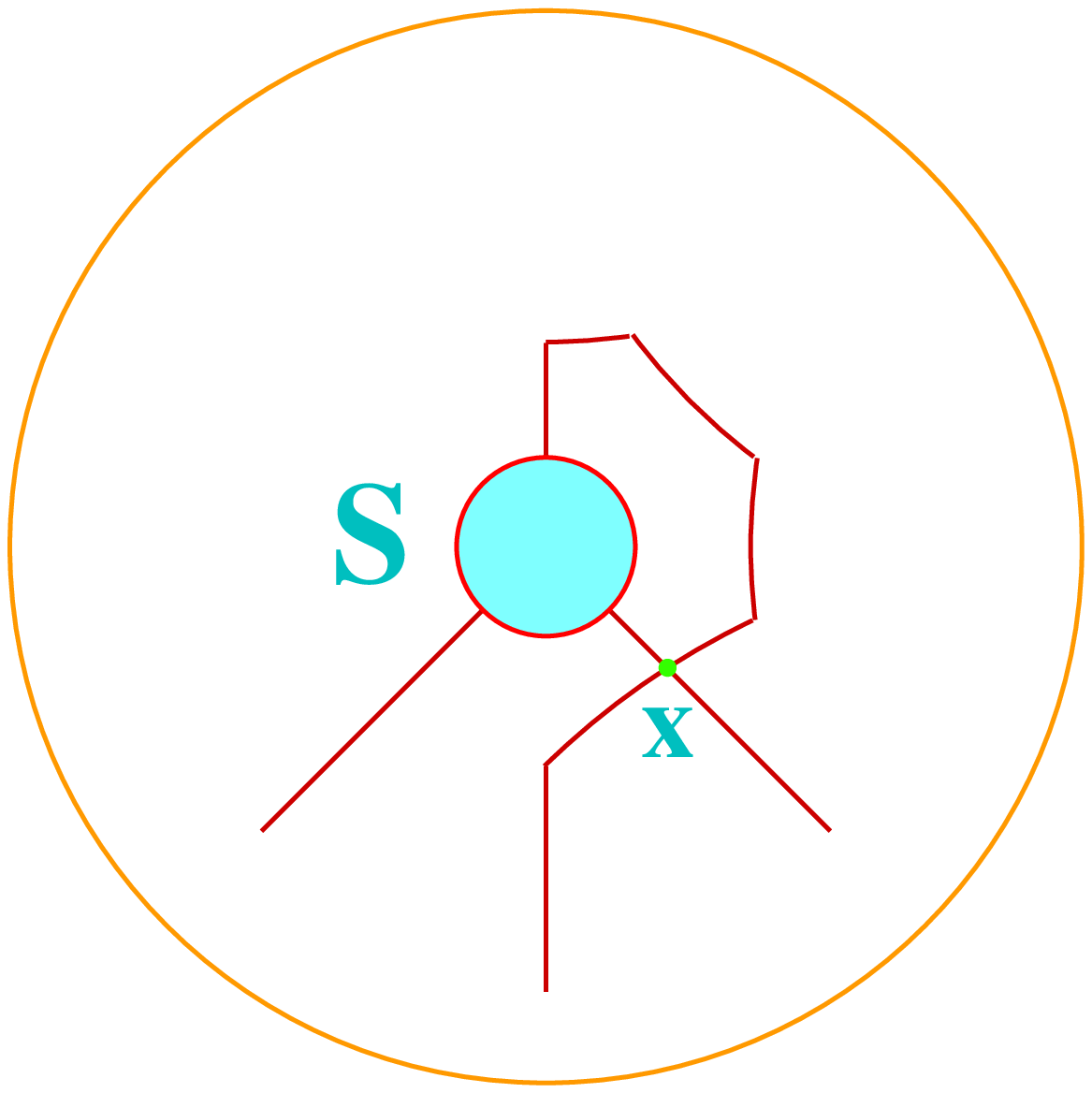}}
\hfill}
\begin{fig}\label{fschflfl}
\leurre
Scheme of the implementation of a flip-flop switch with, to right, a zoom on~{\bf S}.
In the zoom, note the crossing~{\bf x} of a leaving track by the deviated route 
to~{\bf S}.
\end{fig}
}

The locomotive arrives through a segment of a straight line by~$C$ where a fork sits.
Accordingly, two locomotives leave~$C$, one of them towards~$L$, the other towards~$R$.
At~$L$ a controller sits and, on the figure, it let the locomotive go further on the
segment of straight line. Note that the path from~$C$ to~$L$ is also a segment of a
straight line on the figure, which is conformal to the implementation. Now, the structures
which are later involved make the length of that segment to be huge. At~$R$ too a 
controller is sitting but, on the figure, it kills the locomotive which is thus prevented
to go outside the switch. On the way from~$C$ to~$R$ the locomotive meets another fork
at~$A$. The fork sends one of the new locomotives to~$R$ where it is stopped in the 
situation illustrated by the figure and the other is sent to~$S$. There a fork is 
sitting too which sends two locomotives, one of them to~$L$, the other to~$R$. When they
reach their goal, through another face of the controller, the locomotives change the 
configuration of the controller which is sitting there. The controller which let the 
locomotive go will further stop it while the one which stopped it will further let it go.
Accordingly, after the passage of a locomotive and after a certain time, the 
configuration of the flip-flop switch is that we described in Sub-section~\ref{railway}.
We may arrange the circuit so that a new passage of the locomotive at that switch will
happen a long time after the change of function of its controllers is completely changed.

We have now to clarify the implementation of the controller. It is illustrated by 
Figure~\ref{fctrl}.

\vskip 10pt
\vtop{
\ligne{\hfill
\includegraphics[scale=1]{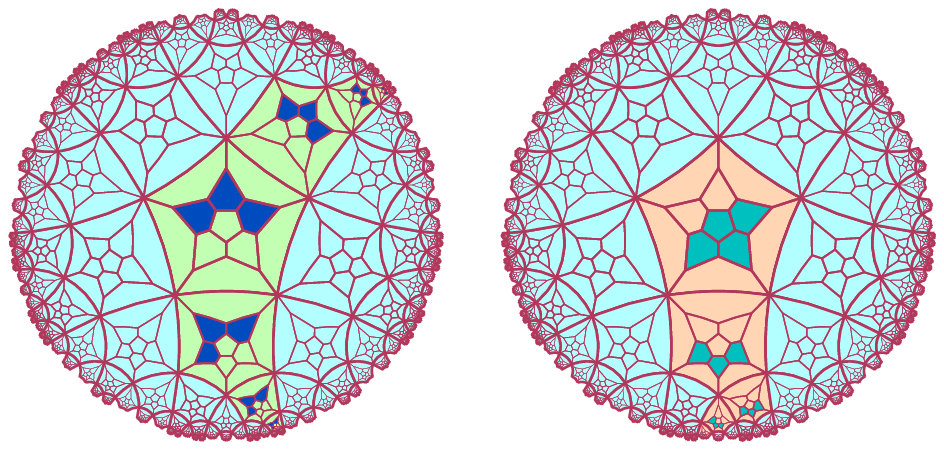}
\hfill}
\begin{fig}\label{fctrl}
\leurre
The idle configuration of the controller. To left, upon~\HH, the track; to right, 
below~\HH, the controller and its access by a locomotive.
\end{fig}
}

Here again, the central cell below~\HH{} is not an element of the track. Its decoration
consists of four blue dodecahedrons which are placed on the faces~6, 9, 10 and~11 of
the tile.

   Figure~\ref{fschflfl} requires the explanation of the zoom: three tracks abut the 
point~{\bf S} which is supposed to behave like a fork. However, the configuration of 
the tracks abutting the central tile of a fork is different. It is the reason for which 
the track arriving to~{\bf S} from~{\bf A} is deviated near~{\bf S} as indicated in the
zoom so that the new configuration is conformal to that of a fork. Note that the pieces
constituting the deviation of the track are segments of straight line in Poincar\'e's
disc model. The price to pay is a crossing at~{\bf x} on the picture illustrating the 
zoom. 

   The crossing happens to be a burden in the hyperbolic plane requiring several complex
structures we do not need in the hyperbolic $3D$-space. The third dimension offers
two possible ways to easily realize a crossing: the bridge or the tunnel. In the present
paper, I have chosen the tunnel, illustrated by Figure~\ref{ftunnel}. Two tracks follow
a line: a red one going from right to left on the three pictures and a blue line, from
bottom to top on the left-hand side picture only.

\vskip 10pt
\vtop{
\ligne{\hfill
\includegraphics[scale=0.8]{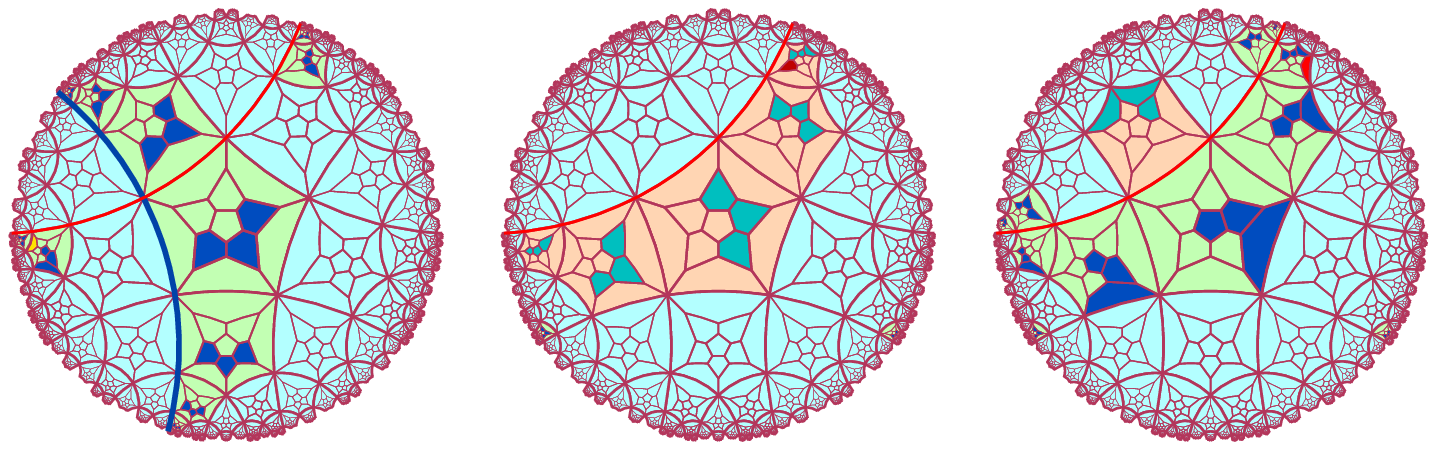}
\hfill}
\begin{fig}\label{ftunnel}
\leurre
The idle configuration of the tunnel. 
\end{fig}
}

The leftmost picture of the figure is the standard projection upon~\HH, where each 
dodecahedron is projected within the face it shares with~\HH. In the picture, the faces
of the elements of the track which lie on~\HH{} are faces~0. Using the numbering of the 
tiles of~\HH, the elements of the track following the blue line, going from bottom to 
top on the picture are, in this order : sector 3, tiles 21, 8, 3 and~1; the central tile, then, in sector~1, tiles~1, 4, 12 and 33. Six other tiles can be seen : in sector~2,
tiles 3, 8 and 20 and, in sector~5, 33, 12 and~4 as the track goes from
right to left, following the red line~$\ell$. Three tiles are missing for that track: 
tile~1 of sector~5, the central tile and tile~1 of sector~2. Those tiles are not 
upon~\HH{} but below that plane.  They are illustrated by the middle picture where $\ell$
is again represented. It is the part of the crossing track which passes under the track 
illustrated by the leftmost picture. In the middle picture we can see those elements of 
track as if \HH{} were translucent. Two additional tiles are indicated in the middle 
picture: tile~4 of sector~5 and tile~3 of sector~2. Those tiles correspond to the tiles 
with the same numbers in the same sectors which lie upon~\HH. Each of those tiles upon 
and below~\HH{} allows a locomotive going upon~\HH{} to go below that plane in order 
to cross the other track. Those elements require the locomotive to enter through a 
face~1 and to exit through a face~2.

The rightmost picture of Figure~\ref{ftunnel} shows us the projection of the tunnel
on the plane~\VV{} which is orthogonal to~\HH, cutting that latter plane along the 
line~$\ell$.  We represent on~\VV{} the tiles which have a face on it only. Those which 
are on the same side of~\VV{} as the central tile in the middle picture are seen in direct
projections. Those which are on the other side are seen as if \VV{} were translucent.

Figure~\ref{fprojs} illustrates the possible projections of an element of the track 
depending on whether it is seen upon~\HH, upon~\VV{} or through those planes by 
transparency, also depending on which face is on~\HH{} and which one on~\VV.

Consider the central tile of the leftmost picture of Figure~\ref{ftunnel}. It is
projected on its face~0 on~\HH{} and the locomotive goes upward, entering through face~5 
and leaving through face~2. Consider the entrance into the tunnel. We can see that
the entry of the locomotive through face~1 and its leaving through face~2 cannot be
used on both ends of the tunnel: the reason is that, at the entry, the neighbour~6 of the
tile below~\HH{} would be below face~0 of the tile of the track arriving to the entry
over~\HH: that would stop the locomotive. That problem does not occur for the exit,
so that we can use that way of working for the exit. For the entry, we need to change
the tile in order to induce rules which would be rotationally compatible with the others
and this time the entry into the tile below~\HH{} is through face~7 and the exit still 
through face~2. That tile differs from a tile of the track by the occurrence of red 
neighbour on its face~3.

\vskip 10pt
\vtop{
\ligne{\hfill
\includegraphics[scale=0.4]{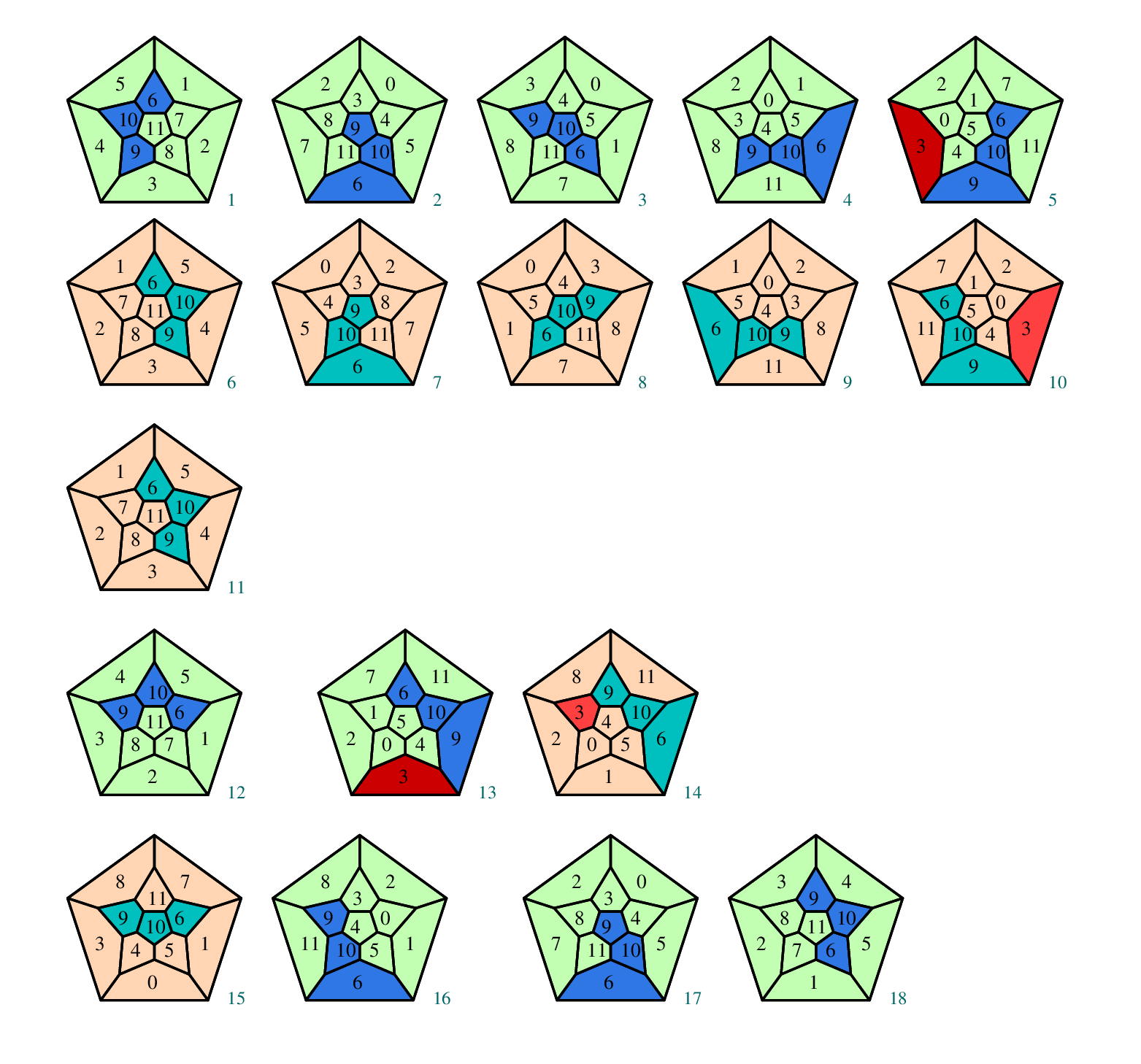}
\hfill}
\begin{fig}\label{fprojs}
\leurre
Projection of various positions of the element track upon~\HH, \VV{} and also of 
elements below \HH.
\end{fig}
}

Figure~\ref{fprojs} gives various views on the tiles we just mentioned. The first
row of the picture indicates the projections over \HH{} of tiles of the track
depending on the face which lies on ~\HH: face~0, face~1, face~2 and face~7
for the pictures~1, 2, 3 and~4 respectively on the first row of the figure. Picture~5
is the projection of the tile of the entry below~\HH, projected above~\HH{} on its face~7.
The second row of the figure, pictures~6 up to~10 gives the same tiles viewed through
a translucent \HH. Picture~11, third row of the figure, illustrates an element of the 
tunnel below~\HH{} seen through a translucent \HH. The fourth row illustrates the tiles
for the entry into the tunnel: picture~12 for the tile above~\HH{} and pictures~13
and~14 for the entry below~\HH{} when it is projected on~\VV, picture~13, on its face~8,
and when it is viewed through a translucent~\HH, picture~14, through its face~7.
The last row of the figure illustrates the tiles for the exit from the tunnel. 
Pictures~15 and~16 illustrate the exit tile below~\HH: viewed through a translucent 
face~2 being on~\HH, while picture~16 shows us the projection on its face~0 over~\VV.

   To conclude with the tunnel, the occurrence of a locomotive in the central tile of
the tunnel will not stop a locomotive on the upper way: first, the locomotive never
stops on an element of the track and, more over, in such a crossing there is
a single locomotive in a window around the central point of the crossing: if it is
present on one path, it cannot be present on the other one.
\vskip 10pt
   We are now in position to deal with the memory switch, active and passive parts.

   First, we deal with the active part. It looks like the flip-flop switch with this 
difference that there is no fork~$A$ in between the path from the initial fork~$C$
to~$R$, one of the controllers. Figure~\ref{fmemo} illustrates both parts of the 
memory switch: to left of the figure, the active part of the switch, to right, its passive
part.

\vskip 10pt
\vtop{
\ligne{\hfill
\includegraphics[scale=0.4]{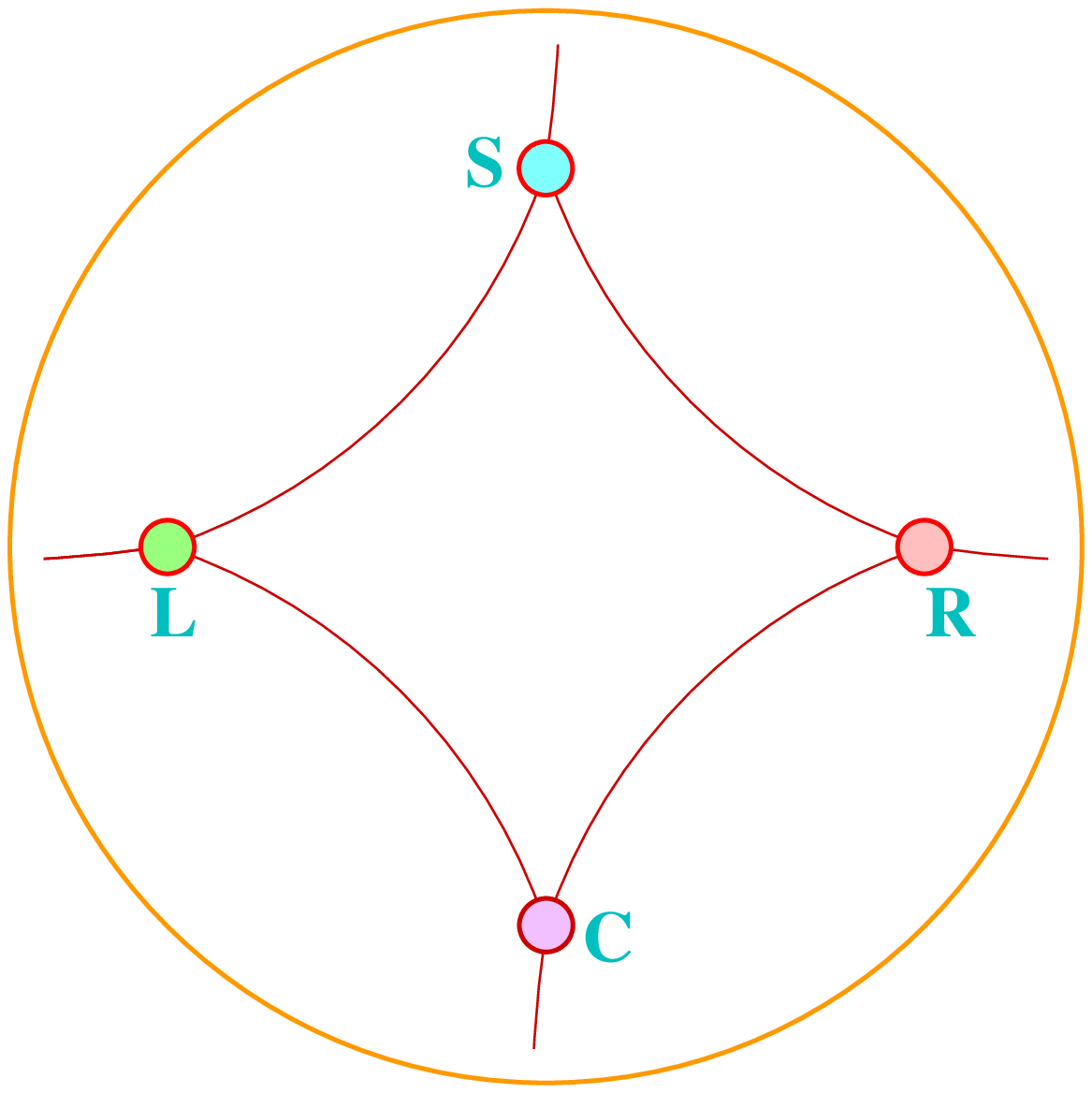}
\includegraphics[scale=0.4]{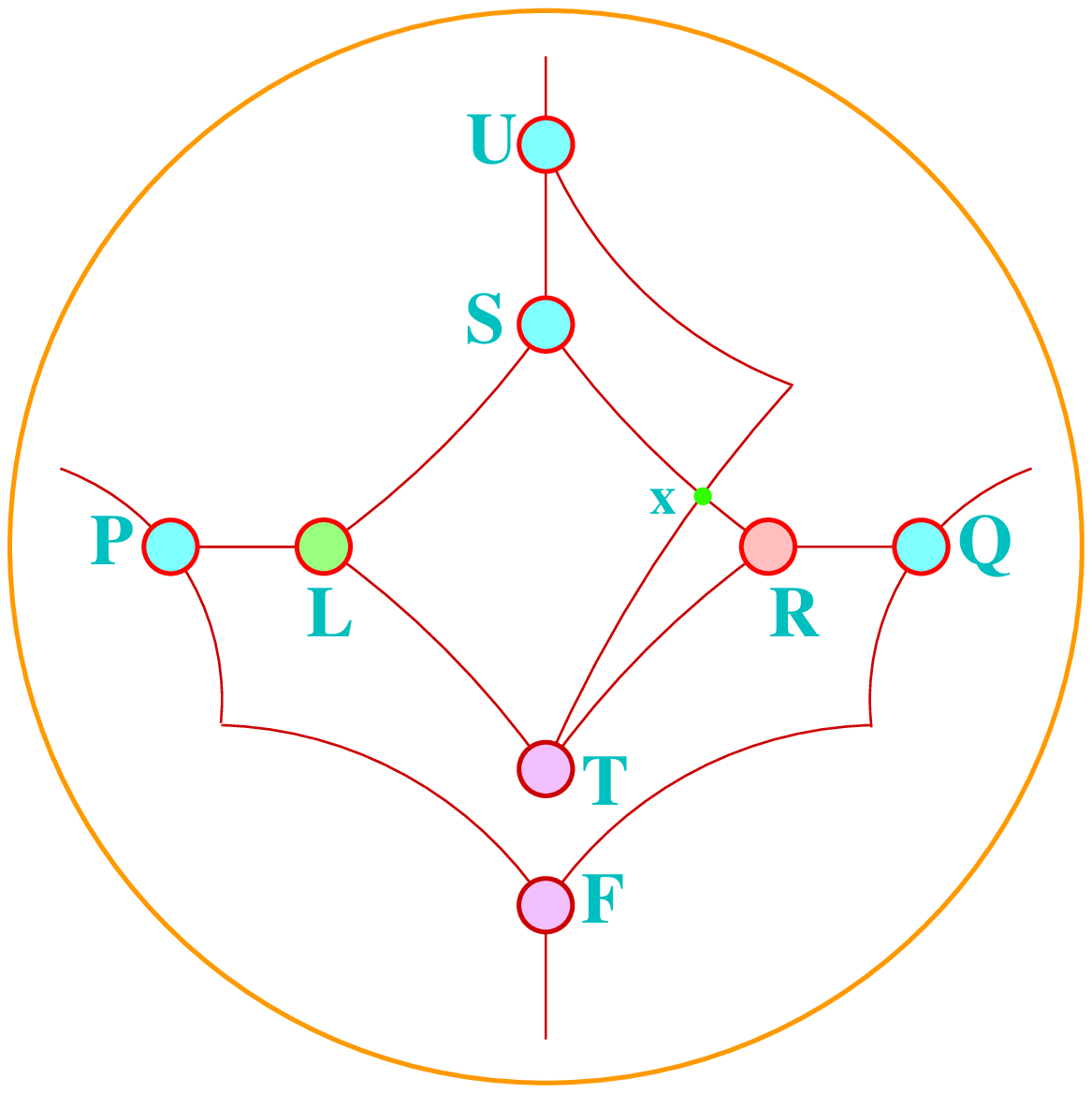}
\hfill}
\begin{fig}\label{fmemo}
\leurre
To left, the active memory switch, to right, the passive one.
\end{fig}
}

In the active part of the switch, left-hand side picture of Figure~\ref{fmemo}, 
the locomotive arrives to a fork sitting at~$C$. From there two locomotives are sent,
one to~$L$, the other to~$R$ and the working of the switch at this point is alike that
of a flip-flop switch. The difference lies in the fact that the passage of the locomotive
does not trigger the exchange of the roles between the controllers. That change is 
triggered by the passage of a locomotive through the non-selected track of the passive
part of the switch. When it is the case, a locomotive is sent from the passive part
to the active one. That locomotive arrives at the fork which is sitting at~$S$. The fork
creates two locomotives which are sent to~$L$ and~$R$ in order to change the permissive
controller to a blocking one and to change the blocking one into a permissive one.

Let us look at the working of the passive part. The locomotive arrives to the switch 
through~$P$ or through~$Q$. Assume as it is through~$P$. A fork sitting at~$P$ sends a
locomotive to the fixed switch~$F$ which let the locomotive leave the switch. The other
locomotive sent by~$P$ goes to~$L$. If that side is that of the selected track, the
controller sitting at~$L$ blocks the locomotive so that no change is performed, neither in
the passive switch, nor in the active one. Accordingly, the selected track of the active
switch is controlled by a permissive controller while the corresponding selected track
of the passive switch is controlled by a blocking controller. Presently, assume that
the side of~$L$ is not that of the selected track. It means that $L$ let the locomotive
go to~$T$ where a fixed switch sends the locomotive to a fork at~$U$. That fork
sends a locomotive to the fork~$S$ of the active switch and the other locomotive is sent
to~$S$ of the passive switch. At that point~$S$, a fork sends a locomotive to~$L$ and
another one to~$R$ in order to change the working of both controllers to the
opposite task. As a parallel change occurs in the active switch the selected track
is redefined in both parts of the switch.

   That working raises several remarks. First, the role of the controllers in the active
and in the memory switches are opposite. Nevertheless, in both cases, the same
programmable controller is used exactly because it is programmable in the way we just
described. The second remark is that all the tracks indicated in the pictures are pieces
of straight lines of the hyperbolic plane. A last remark is that we used one crossing, two
fixed switches, at $F$ and at~$T$ and four forks, at~$P$, $Q$, $U$ and~$S$. Each
structure requires some space, at least a disc whose radius is the length of four tiles
aligned along a straight line. Consequently, the passive memory switch requires a huge
amount of tiles. Again, we may assume that a new passage of the locomotive to the
switch happens after the changes has been performed when it is the case they should 
occur.

\subsubsection{The one-bit memory}\label{sbbunit}

\def\WW{{\bf W}}
\def\RR{{\bf R}}
\def\EE{{\bf E}}
\def\bbz{{\bf b0}}
\def\bbu{{\bf b1}}
\def\zz{{\bf 0}}
\def\uu{{\bf 1}}
It is now time to implement the one-bit memory. Figure~\ref{fonebit} illustrates the
construction for the implementation of Figure~\ref{basicelem} in the dodecagrid.

We can see the active memory switch at~$R$ and the passive one at~$E$. The dark letters
which stand by the blue circle indicate {\bf gates} of the one-bit memory: \WW, \RR, \EE,
\bbz{} and \bbu. We can easily see that if the locomotive enters the unit through 
the gate~\RR, then it leaves the memory through the gate~\bbz{} or through the gate~\bbu{}
depending on the information stored in the memory: that information is provided the unit
by the positions of the switch at~\WW{} and those at~\RR{} and~\EE. Note that the
positions at~\RR{} and at~\EE{} are connected by the path from~\EE{} to~\RR, see the
figure.

When the locomotive enters the memory through the gate~\WW{} where a flip-flop switch is
sitting, it goes to~\RR{} through one of both tracks leaving the switch. If it goes
through the track marked by~\zz, \uu, it arrives to~\EE{} by the track marked with the
opposite symbol, \uu, \zz{} respectively. Indeed, when the locomotive crosses~\WW,
the passing makes the selected track to be changed so, if it went through one track, after
the passage, in particular when the locomotive arrives at~\EE, the new selected track
at~\WW{} is the track through which the locomotive did not pass. So that the
track marked by one symbol at~\WW{} should be marked by the opposite one at~\EE. The
selection observed at~\EE{} is transferred to~\RR{} thanks to the path connecting~\EE{}
to~\RR.

\vskip 10pt
\vtop{
\ligne{\hfill
\includegraphics[scale=0.5]{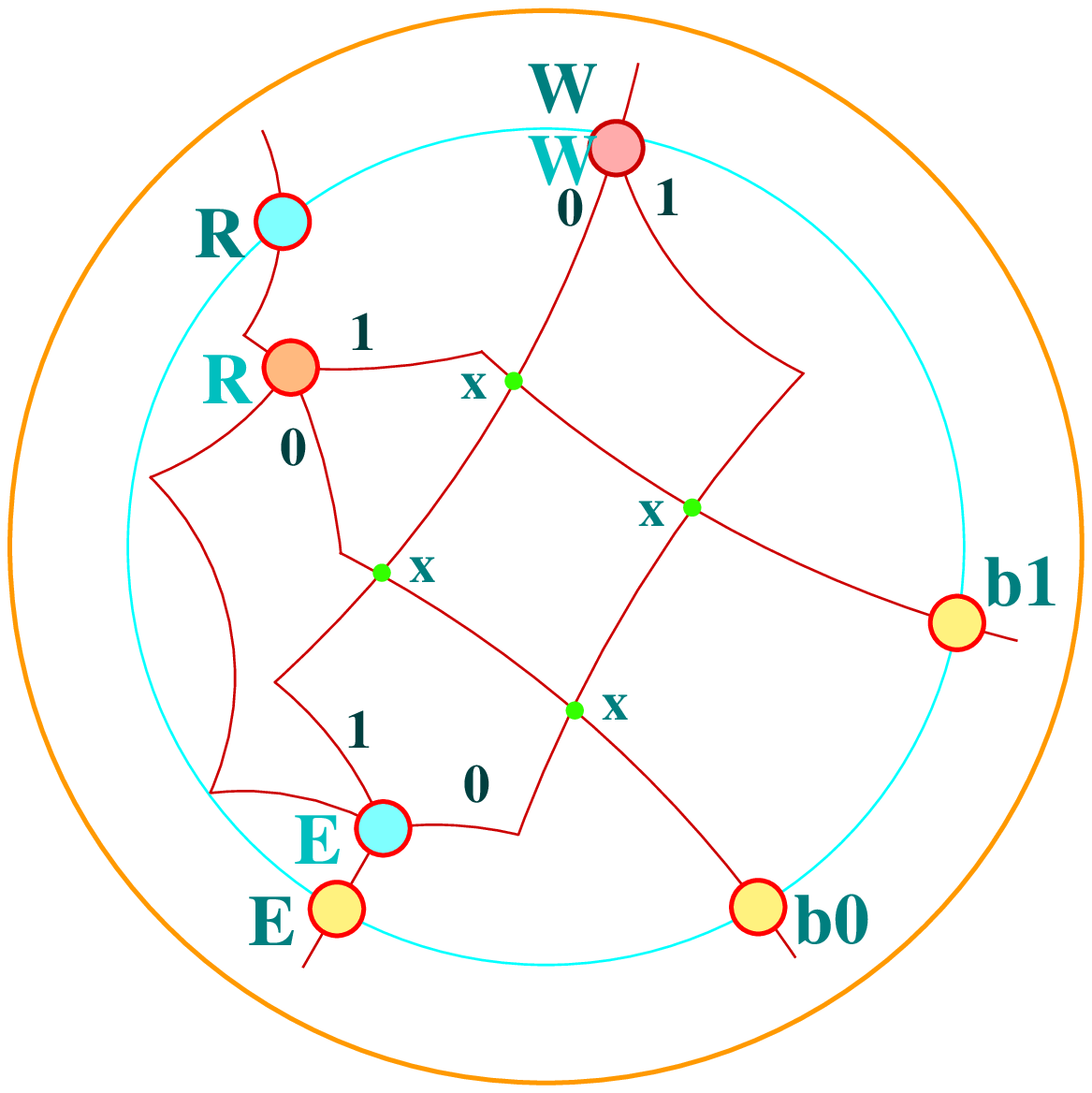}
\hfill}
\begin{fig}\label{fonebit}
\leurre
The idle configuration of the one-bit memory. Note the four crossings in the 
implementation. Note that the connection from~\EE{} to~\RR{} is realized by three 
segments of straight lines.
\end{fig}
}

   As the one-bit memory will be used later, we introduce a simplified notation:
in Figure~\ref{fonebit}, the memory structure is enclosed in a blue circle. At its 
circumference the gates are repeated by the same symbols. In the next figures, when a
one-bit memory will be used, we shall indicate it by a disc with, at its border, the
five gates mentioned in Figure~\ref{fonebit}.

\subsubsection{From instructions to registers and back}\label{sbbregdisp}

   As will be explained in Sub-subsection~\ref{sbbreg}, the locomotive arrives at a 
register at the same element, but not through the same face: face~5 is devoted to
the locomotive coming from an incrementing instruction, face~4 is devoted to the case
when the locomotive comes from a decrementing instruction. As will be seen later,
the blue locomotive coming to decrement the register is changed into a red one just before
arriving at the register. That difference will be justified when we study the operations
on a register, in Section~\ref{sbbreg}. After it performed its
operation, the locomotive goes back through a particular track corresponding to the
operation it performed. But, as far as all incrementations on a given register make the
locomotive arrive at the same point of the register, the return to the appropriate
instruction of the program requires that the instruction yielding the operation should
be memorised in some way.

\def\DDI{{${\mathbb D}_I$}}
\def\DDD{{${\mathbb D}_D$}}
   To that goal we define two structures~\DDI {} and \DDD {} for incrementing and 
decrementing instructions respectively. Each structure consists of as many units
as there are instructions of the corresponding type operating on the same register.
Accordingly, each register is dotted with specific \DDI {} and \DDD. We can imagine
that the small orange and blue boxes of Figure~\ref{fglobconfig} contain, each one,
a copy of \DDI {} and a copy of~\DDD.

First, we consider the case of~\DDI. Each unit is based on a one-bit memory which is
illustrated by Figure~\ref{fdispinc}. The working of~\DDI {} is the following. An 
instruction for incrementing $R$ is connected through a path to a specific unit 
of~\DDI. The path goes from the program to the gate~\WW{} of that unit. At the initial 
time, the configuration of~\DDI {} is such that all its unit contain the bit~0: the 
switches of the one-bit memory are in a position which, by definition defines bit~0. 
Accordingly, when the locomotive enters the unit, it will change the flip-flop and the 
memory switches so that, by definition, the memory contains the bit~1. The locomotive 
leaves the memory through the gate~\EE{} and it meets a flip-flop switch at~$A$ which, 
in its initial position, sends the locomotive to~$R$.

When the locomotive completed the incrementation of~$R$ it goes back to the program.
In order to find the appropriate way, it visits the units of the~\DDI {} attached to~$R$
until it finds the unit whose one-bit memory is set to~1. Indeed, at~\DDI, the 
locomotive enters the memory of the first unit through its ~\RR-gate. If it reads~0, 
it leaves the memory through~\bbz{} and the path sends it to the next unit. So that we are
led to the situation when entering through~\RR, the locomotive reads~1. Accordingly 
it leaves the unit through~\bbu{} which leads it to~\WW. Consequently, the locomotive 
rewrites the bit, turning it to~0 and again exits through the gate~\EE. It again meets 
the flip-flop switch at~$A$ which sends the locomotive on its other path leading the
locomotive back to the program. That new visit of the flip-flop switch at~$A$ make the 
switch again select the track leading to~$R$. Accordingly, when the locomotive 
leaves~\DDI {} the structure recovered its initial configuration.
Accordingly, that scheme
allows us to correctly simulate the working of the units of~\DDI.

\vskip 10pt
Secondly, we examine the structure of \DDD {} which plays for the 
decrementing instruction the role which \DDI {} plays for the incrementing ones. The 
structure is more complex for the following reason. An incrementing instruction is 
always performed which is not necessarily the case for a decrementing instruction. 
Indeed, if the register contains the value~0, it cannot be decremented. We say that 
the register is empty. In that case, the next instruction to be performed is not the 
next one in the program. It is the reason why the case of an empty register requires to 
be differently dealt with. Concretely, it means that the return track of the locomotive 
depends on whether the register was empty or not at the arrival of the locomotive.

Accordingly we need two one-bit memories instead of one. More over, as far as the bit~1
locates the unit of~\DDD {} which will send the locomotive to the right place in the 
program, the content of the memories should be the same: 0, if the locomotive did not
visited that unit at its arrival to~\DDD, 1 if it visited the required unit.
There are two return tracks after decrementation: the $D$-track when the 
decrementation could be performed, the $Z$-track when the decrementing locomotive found 
an empty register. We call $D$-, $Z$-{\bf memory} the one-bit memory visited by $D$-, 
$Z$-track respectively. 

\vskip 10pt
\vtop{
\ligne{\hfill
\includegraphics[scale=0.5]{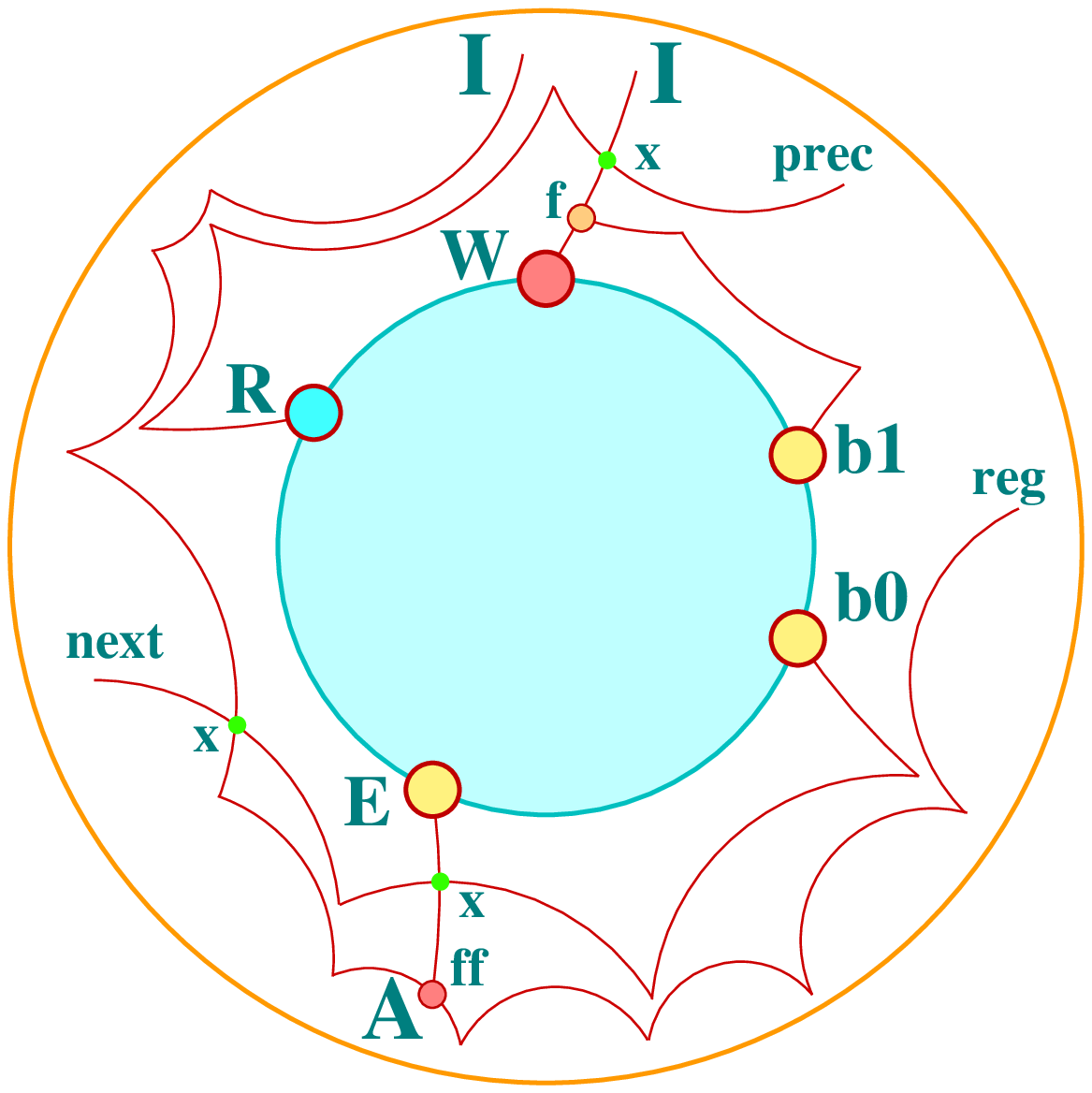}
\hfill}
\begin{fig}\label{fdispinc}
\leurre
The idle configuration of a unit of the structure which memorises the right incrementing
instruction. Note the three crossings in the 
implementation. Also note that the intensive use of sequences of connected segments of
straight lines.
\end{fig}
}

The complication entailed by that organisation make a 
representation of a unit in a single window hardly readable.
It is the reason why it was split into three windows as illustrated by 
Figure~\ref{fdispdec}.

   The $D$-memory is represented by the left-hand side picture of
the figure. A locomotive sent by the program for decrementing a register~$R$ arrives
at the appropriate unit of the \DDD {} attached to~$R$ by the track marked by $D$ in 
the picture. As far as the locomotive enters the memory through its gate~\WW, it rewrites 
its content from~0 to~1. Now, it has to mark the $Z$-memory which is represented on 
the right-hand side picture of the figure. To that goal, the track leaving the $D$-memory
through its gate~\EE, goes to the point~$c$, that italic letter marking a small yellow 
disc close to the border of the window. The track is continued from a small yellow disc 
close to the border of the right-hand side window marked with the same letter~$c$. Other 
such discs in the figure are marked by other letters which are pairwise the same in two 
different windows. From that second $c$, the track leads the locomotive to the 
gate~\WW{} of the $Z$-memory whose content, accordingly, will be changed from~0 to~1. 
Leaving the $Z$-memory through its gate~\EE, the locomotive is lead to a small disc~$d$ 
so that the corresponding track is continued by the small disc~$d$ we find in the middle 
window of Figure~\ref{fdispdec}. From there, the locomotive is lead to a flip-flop 
switch sitting at~$F$ which, in its initial configuration, selects the track leading 
to~$R$. Now, after the switch is passed by the locomotive, its selection is changed, 
indicating the track leading to another switch sitting at~$A$.

   Consider the case when the locomotive returns from~$R$ after a successful 
decrementation. It one by one visits the units of~\DDD. It arrives to the gate~\RR{}
of the $D$-memory of a unit. If it reads~0, the track issued from~\bbz{} of the 
$D$-memory leads the locomotive to the \RR-gate of the $D$-memory of the next unit.
So that such a motion is repeated until the locomotive reads~1{} in the $D$-memory.
When it is the case, the locomotive leaves the memory through its gate~\bbu{} which
leads the locomotive to a small disc~$a$, so that the track is continued from the
small disc~$a$ we can see in the middle picture of Figure~\ref{fdispdec}. From there,
the locomotive arrives to~$P$, the passive part of the memory switch whose active part
lies at~$A$. The locomotive is sent from~$P$ to a small disc~$e$ whose track arriving
there is continued by the track issued from the small disc~$e$ of the left-hand side
picture of the figure. That track leads the locomotive back to the \WW-gate of the
$D$-memory. Consequently, the content~1 of the memory is returned to~0 and leaving
the memory through its gate~\EE, the locomotive again arrives to the \WW-gate of the
$Z$-memory through the small discs~$c$. Again the \EE-gate of the $Z$-memory sends
the locomotive to the small disc~$d$ which, through the other such disc in the middle
window, again arrives at $A$. As far as it took much more time for the locomotive to 
rewrite both memories of the unit than for another locomotive to go from~$P$ to~$A$
in order to set the selection of the active switch, the switch at~$A$ indicates the
track corresponding to the branch $P$$a$ of the passive switch. That track leads to
the right decrementing instruction in the program and after that passage of the 
locomotive, the switch at~$F$, indicates the path to~$R$: the structure recovered its
initial configuration.

\vskip 10pt
\vtop{
\ligne{\hfill
\includegraphics[scale=0.45]{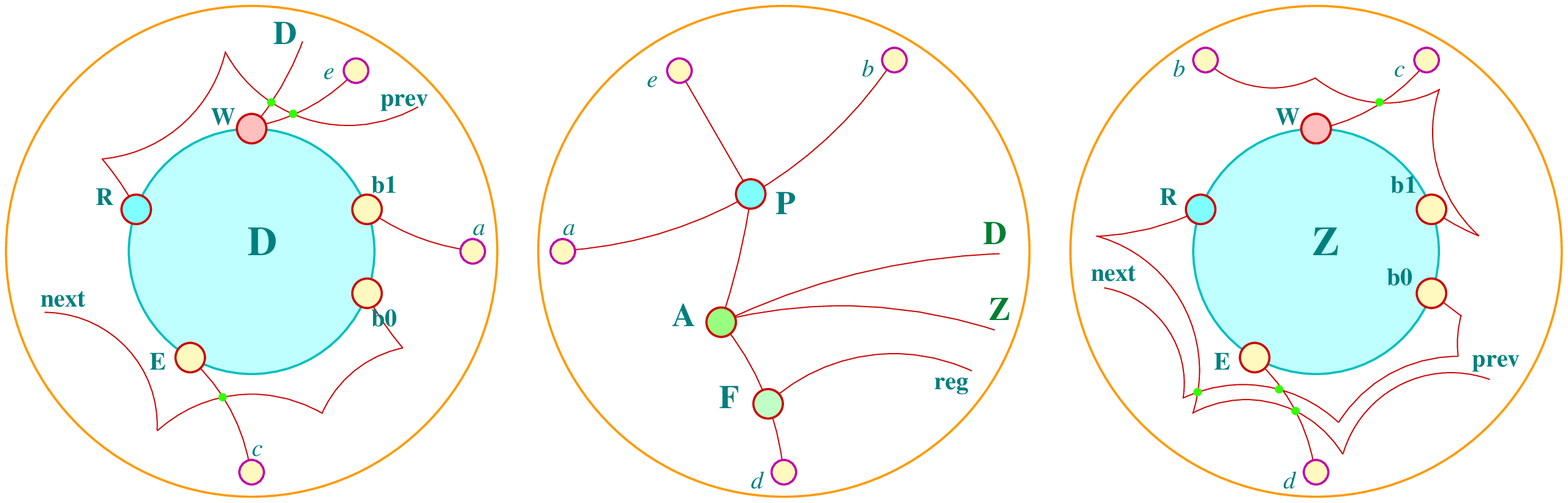}
\hfill}
\begin{fig}\label{fdispdec}
\leurre
The three windows illustrating the idle configuration of a unit of the structure which 
memorises the right decrementing instruction. Note the small discs accompanied with
small italic letters allowing the locomotive to pass from one window to the appropriate 
one. Also note the sketchy representation of the one-bit memories.
\end{fig}
}

   Presently, consider the case when the locomotive returns from an empty $R$.
It returned through the $Z$-track and it arrives to the~\DDD {} attached to~$R$.
There it visits the units of the structure reading the content of its $Z$-memory through
the track leading to the \RR-gate of that memory in the unit. So that the locomotive 
visits the units until the single one whose $Z$-memory contains~1. When it is the case,
the locomotive leaves the $Z$-memory through its gate~\bbu{} which sends it to the 
\WW-gate of the $D$-memory through the small disc~$b$. But the continuation happens by~$b$
of the middle picture which leads the locomotive to~$P$. The passive switch sends the 
locomotive to~$e$ and from there to the \WW-gate of the $D$-memory. Accordingly, what
we seen in a visit through a unit thanks to the $D$-track occurs once again. 
Accordingly, the locomotive
rewrites the contents of both the $D$- and the $Z$-memories, returning them from~1 to~0.
Then, the locomotive leaves the $Z$-memory through its \EE-gate so that, via the disc~$d$,
it arrives to~$F$ where the flip-flop switch sitting there sends it to the switch
sitting at~$A$. Now, during its trip for the \EE-gate of the $Z$-memory to the \WW-gate
of the $D$-memory, the locomotive arrived to~$P$ through the branch $P$$b$ of that
passive switch. Accordingly, the switch selected that track and, in the meanwhile, sent
another locomotive to~$A$ in order to select the appropriate track, that which leads
to $Z$ in the program. As the locomotive passed for a second time through~$F$, the
switch selects presently the track leading to~$R$, its initial configuration. Accordingly,
the scheme illustrated by the picture performs what is expected.

   At last and not the least, we have to look at what happens when the locomotive arrives
to the program after performing its operation on the register. Whatever the track arriving
to the program, it leads to the appropriate instruction: the next one in the program if
the just operated one was an incrementation or a successful decrementation, a specified
one if the decrementation was unsuccessful. We have to remember, when we shall
discuss the rules that when it leaves a register, whatever the return path, the locomotive
must be blue.

\subsubsection{Constitution of a register}\label{sbbreg}

   The implementation of the register requires a special examination. As the computation
is supposed to start from a finite configuration, at initial time, each register has
only finitely many non blank tiles. So that incrementation and decrementation necessarily
change the length of the register which is always finite. 

    Our choice is to make the register grow as long as the computation is not completed.
The operations of incrementing and decrementing can be implemented by a change of colour
at the beginning end of the register: its content is the number of cells of that 
particular colour. The continuation of the register is made at speed 
$\displaystyle{1\over2}$ for a reason we soon indicate. It happens independently of the
operations performed on the register. As far as the length of the track from the register
and back is enormous, the other end of the register is farther and farther from its 
beginning end. So, to stop the register when the stopping instruction is reached, the
program sends a locomotive to all registers. Each locomotive crosses its register at 
speed~1. That condition guarantees that the stopping signal will eventually reach the 
growing end of the register so that it can stop the continuation.

\def\ftt #1 {{\footnotesize\tt#1}}
\def\hhzz{\hskip-0.5pt}
\def\sww{{\ftt W }\hhzz }
\def\sbb{{\ftt B }\hhzz }
\def\srr{{\ftt R }\hhzz }
\def\sgg{{\ftt G }\hhzz }
\def\syy{{\ftt Y }\hhzz }
    The register consists of four sequences of tiles following a line~$\ell$ on~\HH.
 We say that such a sequence is a {\bf strand} of the register.
Two strands are upon~\HH{} and the two others are below~\HH. One strand over~\HH{}
contains the content of the register, we denote it by \RR$_c$. That strand is blue, 
denoted~\sbb {} and the content is blank which we denote by~\sww. The strand below~\RR$_c$
is red, denote by~\srr, and it is the return path for an incrementing instruction. We 
denote it by~\RR$_i$. The other strand over~\HH{} is yellow, denoted by~\syy {} and it 
is the return path for a decrementing instruction. We denote it by~\RR$_d$. The 
fourth strand, hence below~\HH, is blue and is the path for the stopping signal. We 
denote it by~\RR$_s$. The growing end consists of the last tiles of each sequence,
those tiles being green, denoted by~\sgg{}. We shall see more precisely how it is 
performed when we shall study the rules for implementing the simulation. In order to 
facilitate our explanations, if $\mathcal S$ is one of the four strands we depicted, 
$\mathcal S$(0) denotes its first cell, more generally, $\mathcal S$$(i)$, 
with \hbox{$i\in{\mathbb N}^+$}, denotes the $i^{\rm th}$ element on the strand. As we 
also need to define the cell which has a side on~$\ell$ and which is continued by 
$\mathcal S$, we denote that cell by $\mathcal S$(-1). We define a numbering for
the tiles of $\mathcal S$ as illustrated by Figure~\ref{fles4} and as explained in our
discussion about the neighbours of a tile. The motion of a locomotive from a face~5 to a
face~2 allows us to see that on two strands, \RR$_c$ and \RR$_s$, the motion goes 
on $\mathcal S$($n$) by following increasing values of~$n$ while of the two other ones, 
\RR$_i$ and \RR$_d$, it goes by following decreasing values of~$n$.

\section{The rules}\label{srules}

   Let us remind the reader that cellular automata are a model of massive parallelism.
The base of a cellular automaton is a cell. The set of cells is supposed to be homogeneous
in several aspects: the neighbours of each cell constitute subsets which have the same
structure; the cell changes its state at each tip of a discrete clock according to the
states of its neighbours and its own state. The change is dicted by a finite automaton 
which is the same for each cell. A tessellation is an appropriate space for implementing
cellular automata. Let $T$ be a tile and let $N(T)$ be the set of its neighbours. In a 
tessellation, $N(T)$ is the same for any~$T$ and two tiles of the tessellation are 
isomorphic with respect to the geometry of the space on which the tessellation is defined.
The dodecagrid is a tessellation of the hyperbolic $3D$-space. Moreover, there is an 
algorithm to locate the tiles which is cubic in time and in the size of the code 
attached to each tile, see \cite{mmbook2} for instance. However, we do not need that 
location system. The main reason is that we work on the basis of~\HH{} where there is a 
linear algorithm to locate the cells, see~\cite{mmbook2} too, and our incursions in the 
third dimension does not drive us far from~\HH. The tunnel and the growth of 
the registers indicated us how far it can be.

\def\PP{$\mathbb P$}
    The way the automaton manages the change of states can be defined by a finite set
of {\bf rules} we shall call the {\bf program} of the automaton denoted by \PP. We 
organise~\PP{} as a {\bf table} we display by pieces according to the role of the
corresponding instructions with respect to the place of the locomotive in the circuit
it performs.
In Sub-section~\ref{rrotinv}
we shall define the format of the rules, what means their rotation invariance and how
we check it. In Sub-section~\ref{rstopgen}, we look at the rules regarding neighbours 
of generations~2 and~3.
In Sub-section~\ref{rtracks} we look at the application of the rules for 
the tracks, including those for the crossings. In the same sub-section
we do the same for the rules managing the working of the switches, mainly their control 
structures.
In Sub-section~\ref{rreg} we study the rules for the register, separately considering 
the growth of the register, its incrementation and its decrementation, the case of an 
empty register being included. Sub-section~\ref{srstop} gives us the rules for stopping
the computation.

\subsection{Format of the rules and rotation invariance}\label{rrotinv}

A cell of the dodecagrid in the cellular automaton we define with the set of rules
studied in this section consists of a tile of the tiling, we call it the {\bf support} 
of the cell, together with the finite automaton defined by the rules of the present 
section. However, we shall use the words cells and tiles as synonyms for the sake of
simplicity. The neighbour of a cell~$c$ whose support is~$\Delta$ is a cell whose support 
is one of the $\Delta_i$, \hbox{$i\in\{0..11\}$}, where $\Delta_i$ shares the face~$i$
of~$\Delta$. Accordingly, $\Delta$ can see each $\Delta_i$ and it is seen by that latter
cell but $\Delta_i$ and $\Delta_j$ with $i\not=j$ cannot see each other, as already known.
We also say that $\Delta_i$ is the {\bf $i$-neighbour} of $\Delta$. The index
of $\Delta_i$ refers to a numbering of the faces. We use the one we described
in Section~\ref{intro} with the help of Figure~\ref{fdodecs}. As long as it will be
possible, cells on~\HH{} or below that plane will have their face~0 on~\HH. We already 
fixed the face~1{} of the strands in the previous section.

   As usual, we call {\bf alphabet} of our cellular automaton the finite set of its 
possible states. The alphabet of our cellular automaton consists of \sww, \sbb, \srr,
\sgg{} and \syy {} we already met. We also call \sww{} the {\bf blank} as far as it is 
the quiescent state of automaton and we remind the reader that the cells of the 
dodecagrid are blank, except finitely many of them at whichever time. Let $c$ be a 
cell of the dodecagrid whose state is {\ftt S$_o$ }
and whose support is $\Delta$, and let {\ftt S$_i$ } be the state of its $i$-neighbour.
Let {\ftt S$_n$ } be the {\bf new} state of~$c$, {\it i.e.} the state taken by~$c$
when it is the state associated with {\ftt S$_o$ } and the {\ftt S$_i$ } by the
automaton. We write this as follows:

\def\llrule #1 #2 #3 #4 #5 #6 #7 {%
\hbox{\ftt{#1.#2#3#4#5#6#7} }
}
\def\rrrule #1 #2 #3 #4 #5 #6 #7 {%
\hbox{\ftt{#1#2#3#4#5#6.#7} }
}
\vskip 5pt
\ligne{\hfill
\llrule {S$_o$} {S$_0$} {S$_1$} {S$_2$} {S$_3$} {S$_4$} {S$_5$} 
\hskip-4pt\rrrule {S$_6$} {S$_7$} {S$_8$} {S$_9$} {S$_{10}$} {S$_{11}$} {S$_n$} {}
\hfill(\numerrel)\hskip 10pt}
\vskip 5pt
\noindent
and we say that {\ftt S } is a {\bf rule} of the automaton. All rules of the automaton
we give in the present section obey the format defined by~$(1)$. Of course, {\ftt S$_o$ }
and the {\ftt S$_i$ } belong to the alphabet 
\hbox{$\mathbb A = \{$\sww, \sbb, \srr, \sgg, \syy$\}$}.
From the remark on $\mathbb A$, we can read a rule as a word on $\mathbb A$: it is 
enough to remove the dots which occur in(2).

   Let $\rho$ be a rotation leaving the support of~$c$ globally invariant. Let 
\hbox{{\ftt S } $\in$ \PP}. We call {\bf rotated image of {\ftt S } under $\rho$} the 
rule:
\vskip 5pt
\ligne{\hfill 
\llrule {S$_o$} {S$_{\rho(0)}$} {S$_{\rho(1)}$} {S$_{\rho(2)}$} {S$_{\rho(3)}$} 
{S$_{\rho(4)}$} {S$_{\rho(5)}$} 
\hskip-4pt\rrrule {S$_{\rho(6)}$} {S$_{\rho(7)}$} {S$_{\rho(8)}$} {S$_{\rho(9)}$} 
{S$_{\rho(10)}$} {S$_{\rho(11)}$} {S$_n$} {} 
\hfill\hskip 10pt}
\vskip 5pt
\noindent
which we denote \ftt{S}$_\rho$ and we also say that \ftt{S}$_\rho$ {} as a word is a 
{\bf rotated form} of \ftt{S}.

A rule {\ftt S } is said {\bf rotation invariant} if and only if for any rotation 
$\rho$ leaving the support of the cell globally invariant we have
\vskip 5pt
\ligne{\hfill {\ftt S } $\in$ \PP{} $\Rightarrow$ {\ftt S }$_\rho$ $\in$ \PP\hfill
(\numerrel)\hskip 10pt}
\vskip 5pt
It is important to note that the rotation invariance requires the invariance in all
rotations leaving the dodecahedron globally invariant. The set $\mathcal R$ of those 
rotations cannot be replaced by a set of generators. For example, we have seen that 
the set of rotations around
a face leaving the face invariant defines six generators. Now, if {\ftt S } is
rotation invariant in those generators, it does not involve that it would be invariant
under products of those generators. In other words, if $g_1$ and $g_2$ are such
generators, \hbox{{\ftt S }$_{g_1}$, {\ftt S }$_{g_2}$ $\in$ \PP} does not involve
that \hbox{{\ftt S }$_{g_1\circ g_2 }$} should be in \PP{} as far as {\ftt S }$_{g_1}$
is generally formally different from {\ftt S } even if {\ftt S }$_{g_1}$ were in \PP.
Accordingly, rotation invariance in the case of a cellular automaton in the dodecagrid
is a huge constraint: it requires to check the application of sixty rotations to each rule
of its program. By comparison, the rotation invariance in the heptagrid, for instance,
requires seven conditions only. In the Euclidean frame, the rotation invariance for
a square grid requires four conditions and for the cube twenty four of them. Sixty is 
twice and a half twenty four. This is why the result of Theorem~\ref{letheo} is 
significant: the reduction from 10 to~5 is a relevant effort.

    There is a simple way to check rotation invariance for the rules of~\PP. We already 
noticed that a rule in the format (2) can be read as a word on $\mathbb A$ and we have
also read \ftt{S}$_\rho$ {} as a word
for any $\rho$ in $\mathcal R$. If we lexicographically order the rotation forms of
\ftt{S } for all $\rho$ in $\mathcal R$,
we can choose the smallest one as representative of
all the others. Call that word the {\bf minimal form} of the rotated forms of 
{\ftt S } and denote it by min$_{\mathcal R}$({\ftt S }). We get the following lemma:

\begin{lem}\label{minimal}
Let {\ftt U } and {\ftt V } be two rules of \PP. Then {\ftt V } is a rotated image
of {\ftt U } if and only if their minimal forms are identical.
\end{lem}

\noindent
Proof. Denote {\ftt S }$_{\mathcal R}$ the set of all rotated forms of {\ftt S }. Then,
clearly, \hbox{min$_{\mathcal R}${\ftt S } = {\ftt S }$_\rho$} for some 
$\rho\in$~$\mathcal R$.
Also note that for two rotations $\rho_1$ and $\rho_2$ in $\mathcal R$ and for any rule 
{\ftt S }, we have 
\hbox{{\ftt S }$_{{\rho_1}_{\rho_2}}$ = {\ftt S }$_{\rho_1\circ\rho_2}$}. Accordingly,
{\ftt U } $\in$ {\ftt V }$_{\mathcal R}$ if and only if {\ftt V } $\in$ 
{\ftt U }$_{\mathcal R}$ and also
if and only if {\ftt U }$_{\mathcal R}$ = {\ftt V }$_{\mathcal R}$. Which proves the 
lemma.\hfill $\Box$

\def\MM{$\mathbb M$}
\def\TT{$\mathbb T$}

   The condition given by the lemma induces a criterion which is hardly manageable 
for handy computations. Fortunately, the criterion can easily be performed by an 
algorithm we leave to the reader. We programmed such an algorithm to check the
rotation invariance of the set of rules we define in the paper. Also note that the
lemma gives us the way to simplify the presentation of~\PP: if we strictly apply
(3), as far as for each {\ftt S } in \PP{} and for each $\rho$ in $\mathcal R${} 
{\ftt S }$_\rho$
should also be in \PP, the number of rules in \PP{} is a multiple of sixty. Accordingly,
if in \PP{} we replace all rules which are rotations image of each other by their
minimal forms, we get a new set of rules \MM{} whose number of elements is that of~\PP{}
divided by sixty. The set of rules we shall present is not exactly \MM. Instead of
replacing the rules in {\ftt S }$_{\mathcal R}$ by the minimal form, we take a more 
understandable
element which will appear from our investigations. It is that set of rules, whose
cardinality is that of \MM{} we present and we denote it by \TT. Accordingly
for two rules {\ftt U } and {\ftt V } in \TT, we have that 
\hbox{min$_{\mathcal R}${\ftt U } $\not=$ min$_{\mathcal R}${\ftt V }}. In that case, 
we say that {\ftt U }
and {\ftt V } are {\bf rotationally independent}. If the rules of \TT{} are 
pairwise rotationally independent, we say that \TT{} is {\bf rotationally coherent}.
The computation program I devised in establishing \TT{} also checked the rotational
coherence of that set of rules.

As an example, the rule 
\hbox{\llrule {W} {W} {W} {W} {W} {W} {B} 
\hskip-3.5pt\rrrule {B} {W} {W} {B} {B} {W} {B} } is rotationally equivalent to the 
following rule:
\hbox{\llrule {W} {W} {W} {W} {W} {B} {W} 
\hskip-3.5pt\rrrule {B} {W} {W} {B} {B} {W} {B} }. Their common rotation form
is 
\hbox{\llrule {W} {W} {W} {W} {W} {W} {W} 
\hskip-3.5pt\rrrule {W} {B} {W} {B} {B} {B} {B} }. We get that form from the first rule 
thanks to a rotation around face~1, while we get it from the other rule thanks to a 
rotation around the vertex 6-10-11. As already mentioned, the rule rotationally 
equivalent to the second one does not occur in \TT.

We often considered what we call an idle configuration. In such a condition, the 
configuration must remain unchanged as long as the locomotive is not in the window which 
defines the configuration. Such rules are called {\bf conservative}. When the locomotive 
falls within the window, some cells are changed while the others remain unchanged. The
rules which change a cell
are called {\bf motion rules}. Among those which do not change a cell, some 
of them have the locomotive in their neighbourhood, in general for one step of the 
computation. Although the cell is not changed, we say that it can see the locomotive so 
that it is a witness of its passage so that we call such rules {\bf witness rules}. 
They appear to be an intermediate class between the conservative rules to which they 
belong as far as the new state is the same as the previous one but their neighbourhood 
is changed so that they are in some sense affected by the motion to which they 
contribute by their conservative function. In our discussion, we shall make the 
distinction between conservative and motion rules as long as it will be needed.
The rules are numbered to facilitate their references in the text.

\subsection{Rules regarding generations~2 and~3}\label{rstopgen}

   In our discussion leading to Proposition~\ref{pneighgen} and to Lemma~\ref{ldistneigh},
we considered neighbours of generation~1, generation~2 and generation~3. Of course, 
the rules of~\TT{} are potentially applied to any cell. However, the question of what 
can be said about that distinction among neighbours is important for establishing a 
simulating program. It is know that for hyperbolic spaces the 
number of tiles within a ball of radius~$k$ is exponential in~$k$. As an example,
the number of cells which are at a distance 10 from a tile is 1528. For each of these
cells at least 6 of their neighbours are not shared with the other cells of the same 
sphere. And that deals with a single while many configurations involve around ten cells,
possibly forty eight for the registers.  Such a heavy number is not manageable even by a 
program.  It is desirable to reduce the number of neighbours to be checked. 

In the discussion of Proposition~\ref{pneighgen} and Lemma~\ref{ldistneigh}, we 
considered generations~1, 2 and~3 of a given cell and we considered the connections between
the spheres around a cell. It is possible to define the notion of neighbour of a cell
of generation~$n$ for any positive~$n$: the neighbours of the generation~$n$+1 of a 
tile~$T$ are the neighbours of those of the generation~$n$ which do not belong to a 
generation~$m$ with \hbox{$m<n$}. We noticed that a neighbour of $T$ of generation~3 has 
only a vertex in common with~$T$. Clearly, the neighbours of generation~4 are separated 
from~$T$ by at least one tile. Repeating the argument, the tiles of generation~$4n$ are 
further and further from~$T$ as $n$ is increasing, a property we already mentioned.

\ifnum 1=0 {
Consider the following assertion:

\begin{requ}\label{rreq}
Let $U$ be a neighbour of generation~$n$ of a cell~$C$  of the initial configuration,
with \hbox{$n\geq2$}. If $U$ is white at the initial time, it should remain white 
during the computation.
\end{requ}

By definition, outside a large ball
all cells are blank so that, at initial time the needed rule for those cells is the
rule 
\hbox{\llrule {W} {W} {W} {W} {W} {W} {W} 
\hskip-3.5pt\rrrule {W} {W} {W} {W} {W} {W} {W} } .
That rule is called the quiescent rule and \sww{} is often called the quiescent state.
Here, as already indicated, we say indifferently white or blank state.
}
\fi

Consider the following rules:

\newcount\compteregle\compteregle= 0
\def\gzzz{\global\advance\compteregle by 1}
\def\vszz{\vskip-2pt}
\newdimen\largeouille\largeouille=87pt
\vtop{
\vrule height 0.4pt depth 0.4pt width 280pt
\vskip 3pt
\ligne{\hfill
$\vcenter{
\hbox{\vtop{\leftskip 0pt\parindent 0pt\hsize=\largeouille
\ligne{\hfill\gzzz\ftt{\the\compteregle \ }  
\llrule {W} {W} {W} {W} {W} {W} {W} 
\hskip-3.5pt\rrrule {W} {W} {W} {W} {W} {W} {W} {}\hfill}
\vskip-3pt
\ligne{\hfill\gzzz\ftt{\the\compteregle \ }  
\llrule {W} {B} {W} {W} {W} {W} {W} 
\hskip-3.5pt\rrrule {W} {W} {W} {W} {W} {W} {W} {}\hfill}
\vskip-3pt
\ligne{\hfill\gzzz\ftt{\the\compteregle \ }  
\llrule {W} {R} {W} {W} {W} {W} {W} 
\hskip-3.5pt\rrrule {W} {W} {W} {W} {W} {W} {W} {}\hfill}
\vskip-3pt
\ligne{\hfill\gzzz\ftt{\the\compteregle \ }  
\llrule {W} {G} {W} {W} {W} {W} {W} 
\hskip-3.5pt\rrrule {W} {W} {W} {W} {W} {W} {W} {}\hfill}
\vskip-3pt
\ligne{\hfill\gzzz\ftt{\the\compteregle \ }  
\llrule {W} {Y} {W} {W} {W} {W} {W} 
\hskip-3.5pt\rrrule {W} {W} {W} {W} {W} {W} {W} {}\hfill}
\ligne{\hfill\gzzz\ftt{\the\compteregle \ }  
\llrule {W} {B} {B} {W} {W} {W} {W} 
\hskip-3.5pt\rrrule {W} {W} {W} {W} {W} {W} {W} {}\hfill}
\vskip-3pt
\ligne{\hfill\gzzz\ftt{\the\compteregle \ }  
\llrule {W} {G} {G} {W} {W} {W} {W} 
\hskip-3.5pt\rrrule {W} {W} {W} {W} {W} {W} {W} {}\hfill}
\vskip-3pt
\ligne{\hfill\gzzz\ftt{\the\compteregle \ }  
\llrule {W} {R} {R} {W} {W} {W} {W} 
\hskip-3.5pt\rrrule {W} {W} {W} {W} {W} {W} {W} {}\hfill}
\vskip-3pt
\ligne{\hfill\gzzz\ftt{\the\compteregle \ }  
\llrule {W} {Y} {Y} {W} {W} {W} {W} 
\hskip-3.5pt\rrrule {W} {W} {W} {W} {W} {W} {W} {}\hfill}
\vskip-3pt
\ligne{\hfill\gzzz\ftt{\the\compteregle \ }  
\llrule {W} {R} {B} {W} {W} {W} {W} 
\hskip-3.5pt\rrrule {W} {W} {W} {W} {W} {W} {W} {}\hfill}
\vskip-3pt
\ligne{\hfill\gzzz\ftt{\the\compteregle \ }  
\llrule {W} {R} {Y} {W} {W} {W} {W} 
\hskip-3.5pt\rrrule {W} {W} {W} {W} {W} {W} {W} {}\hfill}
\vskip-3pt
\ligne{\hfill\gzzz\ftt{\the\compteregle \ }  
\llrule {W} {G} {B} {W} {W} {W} {W} 
\hskip-3.5pt\rrrule {W} {W} {W} {W} {W} {W} {W} {}\hfill}
\vskip-3pt
\ligne{\hfill\gzzz\ftt{\the\compteregle \ }  
\llrule {W} {G} {R} {W} {W} {W} {W} 
\hskip-3.5pt\rrrule {W} {W} {W} {W} {W} {W} {W} {}\hfill}
\vskip-3pt
\ligne{\hfill\gzzz\ftt{\the\compteregle \ }  
\llrule {W} {G} {Y} {W} {W} {W} {W} 
\hskip-3.5pt\rrrule {W} {W} {W} {W} {W} {W} {W} {}\hfill}
\vskip-3pt
\ligne{\hfill\gzzz\ftt{\the\compteregle \ }  
\llrule {W} {Y} {B} {W} {W} {W} {W} 
\hskip-3.5pt\rrrule {W} {W} {W} {W} {W} {W} {W} {}\hfill}
\vskip-3pt
}
\hfill
\vtop{\leftskip 0pt\parindent 0pt\hsize=\largeouille
\ligne{\hfill\gzzz\ftt{\the\compteregle \ }  
\llrule {W} {B} {B} {B} {W} {W} {W} 
\hskip-3.5pt\rrrule {W} {W} {W} {W} {W} {W} {W} {}\hfill}
\vskip-3pt
\ligne{\hfill\gzzz\ftt{\the\compteregle \ }  
\llrule {W} {R} {R} {R} {W} {W} {W} 
\hskip-3.5pt\rrrule {W} {W} {W} {W} {W} {W} {W} {}\hfill}
\vskip-3pt
\ligne{\hfill\gzzz\ftt{\the\compteregle \ }  
\llrule {W} {G} {G} {G} {W} {W} {W} 
\hskip-3.5pt\rrrule {W} {W} {W} {W} {W} {W} {W} {}\hfill}
\vskip-3pt
\ligne{\hfill\gzzz\ftt{\the\compteregle \ }  
\llrule {W} {Y} {Y} {Y} {W} {W} {W} 
\hskip-3.5pt\rrrule {W} {W} {W} {W} {W} {W} {W} {}\hfill}
\ligne{\hfill\gzzz\ftt{\the\compteregle \ }  
\llrule {W} {B} {B} {R} {W} {W} {W} 
\hskip-3.5pt\rrrule {W} {W} {W} {W} {W} {W} {W} {}\hfill}
\vskip-3pt
\ligne{\hfill\gzzz\ftt{\the\compteregle \ }  
\llrule {W} {B} {B} {G} {W} {W} {W} 
\hskip-3.5pt\rrrule {W} {W} {W} {W} {W} {W} {W} {}\hfill}
\vskip-3pt
\ligne{\hfill\gzzz\ftt{\the\compteregle \ }  
\llrule {W} {B} {B} {Y} {W} {W} {W} 
\hskip-3.5pt\rrrule {W} {W} {W} {W} {W} {W} {W} {}\hfill}
\vskip-3pt
\ligne{\hfill\gzzz\ftt{\the\compteregle \ }  
\llrule {W} {R} {R} {B} {W} {W} {W} 
\hskip-3.5pt\rrrule {W} {W} {W} {W} {W} {W} {W} {}\hfill}
\vskip-3pt
\ligne{\hfill\gzzz\ftt{\the\compteregle \ }  
\llrule {W} {R} {R} {G} {W} {W} {W} 
\hskip-3.5pt\rrrule {W} {W} {W} {W} {W} {W} {W} {}\hfill}
\vskip-3pt
\ligne{\hfill\gzzz\ftt{\the\compteregle \ }  
\llrule {W} {R} {R} {Y} {W} {W} {W} 
\hskip-3.5pt\rrrule {W} {W} {W} {W} {W} {W} {W} {}\hfill}
\vskip-3pt
\ligne{\hfill\gzzz\ftt{\the\compteregle \ }  
\llrule {W} {G} {G} {B} {W} {W} {W} 
\hskip-3.5pt\rrrule {W} {W} {W} {W} {W} {W} {W} {}\hfill}
\vskip-3pt
\ligne{\hfill\gzzz\ftt{\the\compteregle \ }  
\llrule {W} {G} {G} {R} {W} {W} {W} 
\hskip-3.5pt\rrrule {W} {W} {W} {W} {W} {W} {W} {}\hfill}
\vskip-3pt
\ligne{\hfill\gzzz\ftt{\the\compteregle \ }  
\llrule {W} {G} {G} {Y} {W} {W} {W} 
\hskip-3.5pt\rrrule {W} {W} {W} {W} {W} {W} {W} {}\hfill}
\vskip-3pt
\ligne{\hfill\gzzz\ftt{\the\compteregle \ }  
\llrule {W} {Y} {Y} {B} {W} {W} {W} 
\hskip-3.5pt\rrrule {W} {W} {W} {W} {W} {W} {W} {}\hfill}
\vskip-3pt
\ligne{\hfill\gzzz\ftt{\the\compteregle \ }  
\llrule {W} {Y} {Y} {R} {W} {W} {W} 
\hskip-3.5pt\rrrule {W} {W} {W} {W} {W} {W} {W} {}\hfill}
\vskip-3pt
\ligne{\hfill\gzzz\ftt{\the\compteregle \ }  
\llrule {W} {Y} {Y} {G} {W} {W} {W} 
\hskip-3.5pt\rrrule {W} {W} {W} {W} {W} {W} {W} {}\hfill}
}
\hfill
\vtop{\leftskip 0pt\parindent 0pt\hsize=\largeouille
\vskip-3pt
\ligne{\hfill\gzzz\ftt{\the\compteregle \ }  
\llrule {W} {B} {R} {G} {W} {W} {W} 
\hskip-3.5pt\rrrule {W} {W} {W} {W} {W} {W} {W} {}\hfill}
\vskip-3pt
\ligne{\hfill\gzzz\ftt{\the\compteregle \ }  
\llrule {W} {B} {R} {Y} {W} {W} {W} 
\hskip-3.5pt\rrrule {W} {W} {W} {W} {W} {W} {W} {}\hfill}
\vskip-3pt
\ligne{\hfill\gzzz\ftt{\the\compteregle \ }  
\llrule {W} {B} {G} {R} {W} {W} {W} 
\hskip-3.5pt\rrrule {W} {W} {W} {W} {W} {W} {W} {}\hfill}
\vskip-3pt
\ligne{\hfill\gzzz\ftt{\the\compteregle \ }  
\llrule {W} {B} {G} {Y} {W} {W} {W} 
\hskip-3.5pt\rrrule {W} {W} {W} {W} {W} {W} {W} {}\hfill}
\vskip-3pt
\ligne{\hfill\gzzz\ftt{\the\compteregle \ }  
\llrule {W} {B} {Y} {R} {W} {W} {W} 
\hskip-3.5pt\rrrule {W} {W} {W} {W} {W} {W} {W} {}\hfill}
\vskip-3pt
\ligne{\hfill\gzzz\ftt{\the\compteregle \ }  
\llrule {W} {B} {Y} {G} {W} {W} {W} 
\hskip-3.5pt\rrrule {W} {W} {W} {W} {W} {W} {W} {}\hfill}
\vskip-3pt
\ligne{\hfill\gzzz\ftt{\the\compteregle \ }  
\llrule {W} {R} {G} {Y} {W} {W} {W} 
\hskip-3.5pt\rrrule {W} {W} {W} {W} {W} {W} {W} {}\hfill}
\vskip-3pt
\ligne{\hfill\gzzz\ftt{\the\compteregle \ }  
\llrule {W} {R} {Y} {G} {W} {W} {W} 
\hskip-3.5pt\rrrule {W} {W} {W} {W} {W} {W} {W} {}\hfill}
}}
}$
\hfill(\numerrel)\hskip 10pt}
\vskip 3pt
\vrule height 0.4pt depth 0.4pt width 280pt
}
\vskip 5pt
Those rules obey the following patterns:

\vskip 5pt
\ligne{\hfill
$\vcenter{
\vtop{\leftskip 0pt\parindent 0pt\hsize=285pt
\ligne{\hfill
\hbox{$(a)$ \llrule {W} {X} {W} {W} {W} {W} {W} 
\hskip-3.5pt\rrrule {W} {W} {W} {W} {W} {W} {W} },
\hbox{$(b)$ \llrule {W} {X} {Y} {W} {W} {W} {W} 
\hskip-3.5pt\rrrule {W} {W} {W} {W} {W} {W} {W} },
\hbox{$(c)$ \llrule {W} {X} {Y} {Z} {W} {W} {W} 
\hskip-3.5pt\rrrule {W} {W} {W} {W} {W} {W} {W} },
\hfill}
\ligne{\hfill with \hbox{\ftt{X} {}, \syy {}, \ftt{Z} {}  $\in$ 
$\{$\sbb, \srr, \sgg, \syy$\}$}
\hfill}
}}$
\hfill(\numerrel)\hskip 10pt}
\vskip 5pt


\begin{prop}\label{pdecid}
Consider a cellular automaton $A$ whose program is rotationally coherent and which
contains the rules~$(4)$.
	Then, the halting of the computation of~$A$ starting from a 
finite configuration is decidable.
\end{prop}

\noindent
Proof. Under the hypothesis of the proposition, $A$ contains all possible rules of~(5)
modulo rotation invariance.
Fix a non-blank cell $T_0$ of the initial configuration of a computation of~$A$.
Let $\mathcal B$ be the smallest ball around $T_0$ outside which all cells are blank at 
initial time of the computation. Let $k$ be its radius and let $U$ be a tile at 
distance~$k$ of~$T_0$. Consider~$V$ a neighbour of~$U$ which does not belong 
to~$\mathcal B$. By construction, $V$ is at 
distance~$k$+1 and it is blank. The neighbourhood of~$V$ contains at most three 
neighbours which belong to $\mathcal B$. That assertion is entailed by our study of the 
neighbours of generation~1, 2 and~3 around a tile. Accordingly, at least one of the rules
from~(5) apply to~$V$ which remains blank. That argument can be repeated to any
tile at distance~$k$ from $T_0$, so that the cells outside $\mathcal B$ always remain
blank. Accordingly, the computation remains inside $\mathcal B$. Consequently, either
the computation halts at some time or it becomes periodic after a certain time: both
situation are algorithmically detectable.\hfill$\Box$

   Note that there are 39 rules in~$(4)$. In full generality, there are five rules
of the form~$a$, assuming that \ftt{X} {} is not blank. In the same line there are
16 rules of the form~$(b)$ and 64 of them of the form~$(c)$. Rotation invariance allows
us to reduce the number of rules of the form~$(b)$, $(c$) to 10, 24 ones respectively.
Note that under rotation invariance, rules
\hbox{\llrule {W} {X} {Y} {W} {W} {W} {W} 
\hskip-3.5pt\rrrule {W} {W} {W} {W} {W} {W} {W} } {} and
\hbox{\llrule {W} {Y} {X} {W} {W} {W} {W} 
\hskip-3.5pt\rrrule {W} {W} {W} {W} {W} {W} {W} } {} where \hbox{\ftt{X} {} $\not=$ \syy }
are rotationally identical. Indeed, the neighbours \ftt{X} {} and \syy {} share an 
edge~$s$ and the rotation around the axis passing through the midpoints of~$s$ and of its
opposite side allows us to pass from one rule to the other one. It is also the reason
why we may assume that in the rules of the form~$(c)$, none of \ftt{X} {}, \syy {} 
and \ftt{Z} {} is the blank. If it where the case an appropriate rotation would 
transport such a rule to a rule of the form~$(b)$. Similarly, if two rules of the 
form~$(c)$ have their non-blank patterns as circular permutation of each other, then
those rules have the same minimal form.

    Proposition~\ref{pdecid} is not contradictory with Theorem~\ref{letheo}: indeed,
\TT{} does not satisfy Proposition~\ref{pdecid} as far as it rules out a single rule
of~(4) replacing it by a rule
of the form \hbox{\llrule {W} {X} {W} {W} {W} {W} {W} 
\hskip-3.5pt\rrrule {W} {W} {W} {W} {W} {W} {Y} } , where \ftt{X} {} and \syy {} are
non blank, as will further be seen. Accordingly, withdrawing one carefully chosen rule
from~(4) is enough to establish the strong universality of an appropriate computation. 

However, as far as \TT{} contains all rules of~(4) of the forms~$(b)$ and~$(c)$ and all
rules of~$(a)$ but one, we may consider that a blank cell~$V$ which is a neighbour
of generation~2 or~3 of non-blank cells of the configuration remains blank as long 
as~$V$ remains a neighbour of generation~2 or~3 of those cells. That property allows us
to restrict the simulation to neighbours of generation at most~2 of cells of the 
configuration. We shall always check that such an assumption is sound.

\subsection{The rules for the tracks and the switches}\label{rtracks}

    We have already met the rule which defines the quiescent state: a state which is 
unchanged if the cell and all its neighbours are in that state. In \TT, \sww {} is
the quiescent state and \TT{} contains the corresponding rule, rule~1{} in (4).

   We start our study of the rules and the construction of the rules by those which
manage the tracks. The motion rules are simple as they implement the following abstract 
scheme:
\vskip 5pt
\ligne{\hfill\ftt{
   WWW\hskip 20pt LWW\hskip 20pt WLW\hskip 20pt WWL}
\hfill(\numerrel)\hskip 10pt}
\vskip 5pt

To the left of (6), we have the conservative situation, rule~42{} in 
Table~\ref{ttracks}. Then, the locomotive enters the window and, in the two 
successive steps, rule~43 and~45 it crosses the window. Then, it witnesses the occurrence
of the locomotive in the next element of the track, rule~46 and, which, at the next step,
the cell recovers the conservative condition requiring rule~42. The implementation of 
those rules in the dodecagrid are easy. The first part of Table~\ref{texecvoies}
gives us traces of execution of the automaton along a line as illustrated by 
Figure~\ref{ftracks}. Note that rule~44 converts a red locomotive entering a cell of the
track to become blue when it is in the cell so that rules~45 and~46 apply, allowing the
previously red locomotive to leave the cell as a blue one.

First, let us look at the conservative rules. They mainly deal with the decorations
of the tiles of a track. A cell belonging to the decorations is \sbb{} an it is 
surrounded by white cells so that rule~2 applies. Also, rule~40 applies: the decoration
remains in place during the computation which also requires rule~41: that rules witnesses
the passage of the locomotive in the cell. If we look carefully at Figure~\ref{ftracks},
tiles on faces~6 and~10 do not see each other but, as known from the proof of
Proposition~\ref{pneighgen} they have a common neighbour which is 
white and which must remain white: whence rule~6 is needed.

   In the rules for the tunnel, the rules for the line also apply but the tunnel requires
specific rules. The first reason lies in the decoration of a cell allowing the entry
into the tunnel: it contains a red cell. Consequently, we have rule~47 for the 
conservation of the red cell in idle configurations and rule~48 for the same purpose when
a locomotive crosses the cell, always a blue one. As the red cell shares a side with a 
blue belonging to the decorations, a white cell can see both of those cells so that
rule~10 also applies. We remind the reader the forms
\hbox{\llrule {W} {X} {Y} {W} {W} {W} {W} 
\hskip-3.5pt\rrrule {W} {W} {W} {W} {W} {W} {W} } {} and
\hbox{\llrule {W} {Y} {X} {W} {W} {W} {W} 
\hskip-3.5pt\rrrule {W} {W} {W} {W} {W} {W} {W} } {} where \hbox{\ftt{X} {} $\not=$ \syy }
define the same rule under rotation invariance.

\ligne{\hfill
\vtop{
\begin{tab}\label{ttracks}
\leurre
Table of the rules managing the structures needed for conveying information from
the program to the circuit together with the motion of the locomotive.
\end{tab}
\ligne{\hfill
\vtop{\leftskip 0pt\parindent 0pt\hsize=\largeouille
\ligne{\hfill\ftt{line} \hfill}
\ligne{\hfill\gzzz\ftt{\the\compteregle \ }  
\llrule {B} {W} {W} {W} {W} {W} {W} 
\hskip-3.5pt\rrrule {W} {W} {W} {W} {W} {W} {B} {}\hfill}
\vskip-3pt
\ligne{\hfill\gzzz\ftt{\the\compteregle \ }  
\llrule {B} {B} {W} {W} {W} {W} {W} 
\hskip-3.5pt\rrrule {W} {W} {W} {W} {W} {W} {B} {}\hfill}
\vskip-3pt
\ligne{\hfill\gzzz\ftt{\the\compteregle \ }  
\llrule {W} {W} {W} {W} {W} {W} {W} 
\hskip-3.5pt\rrrule {B} {W} {W} {B} {B} {W} {W} {}\hfill}
\vskip-3pt
\ligne{\hfill\gzzz\ftt{\the\compteregle \ }  
\llrule {W} {W} {W} {W} {W} {W} {B} 
\hskip-3.5pt\rrrule {B} {W} {W} {B} {B} {W} {B} {}\hfill}
\vskip-3pt
\ligne{\hfill\gzzz\ftt{\the\compteregle \ }  
\llrule {W} {W} {W} {W} {W} {W} {R} 
\hskip-3.5pt\rrrule {B} {W} {W} {B} {B} {W} {B} {}\hfill}
\vskip-3pt
\ligne{\hfill\gzzz\ftt{\the\compteregle \ }  
\llrule {B} {W} {W} {W} {W} {W} {W} 
\hskip-3.5pt\rrrule {B} {W} {W} {B} {B} {W} {W} {}\hfill}
\vskip-3pt
\ligne{\hfill\gzzz\ftt{\the\compteregle \ }  
\llrule {W} {W} {W} {B} {W} {W} {W} 
\hskip-3.5pt\rrrule {B} {W} {W} {B} {B} {W} {W} {}\hfill}
\ligne{\hfill\ftt{tunnel} \hfill}
\ligne{\hfill\gzzz\ftt{\the\compteregle \ }  
\llrule {R} {W} {W} {W} {W} {W} {W} 
\hskip-3.5pt\rrrule {W} {W} {W} {W} {W} {W} {R} {}\hfill}
\vskip-3pt
\ligne{\hfill\gzzz\ftt{\the\compteregle \ }  
\llrule {B} {R} {W} {W} {W} {W} {W} 
\hskip-3.5pt\rrrule {W} {W} {W} {W} {W} {W} {B} {}\hfill}
\vskip-3pt
\ligne{\hfill\gzzz\ftt{\the\compteregle \ }  
\llrule {W} {W} {B} {W} {W} {W} {W} 
\hskip-3.5pt\rrrule {B} {W} {W} {B} {B} {W} {B} {}\hfill}
\vskip-3pt
\ligne{\hfill\gzzz\ftt{\the\compteregle \ }  
\llrule {W} {W} {W} {W} {R} {W} {W} 
\hskip-3.5pt\rrrule {B} {W} {W} {B} {B} {W} {W} {}\hfill}
\vskip-3pt
\ligne{\hfill\gzzz\ftt{\the\compteregle \ }  
\llrule {W} {W} {W} {W} {R} {W} {B} 
\hskip-3.5pt\rrrule {B} {W} {W} {B} {B} {W} {B} {}\hfill}
\vskip-3pt
\ligne{\hfill\gzzz\ftt{\the\compteregle \ }  
\llrule {W} {W} {W} {W} {R} {W} {W} 
\hskip-3.5pt\rrrule {B} {B} {W} {B} {B} {W} {B} {}\hfill}
\vskip-3pt
\ligne{\hfill\gzzz\ftt{\the\compteregle \ }  
\llrule {B} {W} {W} {W} {R} {W} {W} 
\hskip-3.5pt\rrrule {B} {W} {W} {B} {B} {W} {W} {}\hfill}
\vskip-3pt
\ligne{\hfill\gzzz\ftt{\the\compteregle \ }  
\llrule {W} {W} {W} {B} {R} {W} {W} 
\hskip-3.5pt\rrrule {B} {W} {W} {B} {B} {W} {W} {}\hfill}
}
\hfill
\vtop{\leftskip 0pt\parindent 0pt\hsize=\largeouille
\ligne{\hfill\ftt{controller} \hfill}
\ligne{\hfill\gzzz\ftt{\the\compteregle \ }  
\llrule {W} {B} {W} {W} {W} {W} {W} 
\hskip-3.5pt\rrrule {B} {W} {W} {B} {B} {W} {W} {}\hfill}
\vskip-3pt
\ligne{\hfill\gzzz\ftt{\the\compteregle \ }  
\llrule {W} {B} {W} {W} {W} {W} {B} 
\hskip-3.5pt\rrrule {B} {W} {W} {B} {B} {W} {W} {}\hfill}
\vskip-3pt
\ligne{\hfill\gzzz\ftt{\the\compteregle \ }  
\llrule {W} {W} {W} {B} {W} {W} {B} 
\hskip-3.5pt\rrrule {B} {W} {W} {B} {B} {W} {B} {}\hfill}
\vskip-3pt
\ligne{\hfill\gzzz\ftt{\the\compteregle \ }  
\llrule {B} {W} {W} {B} {W} {W} {W} 
\hskip-3.5pt\rrrule {B} {W} {W} {B} {B} {W} {W} {}\hfill}
\ligne{\hfill\gzzz\ftt{\the\compteregle \ }  
\llrule {W} {W} {W} {W} {W} {W} {W} 
\hskip-3.5pt\rrrule {B} {W} {W} {B} {B} {B} {W} {}\hfill}
\vskip-3pt
\ligne{\hfill\gzzz\ftt{\the\compteregle \ }  
\llrule {B} {W} {W} {W} {W} {W} {W} 
\hskip-3.5pt\rrrule {B} {W} {W} {B} {B} {B} {B} {}\hfill}
\vskip-3pt
\ligne{\hfill\gzzz\ftt{\the\compteregle \ }  
\llrule {W} {W} {W} {W} {W} {W} {B} 
\hskip-3.5pt\rrrule {B} {W} {W} {B} {B} {B} {B} {}\hfill}
\vskip-3pt
\ligne{\hfill\gzzz\ftt{\the\compteregle \ }  
\llrule {B} {W} {W} {W} {W} {W} {B} 
\hskip-3.5pt\rrrule {B} {W} {W} {B} {B} {B} {W} {}\hfill}
\vskip-3pt
\ligne{\hfill\gzzz\ftt{\the\compteregle \ }  
\llrule {B} {B} {B} {W} {W} {W} {W} 
\hskip-3.5pt\rrrule {W} {W} {W} {W} {W} {W} {B} {}\hfill}
\vskip-3pt
\ligne{\hfill\gzzz\ftt{\the\compteregle \ }  
\llrule {W} {W} {W} {W} {W} {W} {W} 
\hskip-3.5pt\rrrule {B} {R} {W} {B} {B} {W} {W} {}\hfill}
\vskip-3pt
\ligne{\hfill\gzzz\ftt{\the\compteregle \ }  
\llrule {W} {W} {W} {W} {W} {W} {B} 
\hskip-3.5pt\rrrule {B} {R} {W} {B} {B} {W} {W} {}\hfill}
}
\hfill
\vtop{\leftskip 0pt\parindent 0pt\hsize=\largeouille
\ligne{\hfill\ftt{fork} \hfill}
\ligne{\hfill\gzzz\ftt{\the\compteregle \ }  
\llrule {W} {W} {W} {W} {W} {W} {W} 
\hskip-3.5pt\rrrule {B} {B} {B} {B} {W} {R} {W} {}\hfill}
\vskip-3pt
\ligne{\hfill\gzzz\ftt{\the\compteregle \ }  
\llrule {W} {W} {W} {B} {W} {W} {W} 
\hskip-3.5pt\rrrule {B} {B} {B} {B} {W} {R} {B} {}\hfill}
\vskip-3pt
\ligne{\hfill\gzzz\ftt{\the\compteregle \ }  
\llrule {B} {W} {W} {W} {W} {W} {W} 
\hskip-3.5pt\rrrule {B} {B} {B} {B} {W} {R} {W} {}\hfill}
\vskip-3pt
\ligne{\hfill\gzzz\ftt{\the\compteregle \ }  
\llrule {W} {W} {W} {W} {W} {B} {B} 
\hskip-3.5pt\rrrule {B} {B} {B} {B} {W} {R} {W} {}\hfill}
\vskip-3pt
}
\hfill}
}
\hfill}
\vskip 5pt

\newdimen\svhsize
\newdimen\lalongue\lalongue=170pt
\def\traceline #1 #2 {%
        \svhsize=\hsize
        \hsize=\lalongue
	\ligne{\hbox to 20pt{\hfill\ftt{#1} }\hskip 20pt\hbox{\ftt{#2} }\hfill} {}
        \hsize=\svhsize
\vskip -3pt
}
\vskip 10pt
Rule~49 allows the entry into a cell
through its face~1. Rule~50 is the conservative rule for the cell giving access to
the tunnel. Rules 51 and~52 define two possible entries for a locomotive, either through
face~5 or through face~7. Rules~53 and~54 complete the motion rules, playing for that cell
the role of rules~45 and~46 for an ordinary cell of the track. The first part of 
Table~\ref{texecvoies} illustrates the application of those rules by a simulating 
computer program.

Rule~55 keeps the situation of an element of the track which will stop the locomotive
as indicated by rule~46: the locomotive seen through face~5 is not allowed to enter the
cell. A this point, note that rule~43 and 
\hbox{\llrule {W} {W} {W} {W} {W} {B} {W} 
\hskip-3.5pt\rrrule {B} {W} {W} {B} {B} {W} {B} } {} are rotationally identical to their 
minimal form
\hbox{\llrule {W} {W} {W} {W} {W} {W} {W} 
\hskip-3.5pt\rrrule {W} {B} {W} {B} {B} {B} {B} } {}. That allows us to accept the entry
of a locomotive into an element of the track through its face~4. The same can be said 
about rule~56 which is rotationally equivalent to the rule 
\hbox{\llrule {W} {B} {W} {W} {W} {W} {W} 
\hskip-3.5pt\rrrule {B} {W} {W} {B} {B} {W} {W} } {}, their minimal form being
\hbox{\llrule {W} {W} {W} {W} {W} {W} {B} 
\hskip-3.5pt\rrrule {W} {B} {W} {B} {B} {B} {W} } {}, so that the entry through face~4
is also ruled out in that case. 

Rules~57 and~58 deal with the element~$\eta$ of the track which stands by the cell of the 
controller. Both rules concern the case when the controller is blue which forbids the
locomotive to enter a cell. Rule~57 let the locomotive enter~$\eta$, rule~58 says that 
the locomotive which entered~$\eta$ does not stay in~$\eta$. Rule~59 and~60 are the 
conservative rule for the controller whose 
decoration consists of blue cells on its faces~6, 9, 10 and~11. Rule~59 applies to
a blank controller which let a locomotive go while rule~60 applies to a blue 
controller which stops a locomotive. As illustrated by Figure~\ref{fctrl}, the controller
is placed under the element where we want to stop a locomotive. The access to a controller
occurs below~\HH{} when the track is over that plane and it occurs upon it if the track
is below~it. The same tiles as for the entry to a tunnel are used in order to go from
one side of~\HH{} to its opposite side. Rules~61 and~62 show the change of colour in the
controller. 

\vtop{
\begin{tab}\label{texecvoies}
\leurre
Traces of execution of the rules of Table~{\rm\ref{ttracks}} on the different structures
used in the switches and crossings.
\end{tab}
\ligne{\hfill
\vtop{\leftskip 0pt\parindent 0pt\hsize=175pt
\ligne{\hfill the line :\hfill}
\vskip 3pt
\traceline {}   {+  -  -  -  -  +  -  -  -  -  +  -  -  -  -}{}
\traceline {}   {W  B  W  W  W  W  W  W  W  W  W  W  W  W  W}{}  
\traceline {0}  {W  W  B  W  W  W  W  W  W  W  W  W  W  W  W}{}  
\traceline {1}  {W  W  W  B  W  W  W  W  W  W  W  W  W  W  W}{}  
\traceline {2}  {W  W  W  W  B  W  W  W  W  W  W  W  W  W  W}{} 
\traceline {3}  {W  W  W  W  W  B  W  W  W  W  W  W  W  W  W}{} 
\traceline {4}  {W  W  W  W  W  W  B  W  W  W  W  W  W  W  W}{}  
\traceline {5}  {W  W  W  W  W  W  W  B  W  W  W  W  W  W  W}{}  
\traceline {6}  {W  W  W  W  W  W  W  W  B  W  W  W  W  W  W}{}
\traceline {7}  {W  W  W  W  W  W  W  W  W  B  W  W  W  W  W}{}  
\traceline {8}  {W  W  W  W  W  W  W  W  W  W  B  W  W  W  W}{}  
\traceline {9}  {W  W  W  W  W  W  W  W  W  W  W  B  W  W  W}{}  
\traceline {10} {W  W  W  W  W  W  W  W  W  W  W  W  B  W  W}{}  
\traceline {11} {W  W  W  W  W  W  W  W  W  W  W  W  W  B  W}{}  
\traceline {12} {W  W  W  W  W  W  W  W  W  W  W  W  W  W  B}{}  
\vskip 3pt
}
\hfill
\vtop{\leftskip 0pt\parindent 0pt\hsize=155pt
\ligne{\hfill the tunnel\hfill}\lalongue=150pt
\vskip 3pt
\traceline {} {+  -  -  -  -  +  -  -  -  -  +  -  -}{}
\traceline {} {W  W  W  W  W  W  W  W  W  W  W  B  W}{}
\traceline {0} {W  W  W  W  W  W  W  W  W  W  B  W  W}{} 
\traceline {1} {W  W  W  W  W  W  W  W  W  B  W  W  W}{}  
\traceline {2} {W  W  W  W  W  W  W  W  B  W  W  W  W}{}  
\traceline {3} {W  W  W  W  W  W  W  B  W  W  W  W  W}{}  
\traceline {4} {W  W  W  W  W  W  B  W  W  W  W  W  W}{}  
\traceline {5} {W  W  W  W  W  B  W  W  W  W  W  W  W}{}  
\traceline {6} {W  W  W  W  B  W  W  W  W  W  W  W  W}{}  
\traceline {7} {W  W  W  B  W  W  W  W  W  W  W  W  W}{}  
\traceline {8} {W  W  B  W  W  W  W  W  W  W  W  W  W}{}  
\traceline {9} {W  B  W  W  W  W  W  W  W  W  W  W  W}{}  
\traceline {10} {B  W  W  W  W  W  W  W  W  W  W  W  W}{}  
\vskip 3pt
}
\hfill}
\vskip 5pt
\ligne{\hfill
\vtop{\leftskip 0pt\parindent 0pt\hsize=155pt
\ligne{\hfill controller:\hfill}\lalongue=140pt
\traceline {} {+  -  -  -  -  +  -  -  -  -  +  -}{}
\traceline {}  {W  B  W  W  W  W  W  B  W  W  W  W}{}  
\traceline {0} {W  W  B  W  W  W  W  W  B  W  W  W}{}  
\traceline {1} {W  W  W  B  W  W  W  W  W  B  W  W}{}  
\traceline {2} {W  W  W  W  W  W  W  W  W  W  B  W}{}  
\traceline {3} {W  W  W  W  W  W  W  W  W  W  W  B}{}
\traceline {4} {W  W  W  W  W  W  W  W  W  W  W  W}{}
\vskip 3pt
}
\hfill
\vtop{\leftskip 0pt\parindent 0pt\hsize=170pt
\ligne{\hfill changing the control:\hfill}\lalongue=160pt
\traceline {} {+  -  -  -  -  +  -  -  -  -  +  -  -  -}{}
\traceline {}  {W  B  W  W  B  W  W  B  W  W  W  W  W  W}{}  
\traceline {0} {W  W  B  W  B  W  W  W  B  W  W  W  W  W}{}  
\traceline {1} {W  W  W  B  B  W  W  W  W  B  W  W  W  W}{}  
\traceline {2} {W  W  W  W  W  W  W  W  W  W  B  W  W  W}{}  
\traceline {3} {W  W  W  W  W  W  W  W  W  W  B  W  W  W}{}  
\traceline {4} {W  W  W  W  W  W  W  W  W  W  B  W  W  W}{}  
\traceline {5} {W  W  W  W  W  W  W  W  W  W  B  W  W  W}{}
\vskip 3pt
}
\hfill}
\vskip 5pt
\ligne{\hfill
\vtop{\leftskip 0pt\parindent 0pt\hsize=170pt
\ligne{\hfill fork:\hfill}\lalongue=160pt
\traceline {} {+  -  -  -  -  +  -  -  -  -  +  -  -  -}{}
\traceline { }  {W  B  W  W  W  W  W  W  W  W  W  W  W  W}{}  
\traceline {0}  {W  W  B  W  W  W  W  W  W  W  W  W  W  W}{}  
\traceline {1}  {W  W  W  B  W  W  W  W  W  W  W  W  W  W}{}  
\traceline {2}  {W  W  W  W  B  W  W  W  W  W  W  W  W  W}{}  
\traceline {3}  {W  W  W  W  W  B  W  W  W  B  W  W  W  W}{}  
\traceline {4}  {W  W  W  W  W  W  B  W  W  W  B  W  W  W}{}  
\traceline {5}  {W  W  W  W  W  W  W  B  W  W  W  B  W  W}{}  
\traceline {6}  {W  W  W  W  W  W  W  W  B  W  W  W  B  W}{} 
\vskip 3pt
}
\hfill
\vtop{\leftskip 0pt\parindent 0pt\hsize=155pt
\ligne{\hfill arc of a circle:\hfill}\lalongue=140pt
\traceline {} {+  -  -  -  -  +  -  -  -  -  +  -}{}
\traceline {}   {W  B  W  W  W  W  W  W  W  W  W  W}{}
\traceline {0}  {W  W  B  W  W  W  W  W  W  W  W  W}{}  
\traceline {1}  {W  W  W  B  W  W  W  W  W  W  W  W}{}  
\traceline {2}  {W  W  W  W  B  W  W  W  W  W  W  W}{}  
\traceline {3}  {W  W  W  W  W  B  W  W  W  W  W  W}{}  
\traceline {4}  {W  W  W  W  W  W  B  W  W  W  W  W}{}  
\traceline {5}  {W  W  W  W  W  W  W  B  W  W  W  W}{}  
\traceline {6}  {W  W  W  W  W  W  W  W  B  W  W  W}{} 
\traceline {7}  {W  W  W  W  W  W  W  W  W  B  W  W}{} 
\traceline {8}  {W  W  W  W  W  W  W  W  W  W  B  W}{}  
\traceline {9}  {W  W  W  W  W  W  W  W  W  W  W  B}{}  
\traceline {10} {W  W  W  W  W  W  W  W  W  W  W  W}{}  
\vskip 3pt
} 
\hfill}
}
\vskip 10pt
Rules~64 and 65 define a new element of the track which unconditionally 
stops a locomotive. It was used in the simulating program to facilitate the 
implementation.

The last four rules of the table deal with the fork. Rule~66 is the conservative rule
of the centre of the fork in idle configurations. Note the particular decoration:
blue faces~6, 7 8 and~9 and a red face~11, see Figure~\ref{fstab_fxfk} in 
Sub-section~\ref{newrailway}. Rule~67 show us that the locomotive enters the
fork through its face~2, rule~68 witnesses the crossing and rule~69 witnesses that two
locomotives exited through faces~4 and~5.

   Note that there are no special rules for the switches outside those we indicated 
up to now, even for the fixed switch. Indeed, when the locomotive comes from the 
left-hand side branch, it requires the rule~$\rho$ 
\hbox{\llrule {W} {W} {W} {W} {W} {B} {W} 
\hskip-3.5pt\rrrule {B} {W} {W} {B} {B} {W} {B} } {} and it requires rule~43 when
the locomotive comes from the right-hand side branch. Both rules have the same minimal 
form, namely
\hbox{\llrule {W} {W} {W} {W} {W} {W} {W} 
\hskip-3.5pt\rrrule {W} {B} {W} {B} {B} {B} {B} } {}. For the rule~$\rho$, the rotation 
around faces~1 and~9 gives that minimal form. For rule~43, the rotation around the 
vertices 0-2-3 and 6-10-11 yields the same minimal form.

  The main features of the switches are the organization of their needed 
circuits in the dodecagrid. We have seen that the elements of the track are enough for 
that purpose. The rules for the fork and for the controller complete the needed set of
rules.

\subsection{The rules for the register}\label{rreg}

   The present part of~\TT{} is the biggest as all the remaining rules are devoted to
the registers. 

First, in Sub-subsection~\ref{sbbrstrands} we examine the conservative 
rules for each strand of a register together with the few rules allowing the 
motions in the strand described in Sub-section~\ref{newrailway}. 
Then, in Sub-subsection~\ref{sbbrgrow} we give the rules for the growth of the 
register. Next, we study the decrementation in Sub-section~\ref{sbbrregdec}.
After that, Sub-subsection~\ref{sbbrreginc} manages the incrementation.
At last, Sub-subsection~\ref{srstop} deals with the halting of the
computation which concerns the register as a specific strand is devoted to that
operation.

\subsubsection{The strands of a register}\label{sbbrstrands}

   Before looking at each strand separately, we have to look at the implementation
we give to the strands.

    Here is an example of the constraint raised by the rotation invariance. 
Consider the following rules: 
$(a)$ 
\hbox{\llrule {W} {R} {Y} {W} {W} {R} \hskip-3.5pt \rrrule {W} {W} {W} {W} {W} {W} {R} }
and $(b)$
\hbox{\llrule {W} {R} {R} {Y} {W} {W} \hskip-3.5pt \rrrule {W} {W} {W} {W} {W} {W} {W} }.
For both of them, the neighbourhood has the reduced form: 
$(r)$ \hbox{\ftt{WWWWWWWWWRRY} }.
It can be checked on Figure~\ref{rotadodec3} that the rotation around the axis
through the midpoint of 4.5 and 7.8 transforms $(a)$ onto $(r)$ while a rotation
around faces~4 and~7 transforms $(b)$ onto $(r)$. Rule $(a)$ is 
necessary for the motion of a red locomotive on a strand according to the definition
given to that strand of a register in~\cite{mmarXiv21}. But rule~$(b)$ is needed
according to Proposition~\ref{pneighgen}.
Accordingly the definition of our strands have to be seriously tuned.

    We use the notations introduced in Sub-subsection~\ref{sbbreg}. Let us remind the 
reader that the four strands of a register are \RR$_c$, \RR$_i$, \RR$_d$
and \RR$_s$. Two of them, \RR$_c$ and~\RR$_d$ are over~\HH{} while the two others,
\RR$_i$ and~\RR$_s$ are below that plane. Each $\mathcal S$($n$) has its face~0 on~\HH{}
and for each of them, its face~1 has a side on a line~$\ell$ of~\HH{} attached to that
register. That fixes the numbering as already mentioned. Note that for strands 
over~\HH{} the numbering is increasing while clockwise turning around the projection of 
the cell. For the strands seen through a translucent \HH{} the numbering is 
counterclockwise increasing.

A cell has at least four important neighbours : a cell of the same coordinate belonging 
to another strand seen from its faces~0 and~1 and two cells of the same strand seen 
from its faces~5 and~2. We append two decorations to each $\mathcal S$($n$) : a blue 
cell on its faces~4 and~6 for \RR$_i$ and \RR$_d$, on its faces~3 and~7 on \RR$_c$ 
and~\RR$_s$. 

The main reason for those decorations is to ensure the rotation invariance of the rules.
Let us now explain why namely those faces are involved. In order to understand that
it is necessary to describe how the register is growing. We decided that \RR$_c$ is 
\sbb, at least, close to the growing end, that \RR$_s$ also is \sbb{} while \RR$_i$ is
\srr{} and \RR$_d$ is \syy. Let $N_t$ be the coordinate such that, at time~$t$,
$\mathcal S$$(n)$ is \sww{} if $n>N_t$ and that it is the colour of its strand if
$n < N_t$. For $n=N_t$, we decide that $\mathcal S$$(n)$ is \sgg. We also decide
that, at time $2t$, each neighbour of~$\mathcal S$$(n)$ which has a single non-blank
neighbour becomes~\sbb{} while $\mathcal S$$(n)$ remains \sgg. To decide what to do at 
time~$2t$+1, it is necessary to closer look at the situation. Two figures help us to
see the configuration of the end of the register: Figures~\ref{fcut} and~\ref{fregfin}.

The former figure is a cut along a plane~\VV{} which is perpendicular to~$\ell$ and which
contains a face of each $\mathcal S$$(n)$ for some~$n$. On the right-hand side of the
figure, we have the representation on~\HH{} while, on the left-hand side, we have
the projections of the tiles we can see. Remember the condition we have given for 
a neighbour $S_i$ of $\mathcal S$$(n)$: $S_i$ must have eleven white neighbours while
its single non-blank neighbour is $\mathcal S$$(n)$ itself whose colour is~\sgg.
Which faces of $\mathcal S$$(n)$ satisfy such a condition? 

The figure indicates that on all cells, {\it a priori}, we have all faces except faces~0,
and~1, and also face~2 for~\RR$_c$ and \RR$_s$ and face~5 for \RR$_i$ and \RR$_d$. 
But, we have to take into account the fact that $S_7$ and $S_8$ of each
$\mathcal S$($n$) can see $S_6$ and $S_{10}$ respectively on $\mathcal S$$(n$$-$$1)$ 
for \RR$_i$ and \RR$_d$, on $\mathcal S$$(n$+$1)$ for \RR$_c$ and \RR$_s$. Moreover,
$S_3$ of each $\mathcal S$($n$) can see $S_4$ on $\mathcal S$($n$$-$1) for \RR$_i$
and \RR$_d$ or on $\mathcal S$($n$+1) for \RR$_c$ and \RR$_s$. Now, on \RR$_c$ and 
\RR$_s$, $S_7$ of $\mathcal S$($n$) can see $S_6$ of \RR$_d$ and \RR$_i$
respectively for the same coordinate. If we wish to obtain \sgg{} on 
$\mathcal S$($N_t$+1),
we must take advantage of the fact that the $S_2$ of \RR$_c$ and \RR$_s$ can see 
the $S_5$ of both \RR$_i$ and \RR$_d$, similarly for the $S_5$ of \RR$_i$ and \RR$_d$
with the $S_2$ of both \RR$_c$ and \RR$_s$. Accordingly, each of those $S_i$'s have 
two \sbb-neighbours and they all can see a \sgg-cell. We can then decide that
in such a case the \sbb-cell becomes \sgg{} and that the former \sgg-cell takes the
colour of its strand. We remain with what to do with the other \sbb-neighbours of
the \sgg-$\mathcal S$$(N_t)$. From what we just remarked, $S_6$ or $S_7$, and also $S_3$ 
or $S_4$, it depends on which strand the cell is, is blue and can see another blue 
neighbour. We can decide that such a \sbb-cell remains \sbb. 

If we do that, what can be said? In that case, we can see that on \RR$_c$ or \RR$_s$,
an $S_6$ neighbour of the \sgg-cell which is white remains white as far as it can see
the \sgg-cell and a \sbb-decoration on $\mathcal S$($N_t$$-$1). Accordingly, that cell
remains white. We have the same argument with $S_4$ for the same cell. A similar argument
holds for $S_7$ and $S_3$ on the \sgg-cell of \RR$_i$ and of \RR$_d$. Accordingly,
at time $2t$+1 we can decide that the \sbb-cells which can see their \sgg-neighbour as
their unique non-blank neighbour become white. Accordingly, the new cell of the strand
receives the same decoration as its neighbour $\mathcal S$($N_t$$-$1) on the strand.
All other neighbours are white outside its neighbours~0, 1, 2, 5 and outside the 
decorations, neighbours~3 and~7 on \RR$_c$ and on \RR$_s$, neighbours~4 and~6 on \RR$_i$
and on \RR$_d$.

\vskip 10pt
\vtop{
\ligne{\hfill
\includegraphics[scale=1]{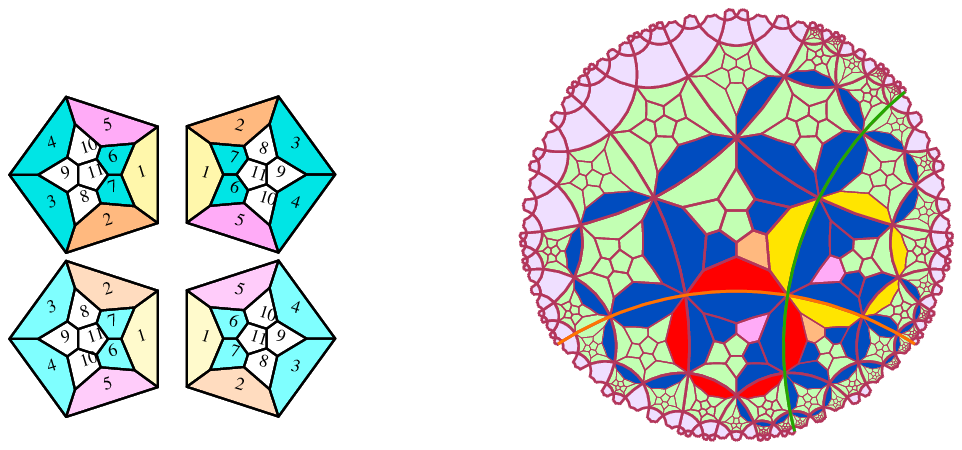}
\hfill}
\begin{fig}\label{fcut}
\leurre
Cut of the strands of a register along a plane~\VV, perpendicular to~$\ell$.
Note the correspondence of colours between the left- and right-hand sides of the figure.
\end{fig}
}
\vskip 5pt
Those conclusions hold for $\mathcal S$($n$) with $n>0$. For $n=0$ it is important that
the cell knows that it is the first one of the strand. As the cell cannot see the 
decoration of its neighbours, that information of being the first cell must be given
in the decoration. For \RR$_c$, it receives one additional blue cell on face~9.
A single blue cell on face~9 is also given to~\RR$_i$. However,
\RR$_d$ receives two additional red cells on faces~9 and~11. Eventually, \RR$_s$ is fitted
with a single blue cell on face~9.

We gather that information on the neighbours of $\mathcal S$($n$) as follows:
\vskip 10pt
\ligne{\hfill
$\vcenter{\vtop{\leftskip 0pt\parindent 0pt\hsize=225pt
	\ligne{\hfill$\mathcal S$($n$), $n>0$:\hfill}
\ligne{\RR$_c$ : 0: \srr, 1: \syy, 2:\sbb$^*$, 3:\sbb, 5:\sbb$^*$, 7: \sbb,
others: \sww;\hfill}
\ligne{\RR$_i$ : 0: \sbb$^*$, 1: \sbb, 2:\srr, 4:\sbb, 5:\srr, 6: \sbb, 
others: \sww;\hfill}
\ligne{\RR$_d$ : 0: \sbb, 1: \sbb$^*$, 2:\syy, 4:\sbb, 5:\syy, 6: \sbb, 
others: \sww;\hfill}
\ligne{\RR$_s$ : 0: \syy, 1: \srr, 2:\sbb, 3:\sbb, 5:\sbb, 7: \sbb, 
others: \sww;\hfill}
	\ligne{\hfill$\mathcal S$(0):\hfill}
\ligne{\RR$_c$ : 0: \srr, 1: \syy, 2:\sbb$^*$, 3:\sbb, 5:\sww, 7: \sbb,
9: \sbb, others: \sww;\hfill}
\ligne{\RR$_i$ : 0: \sbb$^*$, 1: \sbb, 2:\sww, 4:\sbb, 5:\srr, 6: \sbb, 
9: \sbb, others: \sww;\hfill}
\ligne{\RR$_d$ : 0: \sbb, 1: \sbb$^*$, 2:\syy, 4:\sbb, 5:\syy, 6: \sbb, 
9: \srr, 11: \srr, others: \sww;\hfill}
\ligne{\RR$_s$ : 0: \syy, 1: \srr, 2:\sbb, 3:\sbb, 5:\sbb, 7: \sbb, 
9: \sbb, others: \sww;\hfill}
}}$
\hfill(\numerrel)\hskip 10pt}
\vskip 10pt
Note that in (7), the asterisk indicates that \sbb{} may be replaced by \sww: remember
that on~\RR$_c$, the value~$v$ stored in the register is indicated by $v$ contiguous
blank cells starting from \RR$_c$(0) which, consequently, is \sbb{} when $v=0$. That
possibility also concerns \RR$_i$ and \RR$_d$ which can see \RR$_c$ through their face~0
and~1 respectively. It does not concern \RR$_s$ which cannot see \RR$_c$.

Figure~\ref{fregfin} illustrates the configuration of the end of the register
when the \sgg-cells are covered by \sbb-neighbours. We can see the faces which are 
covered with a blank cell. The left-hand side part of the figure illustrates \RR$_c$
and \RR$_d$, the strands above~\HH. The right-hand side part of the figure illustrates
\RR$_i$ and \RR$_s$ which hang below~\HH{} and which are viewed through a 
translucent~\HH. The figure applies the convention for colouring the faces of the
dodecahedrons we already defined. It can be seen that the cells of \RR$_c$ and of \RR$_s$ 
are blue, while those of \RR$_i$ are red and those of \RR$_d$ are yellow.

\vskip 10pt
\vtop{
\ligne{\hfill
\includegraphics[scale=1]{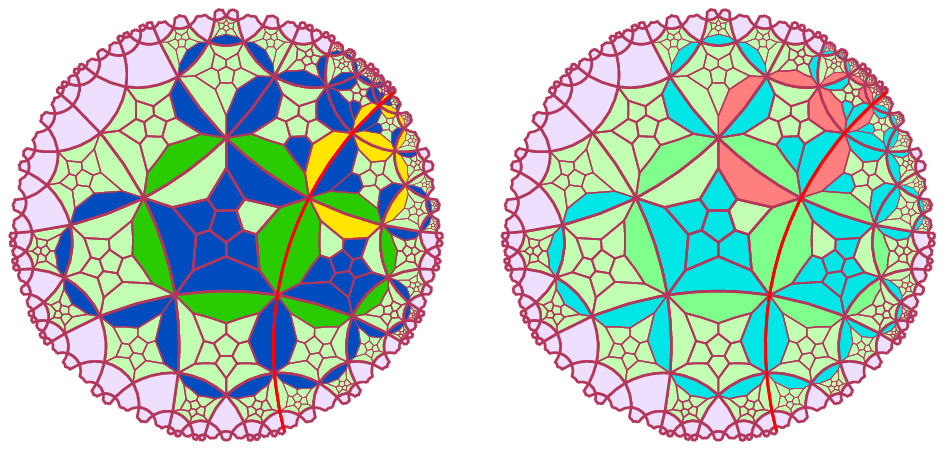}
\hfill}
\begin{fig}\label{fregfin}
\leurre
The idle configuration of the growing end of a register at the time when blue cells cover
the green ones. To left, over~\HH, to right, what is below a translucent~\HH. Note the 
line~$\ell$, in red in the pictures.
\end{fig}
}

\vskip 10pt
\vtop{
\ligne{\hfill
\includegraphics[scale=1]{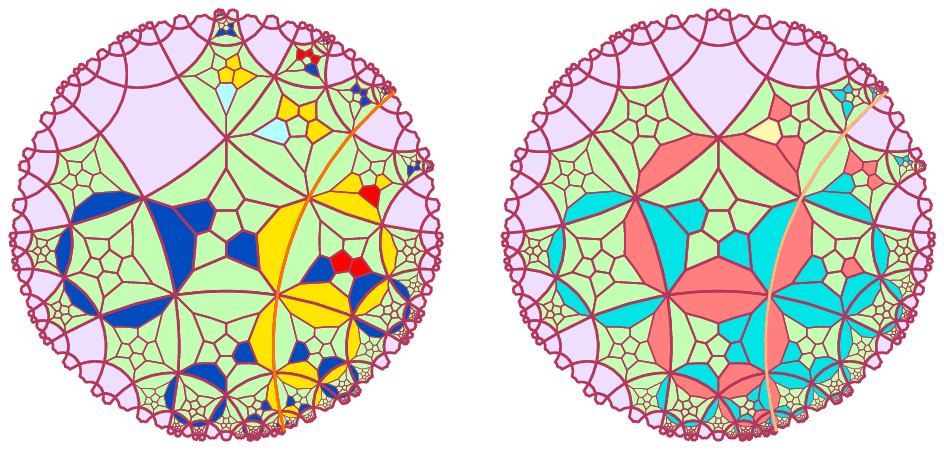}
\hfill}
\begin{fig}\label{fregdeb}
\leurre
The idle configuration of $\mathcal S$$(n)$ for $n\in\{$$-$$1,0..3\}$ for \RR$_c$,
\RR$_i$, \RR$_d$ and \RR$_s$.
To left, the strands over~\HH, to right, those which
hang below a translucent~\HH. Note the line~$\ell$, in orange in the figures.
\end{fig}
}
\vskip 10pt

The orientations for the motion of a locomotive
on a strand are from face~5 to face~2 on all strands. In both \RR$_c$ and \RR$_s$ it means
a motion from $\mathcal S$(0) to the \sgg-cell of $\mathcal S$, while it is the opposite
direction on both \RR$_d$ and \RR$_i$. Those orientations are conformal to the role of
the strands: on \RR$_c$ the locomotive looks after the lowest $c$ such that \RR$_c$($c$) 
is blue, on \RR$_s$ it looks after the \sgg-cell; on \RR$_i$ and \RR$_d$, the locomotive
looks after $\mathcal S$(-1). The cells $\mathcal S$(-1) can be seen on 
Figure~\ref{fregdeb}.

Note the specific decorations of the cells $\mathcal S$(-1). We summarize what is 
illustrated by Figure~\ref{fregdeb} as follow, following the conventions of~(7):

\vskip 10pt
\ligne{\hfill
$\vcenter{\vtop{\leftskip 0pt\parindent 0pt\hsize=160pt
	\ligne{\hfill$\mathcal S$(-1):\hfill}
\ligne{\RR$_c$ : 2, \sbb$^*$, 6, 9, 10: \syy, others: \sww;\hfill}
\ligne{\RR$_i$ : 5: \srr, 6,9: \srr, 10: \syy, others: \sww;\hfill}
\ligne{\RR$_d$ : 2: \syy, 6,9: \syy, 10:\srr, others: \sww;\hfill}
\ligne{\RR$_s$ : 5: \sbb, 6: \sbb, 7: \srr, 9, 10: \sbb, others: \sww;\hfill}
	\ligne{\hfill\RR$_c$(-2):\hfill}
\ligne{$a$ : 6, 9, 10: \sbb, others: \sww;\hfill}
\ligne{$b$ : 6: \srr, 7: \sbb, 9, 10: \srr, others: \sww;\hfill}
\ligne{$c$ : 6, 7, 11: \syy, 9: \sbb, others: \sww;\hfill}
}}$
\hfill(\numerrel)\hskip 10pt}
\vskip 10pt
We indicated three cells \RR$_c$(-2), denoted by~$a$, by~$b$ and by~$c$. The reason
is that three paths abut \RR$_c$(-1): the first path ending with \RR$_c$(-2)$_a$
leads a locomotive from the \DDI{} attached to the register.
The second path ending with \RR$_c$(-2)$_b$ leads the locomotive from the \DDD{}
attached to the register. The third path, starting from \RR$_c$(-2)$_c$, leads
the locomotive from the register to the \DDD{} of that register when the locomotive
failed to perform the decrementation.

It is now time to look at the rules. We start with the conservative rules which allow the
structure to remain unchanged as long as the locomotive is not present. We also append
the motion rules for coming locomotives on \RR$_c$ and~\RR$_s$, for leaving ones 
on \RR$_i$ and \RR$_d$.

\ligne{\hfill
\vtop{
\begin{tab}\label{tregdeb}
\leurre
Tables for the strands of the register, green end being excluded.
\end{tab}
\ligne{\hfill
\vtop{\leftskip 0pt\parindent 0pt\hsize=\largeouille
\ligne{\hfill\ftt{\RR$_c$} \hfill}
\ligne{\hfill\gzzz\ftt{\the\compteregle \ }  
\llrule {W} {R} {Y} {W} {B} {W} {W} 
\hskip-3.5pt\rrrule {W} {B} {W} {B} {W} {W} {W} {}\hfill}
\vskip-3pt
\ligne{\hfill\gzzz\ftt{\the\compteregle \ }  
\llrule {W} {R} {Y} {W} {B} {W} {W} 
\hskip-3.5pt\rrrule {W} {B} {W} {W} {W} {W} {W} {}\hfill}
\vskip-3pt
\ligne{\hfill\gzzz\ftt{\the\compteregle \ }  
\llrule {B} {R} {Y} {B} {B} {W} {W} 
\hskip-3.5pt\rrrule {W} {B} {W} {W} {W} {W} {B} {}\hfill}
\ligne{\hfill\gzzz\ftt{\the\compteregle \ }  
\llrule {W} {R} {Y} {W} {B} {W} {B} 
\hskip-3.5pt\rrrule {W} {B} {W} {W} {W} {W} {B} {}\hfill}
\vskip-3pt
\ligne{\hfill\gzzz\ftt{\the\compteregle \ }  
\llrule {B} {R} {Y} {W} {B} {W} {W} 
\hskip-3.5pt\rrrule {W} {B} {W} {W} {W} {W} {W} {}\hfill}
\vskip-3pt
\ligne{\hfill\gzzz\ftt{\the\compteregle \ }  
\llrule {W} {R} {Y} {B} {B} {W} {W} 
\hskip-3.5pt\rrrule {W} {B} {W} {W} {W} {W} {W} {}\hfill}
\ligne{\hfill\gzzz\ftt{\the\compteregle \ }  
\llrule {W} {R} {Y} {W} {B} {W} {B} 
\hskip-3.5pt\rrrule {W} {B} {W} {B} {W} {W} {B} {}\hfill}
\vskip-3pt
\ligne{\hfill\gzzz\ftt{\the\compteregle \ }  
\llrule {W} {R} {Y} {B} {B} {W} {W} 
\hskip-3.5pt\rrrule {W} {B} {W} {B} {W} {W} {W} {}\hfill}
\ligne{\hfill\gzzz\ftt{\the\compteregle \ }  
\llrule {W} {R} {Y} {W} {B} {W} {R} 
\hskip-3.5pt\rrrule {W} {B} {W} {W} {W} {W} {R} {}\hfill}
\vskip-3pt
\ligne{\hfill\gzzz\ftt{\the\compteregle \ }  
\llrule {R} {R} {Y} {W} {B} {W} {W} 
\hskip-3.5pt\rrrule {W} {B} {W} {W} {W} {W} {W} {}\hfill}
\vskip-3pt
\ligne{\hfill\gzzz\ftt{\the\compteregle \ }  
\llrule {W} {R} {Y} {R} {B} {W} {W} 
\hskip-3.5pt\rrrule {W} {B} {W} {W} {W} {W} {W} {}\hfill}
\ligne{\hfill\gzzz\ftt{\the\compteregle \ }  
\llrule {W} {R} {Y} {W} {B} {W} {R} 
\hskip-3.5pt\rrrule {W} {B} {W} {B} {W} {W} {R} {}\hfill}
\vskip-3pt
\ligne{\hfill\gzzz\ftt{\the\compteregle \ }  
\llrule {W} {R} {Y} {R} {B} {W} {W} 
\hskip-3.5pt\rrrule {W} {B} {W} {B} {W} {W} {W} {}\hfill}
}
\hfill
\vtop{\leftskip 0pt\parindent 0pt\hsize=\largeouille
\ligne{\hfill\ftt{\RR$_i$} \hfill}
\ligne{\hfill\gzzz\ftt{\the\compteregle \ }  
\llrule {R} {W} {B} {W} {W} {B} {R} 
\hskip-3.5pt\rrrule {B} {W} {W} {B} {W} {W} {R} {}\hfill}
\vskip-3pt
\ligne{\hfill\gzzz\ftt{\the\compteregle \ }  
\llrule {R} {B} {B} {W} {W} {B} {R} 
\hskip-3.5pt\rrrule {B} {W} {W} {B} {W} {W} {R} {}\hfill}
\vskip-3pt
\ligne{\hfill\gzzz\ftt{\the\compteregle \ }  
\llrule {R} {B} {B} {R} {W} {B} {R} 
\hskip-3.5pt\rrrule {B} {W} {W} {W} {W} {W} {R} {}\hfill}
\ligne{\hfill\gzzz\ftt{\the\compteregle \ }  
\llrule {R} {W} {B} {R} {W} {B} {B} 
\hskip-3.5pt\rrrule {B} {W} {W} {W} {W} {W} {B} {}\hfill}
\vskip-3pt
\ligne{\hfill\gzzz\ftt{\the\compteregle \ }  
\llrule {B} {W} {B} {R} {W} {B} {R} 
\hskip-3.5pt\rrrule {B} {W} {W} {W} {W} {W} {R} {}\hfill}
\vskip-3pt
\ligne{\hfill\gzzz\ftt{\the\compteregle \ }  
\llrule {R} {W} {B} {B} {W} {B} {R} 
\hskip-3.5pt\rrrule {B} {W} {W} {W} {W} {W} {R} {}\hfill}
\vskip-3pt
\ligne{\hfill\gzzz\ftt{\the\compteregle \ }  
\llrule {R} {W} {B} {R} {W} {B} {R} 
\hskip-3.5pt\rrrule {B} {W} {W} {W} {W} {W} {R} {}\hfill}
\ligne{\hfill\gzzz\ftt{\the\compteregle \ }  
\llrule {B} {W} {B} {W} {W} {B} {R} 
\hskip-3.5pt\rrrule {B} {W} {W} {B} {W} {W} {R} {}\hfill}
\vskip-3pt
\ligne{\hfill\gzzz\ftt{\the\compteregle \ }  
\llrule {R} {W} {B} {W} {W} {B} {B} 
\hskip-3.5pt\rrrule {B} {W} {W} {B} {W} {W} {B} {}\hfill}
\vskip-3pt
\ligne{\hfill\gzzz\ftt{\the\compteregle \ }  
\llrule {R} {W} {B} {B} {W} {B} {R} 
\hskip-3.5pt\rrrule {B} {W} {W} {B} {W} {W} {R} {}\hfill}
\ligne{\hfill\ftt{\RR$_d$} \hfill}
\ligne{\hfill\gzzz\ftt{\the\compteregle \ }  
\llrule {Y} {B} {W} {W} {W} {B} {Y} 
\hskip-3.5pt\rrrule {B} {W} {W} {R} {W} {R} {Y} {}\hfill}
\vskip-3pt
\ligne{\hfill\gzzz\ftt{\the\compteregle \ }  
\llrule {Y} {B} {W} {Y} {W} {B} {Y} 
\hskip-3.5pt\rrrule {B} {W} {W} {W} {W} {W} {Y} {}\hfill}
\vskip-3pt
\ligne{\hfill\gzzz\ftt{\the\compteregle \ }  
\llrule {Y} {B} {B} {W} {W} {B} {Y} 
\hskip-3.5pt\rrrule {B} {W} {W} {R} {W} {R} {Y} {}\hfill}
\vskip-3pt
\ligne{\hfill\gzzz\ftt{\the\compteregle \ }  
\llrule {Y} {B} {B} {Y} {W} {B} {Y} 
\hskip-3.5pt\rrrule {B} {W} {W} {W} {W} {W} {Y} {}\hfill}
}
\hfill
\vtop{\leftskip 0pt\parindent 0pt\hsize=\largeouille
\ligne{\hfill\ftt{\RR$_d$, continued} \hfill}
\ligne{\hfill\gzzz\ftt{\the\compteregle \ }  
\llrule {Y} {B} {W} {Y} {W} {B} {R} 
\hskip-3.5pt\rrrule {B} {W} {W} {W} {W} {W} {R} {}\hfill}
\vskip-3pt
\ligne{\hfill\gzzz\ftt{\the\compteregle \ }  
\llrule {R} {B} {W} {Y} {W} {B} {Y} 
\hskip-3.5pt\rrrule {B} {W} {W} {W} {W} {W} {Y} {}\hfill}
\vskip-3pt
\ligne{\hfill\gzzz\ftt{\the\compteregle \ }  
\llrule {Y} {B} {W} {R} {W} {B} {Y} 
\hskip-3.5pt\rrrule {B} {W} {W} {W} {W} {W} {Y} {}\hfill}
\vskip-3pt
\ligne{\hfill\gzzz\ftt{\the\compteregle \ }  
\llrule {Y} {B} {W} {W} {W} {B} {R} 
\hskip-3.5pt\rrrule {B} {W} {W} {R} {W} {R} {R} {}\hfill}
\vskip-3pt
\ligne{\hfill\gzzz\ftt{\the\compteregle \ }  
\llrule {R} {B} {W} {W} {W} {B} {Y} 
\hskip-3.5pt\rrrule {B} {W} {W} {R} {W} {R} {Y} {}\hfill}
\vskip-3pt
\ligne{\hfill\gzzz\ftt{\the\compteregle \ }  
\llrule {Y} {B} {W} {R} {W} {B} {Y} 
\hskip-3.5pt\rrrule {B} {W} {W} {R} {W} {R} {Y} {}\hfill}
\ligne{\hfill\ftt{\RR$_s$} \hfill}
\ligne{\hfill\gzzz\ftt{\the\compteregle \ }  
\llrule {B} {Y} {R} {B} {B} {W} {W} 
\hskip-3.5pt\rrrule {W} {B} {W} {B} {W} {W} {B} {}\hfill}
\vskip-3pt
\ligne{\hfill\gzzz\ftt{\the\compteregle \ }  
\llrule {B} {Y} {R} {B} {B} {W} {B} 
\hskip-3.5pt\rrrule {W} {B} {W} {W} {W} {W} {B} {}\hfill}
\ligne{\hfill\gzzz\ftt{\the\compteregle \ }  
\llrule {B} {Y} {R} {B} {B} {W} {R} 
\hskip-3.5pt\rrrule {W} {B} {W} {B} {W} {W} {R} {}\hfill}
\vskip-3pt
\ligne{\hfill\gzzz\ftt{\the\compteregle \ }  
\llrule {R} {Y} {R} {B} {B} {W} {W} 
\hskip-3.5pt\rrrule {W} {B} {W} {B} {W} {W} {B} {}\hfill}
\vskip-3pt
\ligne{\hfill\gzzz\ftt{\the\compteregle \ }  
\llrule {B} {Y} {R} {R} {B} {W} {W} 
\hskip-3.5pt\rrrule {W} {B} {W} {B} {W} {W} {B} {}\hfill}
\vskip-3pt
\ligne{\hfill\gzzz\ftt{\the\compteregle \ }  
\llrule {B} {Y} {R} {B} {B} {W} {R} 
\hskip-3.5pt\rrrule {W} {B} {W} {W} {W} {W} {R} {}\hfill}
\vskip-3pt
\ligne{\hfill\gzzz\ftt{\the\compteregle \ }  
\llrule {R} {Y} {R} {B} {B} {W} {B} 
\hskip-3.5pt\rrrule {W} {B} {W} {W} {W} {W} {B} {}\hfill}
\vskip-3pt
\ligne{\hfill\gzzz\ftt{\the\compteregle \ }  
\llrule {B} {Y} {R} {R} {B} {W} {B} 
\hskip-3.5pt\rrrule {W} {B} {W} {W} {W} {W} {B} {}\hfill}
}
\hfill}
}
\hfill}
\vskip 10pt
In order to check the rules, it is enough to combine the information given in
(7), (8) and (4), knowing that the concerned neighbours are 0, 1, 3, 4, 6, 7 for (7),
we have to consider also neighbours 9 and 11 for~(8) and, at last, the concerned 
neighbours for~(4) are~2 and~5. Of course, we have to take into account the direction
of the motion: from~5 to~2 on~\RR$_c$ and on~\RR$_s$, while it is from~2 to~5 on both
\RR$_i$ and \RR$_d$.

There are more rules for~\RR$_c$ for two reasons. The first one is that contrarily to the
other strands, the colour of~\RR$_c$ may change during the computation, at least on its
beginning part. It is the reason why we need to duplicate conservative rules. The second
reason is that there are two possible motions for the locomotive, although always from
face~5 to face~2. For an incrementation, a blue locomotive runs on the strand while for a
decrementation, it is a red one. On both cases, the locomotive runs on the blank part of
the strand only. It is the reason while rules~78 to~80 are copied from rules 73 to~74
where \sbb{} is replaced by \srr{} at the appropriate places. Also note that the 
decoration of \RR$_c$(0) requires specific rules, both for the conservative ones as well 
as for the motion ones.

The specificity of $\mathcal S$(0) also concerns the other strands both for the 
conservative rules and for the motion ones. Duplication of a few rules occur also for 
\RR$_i$ and \RR$_d$ as far as those both strand can see \RR$_c$. Accordingly, neighbour~0
for~\RR$_i$ and neighbour~1 for~\RR$_d$ are either \sbb{} or \sww{}. 

We did not mention the rules for $\mathcal S$(-1) and for \RR$_c$(-2)$_b$ and 
for \RR$_c$(-2)$_c$: we postpone that to the study of the decrementation.
Presently, we turn to the growth of the register.

\subsubsection{The growing end of the register}~\label{sbbrgrow}

   We have studied the process in the previous sub-subsection thanks to 
Proposition~\ref{pneighgen} and thank to Figures~\ref{fregdeb} and~\ref{fregfin}.
We revisit the argument given there on checking the rules devoted to that point.

\ligne{\hfill
\vtop{
\begin{tab}\label{treggrow}
\leurre
Rules for the growth of the register.
\end{tab}
\vskip -7pt
\ligne{\hfill
\vtop{\leftskip 0pt\parindent 0pt\hsize=\largeouille
\ligne{\hfill\ftt{conservation} \hfill}
\ligne{\hfill\gzzz\ftt{\the\compteregle \ }  
\llrule {B} {R} {Y} {G} {B} {W} {B} 
\hskip-3.5pt\rrrule {W} {B} {W} {W} {W} {W} {B} {}\hfill}
\vskip-3pt
\ligne{\hfill\gzzz\ftt{\the\compteregle \ }  
\llrule {R} {B} {B} {R} {W} {B} {G} 
\hskip-3.5pt\rrrule {B} {W} {W} {W} {W} {W} {R} {}\hfill}
\vskip-3pt
\ligne{\hfill\gzzz\ftt{\the\compteregle \ }  
\llrule {Y} {B} {B} {Y} {W} {B} {G} 
\hskip-3.5pt\rrrule {B} {W} {W} {W} {W} {W} {Y} {}\hfill}
\vskip-3pt
\ligne{\hfill\gzzz\ftt{\the\compteregle \ }  
\llrule {B} {Y} {R} {G} {B} {W} {B} 
\hskip-3.5pt\rrrule {W} {B} {W} {W} {W} {W} {B} {}\hfill}
\ligne{\hfill\gzzz\ftt{\the\compteregle \ }  
\llrule {G} {G} {G} {W} {W} {W} {B} 
\hskip-3.5pt\rrrule {W} {W} {W} {W} {W} {W} {G} {}\hfill}
\vskip-3pt
\ligne{\hfill\gzzz\ftt{\the\compteregle \ }  
\llrule {G} {G} {G} {R} {W} {W} {W} 
\hskip-3.5pt\rrrule {W} {W} {W} {W} {W} {W} {G} {}\hfill}
\vskip-3pt
\ligne{\hfill\gzzz\ftt{\the\compteregle \ }  
\llrule {G} {G} {G} {Y} {W} {W} {W} 
\hskip-3.5pt\rrrule {W} {W} {W} {W} {W} {W} {G} {}\hfill}
}
\hfill
\vtop{\leftskip 0pt\parindent 0pt\hsize=\largeouille
\ligne{\hfill\ftt{growth : 1$^{\rm st}$ step} \hfill}
\ligne{\hfill\gzzz\ftt{\the\compteregle \ }  
\llrule {W} {G} {W} {W} {W} {W} {W} 
\hskip-3.5pt\rrrule {W} {W} {W} {W} {W} {W} {B} {}\hfill}
\ligne{\hfill\gzzz\ftt{\the\compteregle \ }  
\llrule {G} {G} {G} {B} {B} {W} {B} 
\hskip-3.5pt\rrrule {W} {B} {B} {B} {B} {B} {B} {}\hfill}
\vskip-3pt
\ligne{\hfill\gzzz\ftt{\the\compteregle \ }  
\llrule {G} {G} {G} {R} {W} {B} {B} 
\hskip-3.5pt\rrrule {B} {W} {B} {B} {B} {B} {R} {}\hfill}
\vskip-3pt
\ligne{\hfill\gzzz\ftt{\the\compteregle \ }  
\llrule {G} {G} {G} {Y} {W} {B} {B} 
\hskip-3.5pt\rrrule {B} {W} {B} {B} {B} {B} {Y} {}\hfill}
%
\ligne{\hfill\ftt{growth : 2$^{\rm nd}$ step} \hfill}
\ligne{\hfill\gzzz\ftt{\the\compteregle \ }  
\llrule {B} {G} {B} {B} {W} {W} {W} 
\hskip-3.5pt\rrrule {W} {W} {W} {W} {W} {W} {G} {}\hfill}
\vskip-3pt
\ligne{\hfill\gzzz\ftt{\the\compteregle \ }  
\llrule {B} {G} {B} {W} {W} {W} {W} 
\hskip-3.5pt\rrrule {W} {W} {W} {W} {W} {W} {B} {}\hfill}
\vskip-3pt
\ligne{\hfill\gzzz\ftt{\the\compteregle \ }  
\llrule {B} {G} {W} {W} {W} {W} {W} 
\hskip-3.5pt\rrrule {W} {W} {W} {W} {W} {W} {W} {}\hfill}
}
\hfill
\vtop{\leftskip 0pt\parindent 0pt\hsize=\largeouille
\ligne{\hfill\ftt{growth : 3$^{\rm rd}$ step} \hfill}
\ligne{\hfill\gzzz\ftt{\the\compteregle \ }  
\llrule {R} {B} {B} {R} {W} {B} {R} 
\hskip-3.5pt\rrrule {B} {W} {W} {W} {W} {W} {R} {}\hfill}
\vskip-3pt
\ligne{\hfill\gzzz\ftt{\the\compteregle \ }  
\llrule {Y} {B} {B} {Y} {W} {B} {W} 
\hskip-3.5pt\rrrule {B} {W} {W} {W} {W} {W} {Y} {}\hfill}
\vskip-3pt
\ligne{\hfill\gzzz\ftt{\the\compteregle \ }  
\llrule {Y} {B} {B} {Y} {W} {B} {Y} 
\hskip-3.5pt\rrrule {B} {W} {W} {W} {W} {W} {Y} {}\hfill}
\vskip-3pt
\ligne{\hfill\gzzz\ftt{\the\compteregle \ }  
\llrule {B} {Y} {R} {W} {B} {W} {B} 
\hskip-3.5pt\rrrule {W} {B} {W} {W} {W} {W} {B} {}\hfill}
\vskip-3pt
}
\hfill}
}
\hfill}
\vskip 10pt

   Table~\ref{treggrow} gives us the rules for the growth of the strands of a register.
Our first remark is that the strands grow all together at the same pace. It is important
for the achievement of the process. The first column of the table indicates a
conservative situation: if $g$ is the coordinate of the \sgg-cells, rules~111 to~114
give the rules for the conservation of $\mathcal S$($g$$-$1) for all strands. Then,
rules~115 to~117 give the conservation of~\sgg{} in the first step of the process.
Note that rule~115 concerns both \RR$_c$($g$) and \RR$_s$($g$) as far as for those 
\sgg-cells the non-blank and non-\sgg{} neighbour is \sbb{} in $\mathcal S$($g$$-$1).

At the same time~$t$ when rules~111 up to~117 apply, rule~118 apply to all blank 
neighbours of~$\mathcal S$($g$) which have a single non-blank neighbour: the \sgg-cell
at the coordinate $g$. All those cells become \sbb{} as indicated by rule~118.

The second column of the table contains rule~118 and the rules which apply at time~$t$+1.
We can see on rule~119 which applies to both \RR$_c$($g$) and \RR$_s$($g$) that
the \sgg-cell becomes \sbb. It means that on those strands, the \sgg-cell covered with
blue has taken the colour of its strand, exactly as indicated in 
Sub-subsection~\ref{sbbrstrands}. Rule~119 also indicates that the neighbours~4 and~6 of 
those cells remain white, which is conformal to the decoration of the cells of those
strands. Rule~120, 121 deal with \RR$_i$($g$), \RR$_d$($g$) respectively, also giving 
to that cell the colour of its strand. We also note that here, the neighbours~3 and~7 of
those cells remain white, which is also conformal to the decoration of a cell of those 
strands.

\ligne{\hfill
\vtop{
\begin{tab}\label{texecreggrow}
\leurre
	Traces of execution of the rules of Table~{\rm\ref{treggrow}}.
\end{tab}
\vskip -7pt
\ligne{\hfill
\vtop{\leftskip 0pt\parindent 0pt\hsize=135pt
\lalongue=129pt
\traceline {0} {+ - - - - + - - - -}{}    
\traceline {c}  {B  B  B  B  G  W  W  W  W  W}{}  
\traceline {i}  {R  R  R  R  G  W  W  W  W  W}{}
\traceline {d}  {Y  Y  Y  Y  G  W  W  W  W  W}{}  
\traceline {s}  {B  B  B  B  G  W  W  W  W  W}{}
\traceline {1} {+ - - - - + - - - -}{}    
\traceline {c}  {B  B  B  B  G  B  W  W  W  W}{}
\traceline {i}  {R  R  R  R  G  B  W  W  W  W}{}
\traceline {d}  {Y  Y  Y  Y  G  B  W  W  W  W}{}
\traceline {s}  {B  B  B  B  G  B  W  W  W  W}{} 
\traceline {2} {+ - - - - + - - - -}{}    
\traceline {c}  {B  B  B  B  B  G  W  W  W  W}{}  
\traceline {i}  {R  R  R  R  R  G  W  W  W  W}{}  
\traceline {d}  {Y  Y  Y  Y  Y  G  W  W  W  W}{}  
\traceline {s}  {B  B  B  B  B  G  W  W  W  W}{}  
}\hfill
\vtop{\leftskip 0pt\parindent 0pt\hsize=135pt
\lalongue=129pt
\traceline {3} {+ - - - - + - - - -}{}    
\traceline {c}  {B  B  B  B  B  G  B  W  W  W}{}  
\traceline {i}  {R  R  R  R  R  G  B  W  W  W}{}  
\traceline {d}  {Y  Y  Y  Y  Y  G  B  W  W  W}{}  
\traceline {s}  {B  B  B  B  B  G  B  W  W  W}{}  
\traceline {4} {+ - - - - + - - - -}{}    
\traceline {c}  {B  B  B  B  B  B  G  W  W  W}{}  
\traceline {i}  {R  R  R  R  R  R  G  W  W  W}{}  
\traceline {d}  {Y  Y  Y  Y  Y  Y  G  W  W  W}{}  
\traceline {s}  {B  B  B  B  B  B  G  W  W  W}{}
\traceline {5} {+ - - - - + - - - -}{}    
\traceline {c}  {B  B  B  B  B  B  G  B  W  W}{}  
\traceline {i}  {R  R  R  R  R  R  G  B  W  W}{} 
\traceline {d}  {Y  Y  Y  Y  Y  Y  G  B  W  W}{} 
\traceline {s}  {B  B  B  B  B  B  G  B  W  W}{}  
}
\hfill}
}
\hfill}
\vskip 10pt
The lower part of the column gives us three rules which manage the fate of the \sbb-cells
raised by rule~118. Rule 122 says that if a \sbb-cell can see \sgg- and two \sbb-cells,
those three neighbours as if the concerned cell would be a neighbour of second generation
for another neighbour of those latter cells, then that \sbb-cell becomes \sgg. Note
that on \sgg{} of \RR$_c$ and \RR$_s$, the blue cell on its face~2 can see the blue 
neighbour of the face~5 of the \sgg{} belonging to~\RR$_i$ and that which belongs to
\RR$_d$. The same conclusion for the face~5 of \RR$_i$, \RR$_d$ with both faces~2 of
\RR$_c$ and \RR$_s$. The conclusion is that on each $\mathcal S$($g$+1) we have now
a \sgg-cell, so that the register is now bigger by one cell on each strand.
Rule~124 says that a \sbb-cell whose unique non-blank cell is \sgg{} becomes \sww.
That apply to many cells on \sgg. To better see that, consider rule~123: it says that
if a \sbb-cell on \sgg{} can see a unique \sbb-cell and only \sww-cells outside its 
\sgg- and \sbb-neighbours remains \sbb. That applies to the \sbb-neighbour on face~7 for
\RR$_c$ and \RR$_s$. Indeed, as can be noticed on Figure~\ref{fcut}, that neighbour
can see the neighbour~6 of \RR$_d$ for \RR$_s$ and of \RR$_i$ for \RR$_c$, those both
cells being blue. That neighbour~7 can also see the neighbour~6 of the next neighbour 
of its strand. But that neighbour~6 is blank, so that rule~123 applies to the neighbour~7
of both \RR$_c$($g$) and \RR$s$($g$) which remain blue. That argument can be applied
to the neighbour~3 of those latter cells on $\mathcal S$ and also to the neighbours~
4 and~6 of \RR$_i$($g)$ and \RR$_d$($g$) so that on $\mathcal S$, at time~$t$+2,
the cell of coordinate $g$ has the colour and the decoration of its strand and now,
a bare \sgg-cell stands at the coordinate $g$+1.

The process increases the length of the register by one cell after two tips of 
the clock, whence the announced speed $\displaystyle{1\over2}$ of the growth. 

\subsubsection{The starting end of the register}\label{sbbrstart}

   Figure~\ref{fregdeb} illustrates the idle configuration of a register. 
On that figure, the content of the register is not~0 as far as the cells of~\RR$_c$ we
can see on the left-hand side picture of the figure are blank. We recognize on the 
cells $\mathcal S$(0) of the register the specific decorations indicated in $(7)$.

On the figures we can see that the cell $\mathcal S$(-1) bear the decoration defined
in~$(8)$. In sub-subsections~\ref{sbbrreginc} and~\ref{sbbrregdec} we shall see the use 
of those particular decorations as far as they are connected with the operations on the 
register.

Note that a cell of $\mathcal S$($n$), with $n>0$, a
has mainly four significant neighbours: two neighbours on its own strand
and one neighbour on each strand seen by its own strand. The rules dealing with those
cells are most often recognized by the second and third letter of the word they constitute
as far as those letters correspond to neighbours~0 and~1 of the cell. The patterns
are \ftt{RY }  for \RR$_c$, \ftt{BB } or \ftt{WB } for \RR$_i$, \ftt{BB } or \ftt{BW }
for \RR$_d$ and \ftt{YR } for \RR$_s$. The rules for $\mathcal S$(0) are easily 
recognizable by the least letters of the word which correspond to their neighbours 8 up
to~11. We have the suffix \ftt{WBWw } for~\RR$_c$, \RR$_i$ and \RR$_s$ and \ftt{WRWR } 
for~\RR$_d$.

We have seen on Table~\ref{tregdeb} the rules for the cells of the strands for the
coordinates $n\geq0$ for each strand. Presently, we consider the cells $\mathcal S$(-1)
for all strands and the particular cells \RR$_c$(-2)$_b$ and \RR$_c$(-2)$_c$ which we 
already mentioned in the previous sub-subsection. The rules are given in
Table~\ref{tprereg}.

\ligne{\hfill
\vtop{
\begin{tab}\label{tprereg}
\leurre
Table of the rules for $\mathcal S$$(-1)$ and for \RR$_c(-2)_b$ together with 
\RR$_c(-2)_c$
\end{tab}
\vskip-7pt
\ligne{\hfill
\vtop{\leftskip 0pt\parindent 0pt\hsize=\largeouille
\ligne{\hfill\ftt{\RR$_c$(-1)} \hfill}
\ligne{\hfill\gzzz\ftt{\the\compteregle \ }  
\llrule {W} {W} {W} {W} {W} {W} {W} 
\hskip-3.5pt\rrrule {Y} {W} {W} {Y} {Y} {W} {W} {}\hfill}
\vskip-3pt
\ligne{\hfill\gzzz\ftt{\the\compteregle \ }  
\llrule {W} {W} {W} {W} {W} {R} {W} 
\hskip-3.5pt\rrrule {Y} {W} {W} {Y} {Y} {W} {R} {}\hfill}
\vskip-3pt
\ligne{\hfill\gzzz\ftt{\the\compteregle \ }  
\llrule {R} {W} {W} {W} {W} {W} {W} 
\hskip-3.5pt\rrrule {Y} {W} {W} {Y} {Y} {W} {W} {}\hfill}
\vskip-3pt
\ligne{\hfill\gzzz\ftt{\the\compteregle \ }  
\llrule {W} {W} {W} {R} {W} {W} {W} 
\hskip-3.5pt\rrrule {Y} {W} {W} {Y} {Y} {W} {W} {}\hfill}
}
\hfill
\vtop{\leftskip 0pt\parindent 0pt\hsize=\largeouille
\ligne{\hfill\ftt{\RR$_d$(-1)} \hfill}
\ligne{\hfill\gzzz\ftt{\the\compteregle \ }  
\llrule {W} {W} {W} {W} {W} {W} {Y} 
\hskip-3.5pt\rrrule {Y} {W} {W} {Y} {R} {W} {W} {}\hfill}
\vskip-3pt
\ligne{\hfill\gzzz\ftt{\the\compteregle \ }  
\llrule {W} {W} {W} {W} {W} {W} {R} 
\hskip-3.5pt\rrrule {Y} {W} {W} {Y} {R} {W} {R} {}\hfill}
\vskip-3pt
\ligne{\hfill\gzzz\ftt{\the\compteregle \ }  
\llrule {W} {W} {W} {R} {W} {W} {Y} 
\hskip-3.5pt\rrrule {Y} {W} {W} {Y} {R} {W} {R} {}\hfill}
\vskip-3pt
\ligne{\hfill\gzzz\ftt{\the\compteregle \ }  
\llrule {W} {W} {W} {W} {B} {W} {Y} 
\hskip-3.5pt\rrrule {Y} {W} {W} {Y} {R} {W} {W} {}\hfill}
\ligne{\hfill\ftt{\RR$_i$(-1)} \hfill}
\ligne{\hfill\gzzz\ftt{\the\compteregle \ }  
\llrule {W} {W} {W} {W} {W} {W} {R} 
\hskip-3.5pt\rrrule {R} {W} {W} {R} {Y} {W} {W} {}\hfill}
\vskip-3pt
\ligne{\hfill\gzzz\ftt{\the\compteregle \ }  
\llrule {W} {W} {W} {W} {W} {B} {R} 
\hskip-3.5pt\rrrule {R} {W} {W} {R} {Y} {W} {B} {}\hfill}
\vskip-3pt
\ligne{\hfill\gzzz\ftt{\the\compteregle \ }  
\llrule {B} {W} {W} {W} {W} {W} {R} 
\hskip-3.5pt\rrrule {R} {W} {W} {R} {Y} {W} {W} {}\hfill}
\vskip-3pt
\ligne{\hfill\gzzz\ftt{\the\compteregle \ }  
\llrule {W} {W} {W} {B} {W} {W} {R} 
\hskip-3.5pt\rrrule {R} {W} {W} {R} {Y} {W} {W} {}\hfill}
}
\hfill
\vtop{\leftskip 0pt\parindent 0pt\hsize=\largeouille
\ligne{\hfill\ftt{\RR$_c$(-2)$_b$ and \RR$_s$} \hfill}
\ligne{\hfill\gzzz\ftt{\the\compteregle \ }  
\llrule {W} {W} {W} {W} {W} {W} {W} 
\hskip-3.5pt\rrrule {R} {B} {W} {R} {R} {W} {W} {}\hfill}
\vskip-3pt
\ligne{\hfill\gzzz\ftt{\the\compteregle \ }  
\llrule {W} {W} {W} {W} {W} {W} {B} 
\hskip-3.5pt\rrrule {R} {B} {W} {R} {R} {W} {R} {}\hfill}
\vskip-3pt
\ligne{\hfill\gzzz\ftt{\the\compteregle \ }  
\llrule {R} {W} {W} {W} {W} {W} {W} 
\hskip-3.5pt\rrrule {R} {B} {W} {R} {R} {W} {W} {}\hfill}
\vskip-3pt
\ligne{\hfill\gzzz\ftt{\the\compteregle \ }  
\llrule {W} {W} {W} {B} {W} {W} {W} 
\hskip-3.5pt\rrrule {R} {B} {W} {R} {R} {W} {W} {}\hfill}
\vskip-3pt
\ligne{\hfill\ftt{\RR$_c$(-2)$_c$} \hfill}
\ligne{\ftt{r19 } \hskip 0.5pt
\llrule {W} {W} {W} {W} {W} {W} {W} 
\hskip-3.5pt\rrrule {W} {W} {Y} {Y} {W} {Y} {W} {}\hfill}
}
\hfill}
}
\hfill}
\vskip 10pt
The conservative rules for $\mathcal S$(-1) are 129, 133, 137 and~144. That latter rule
also holds for \RR$_c$(-2)$_b$ which has exactly the same decorations as \RR$_s$.
The other rules are the motion rules for the locomotive for those rules, \RR$_c$(-2)$_c$
being excepted. The rule given for that latter cell is labelled \ftt{r19 } as far as 
its minimal form is exactly that of rule~19:
\hbox{\llrule {W} {W} {W} {W} {W} {W} {W} 
\hskip-3.5pt\rrrule {W} {W} {W} {Y} {Y} {Y} {W} {}}. We obtain that form from rule~19
by the rotation around the axis joining the mid-points of the edges 1-6 and 3-9. 
The rotation around the axis joining the vertices 0-1-2 and 9-10-11 transforms \ftt{r19 }
to the same minimal form. The motion rules for \RR$_c$(-2)$_c$ will be studied in
Sub-subsection~\ref{sbbrregdec} to which we now turn.

\subsubsection{Decrementation of the register}\label{sbbrregdec}

   That operation is the most difficult to solve for the simulation. We formulate a
general principle which we shall as far as possible implement in all cases. 
A red locomotive arrives on the blank part of the register which represents the value
it is storing until it meets the first blue cell of the remaining part of the 
register. As the circuit which is run over by the locomotive is very big, although the
growth of the register happens at speed $\displaystyle{1\over2}$, the blue part of \RR$_c$
is much bigger than the content of the register, whatever large this content could be.

We have three different cases: the general case, when the content~$c$ of the register is
large enough, then when $c=2$, when $c=1$ and, not at all the least, when $c=0$.
By large enough we mean that the detection of the first blue cell on \RR$_c$ happens
on a cell which cannot be seen by \RR$_c$(0). We also require a bit more: that all 
possible motion rules of the locomotive could be tested on that white portion of the
register. The case $c=2$ is particular as the detection of the first blue cell
can be seen from \RR$_c$(0). For the same reason, $c=1$ is a particular cell. For what
is $c=0$, it is the case when the decrementation cannot be performed which, of course, 
requires a special treatment. The general principle is depicted by~(9).

\vskip 10pt
\ligne{\hfill
\vtop{\leftskip 0pt\parindent 0pt\hsize=90pt
$\vcenter{\vbox{
\ftt{
R W W W B B B\vszz
W R W W B B B\vszz
W W R W B B B\vszz
W W W G B B B\vszz
W W W B B B B\vszz
W W W B B B B\vszz
W W W B B B B\vszz
}
}}$
}
\hfill(\numerrel)\hskip 10pt}
\vskip 10pt
In all cases, the locomotive arrives to the register through a path arriving to
\RR$_c$(-2)$_b$ through the face~5. That cell, whose decoration is given in 
Table~\ref{tprereg}, transforms the arriving blue locomotive into a red one, see
again that table. Leaving the cell trough the face~2 of \RR$_c$(-2)$_c$, the locomotive
enters \RR$_c$(-1) through the face~4 of that latter cell as far as its face~5 is 
devoted to the locomotive coming for an incrementation. Consequently, \RR$_c$(-1)
pushes the red locomotive through its face~2 into \RR$_c$(0). The locomotive goes on
the strand \RR$_c$ until a cell~$\nu$ which is blank such that \RR$_c$($\mu$) with
$\mu\leq\nu$ is blank while \RR$_c$($\mu)$ is blue for $\mu>\nu$+1. When this happens,
say at time~$t$, at time~$t$+1 \RR$_c$($\nu$) becomes \sgg{} which is the signal of
decrementation. Of course, we could have chosen another signal. However, \sbb{} is clearly
ruled out. In order to avoid confusion, \srr{} is also ruled out. We remain with \syy.
But a different colour is needed for incrementation. So that I choose \sgg{} for the
decrementation. As (9) shows, the scenario is simple: \sgg{} takes the place of a new
blue cell which will by~1 reduce the value stored in the register. Now, with \sgg{} we
have the phenomenon of a lot of cells becoming blue at the next time.

   Let us carefully look at that point. At time~$t$, rule~145 is applied to \RR$_c$($\nu$)
which become~\sgg. It remains~\sgg{} in order to get rid of the \sbb-cells created among
the neighbours of~\RR$_c$($\nu$). The neighbours of that cell belonging to other strands
are ruled out: neighbours~0 and~1 but also neighbour~2 as it can see a blue cell. 
Neighbour~5 is white but that neighbour can see \RR$_i$($\nu$$-$1) which is red, so that
neighbour~5 remains \sww. The decoration of the previous and the next cell on the strand
prevent neighbours~4 and~6 to become blue, so that they remain \sww. 
Neighbour~3 which is~\sbb{} can see
a blue neighbour of \RR$_i$($\nu)$ and it can see \sgg, of course, so that rule~123
applies to that cell, allowing the decoration to remain unchanged. A similar argument
holds for neighbour~7 which can see the blue neighbour~6 of \RR$_d$($\nu$). Rule~124
applies to the following \sbb-neighbours of \RR$_c$($\nu$): 7, 8, 9, 10 and~11. 
Accordingly, at time~$t$+1, rules~123 and 124 apply so that the decoration remains 
\sbb{} and the \sbb-cells created at time~$t$+1 return to \sww{} at time~$t$+2: it is 
the desired goal. Also, at time~$t$+2, the \sgg-cell turns to~\sbb{} which performs
the decrementation.

\ligne{\hfill
\vtop{
\begin{tab}\label{tregdec}
\leurre
The rules for the decrementation in all the cases.
\end{tab}
\vskip-7pt
\ligne{\hfill the general case and case $c=2$ \hfill}
\ligne{\hfill
\vtop{\leftskip 0pt\parindent 0pt\hsize=\largeouille
\ligne{\hfill\ftt{\RR$_c$} \hfill}
\ligne{\hfill\gzzz\ftt{\the\compteregle \ }  
\llrule {W} {R} {Y} {B} {B} {W} {R} 
\hskip-3.5pt\rrrule {W} {B} {W} {W} {W} {W} {G} {}\hfill}
\vskip-3pt
\ligne{\hfill\gzzz\ftt{\the\compteregle \ }  
\llrule {W} {R} {Y} {G} {B} {W} {W} 
\hskip-3.5pt\rrrule {W} {B} {W} {W} {W} {W} {W} {}\hfill}
\vskip-3pt
\ligne{\hfill\gzzz\ftt{\the\compteregle \ }  
\llrule {B} {R} {Y} {B} {B} {W} {G} 
\hskip-3.5pt\rrrule {W} {B} {W} {W} {W} {W} {B} {}\hfill}
\vskip-3pt
\ligne{\hfill\gzzz\ftt{\the\compteregle \ }  
\llrule {G} {R} {Y} {B} {B} {W} {W} 
\hskip-3.5pt\rrrule {W} {B} {W} {W} {W} {W} {G} {}\hfill}
\vskip-3pt
\ligne{\hfill\gzzz\ftt{\the\compteregle \ }  
\llrule {G} {R} {R} {B} {B} {W} {W} 
\hskip-3.5pt\rrrule {W} {B} {B} {B} {B} {B} {B} {}\hfill}
\vskip-3pt
\ligne{\hfill\gzzz\ftt{\the\compteregle \ }  
\llrule {W} {R} {R} {B} {B} {W} {W} 
\hskip-3.5pt\rrrule {W} {B} {W} {W} {W} {W} {W} {}\hfill}
\ligne{\hfill\gzzz\ftt{\the\compteregle \ }  
\llrule {W} {R} {R} {W} {B} {W} {W} 
\hskip-3.5pt\rrrule {W} {B} {W} {W} {W} {W} {W} {}\hfill}
\vskip-3pt
\ligne{\hfill\gzzz\ftt{\the\compteregle \ }  
\llrule {W} {R} {R} {W} {B} {W} {W} 
\hskip-3.5pt\rrrule {W} {B} {W} {B} {W} {W} {W} {}\hfill}
}
\hfill
\vtop{\leftskip 0pt\parindent 0pt\hsize=\largeouille
\ligne{\hfill\ftt{the other strands} \hfill}
\ligne{\hfill\gzzz\ftt{\the\compteregle \ }  
\llrule {Y} {B} {G} {Y} {W} {B} {Y} 
\hskip-3.5pt\rrrule {B} {W} {W} {W} {W} {W} {R} {}\hfill}
\vskip-3pt
\ligne{\hfill\gzzz\ftt{\the\compteregle \ }  
\llrule {R} {B} {G} {Y} {W} {B} {Y} 
\hskip-3.5pt\rrrule {B} {W} {W} {W} {W} {W} {Y} {}\hfill}
\vskip-3pt
\ligne{\hfill\gzzz\ftt{\the\compteregle \ }  
\llrule {Y} {B} {B} {R} {W} {B} {Y} 
\hskip-3.5pt\rrrule {B} {W} {W} {W} {W} {W} {Y} {}\hfill}
\ligne{\hfill\gzzz\ftt{\the\compteregle \ }  
\llrule {R} {G} {B} {R} {W} {B} {R} 
\hskip-3.5pt\rrrule {B} {W} {W} {W} {W} {W} {R} {}\hfill}
\vskip-3pt
\ligne{\hfill\gzzz\ftt{\the\compteregle \ }  
\llrule {B} {R} {R} {B} {B} {W} {B} 
\hskip-3.5pt\rrrule {W} {B} {W} {W} {W} {W} {B} {}\hfill}
\vskip-3pt
\ligne{\hfill\gzzz\ftt{\the\compteregle \ }  
\llrule {B} {R} {R} {B} {B} {W} {W} 
\hskip-3.5pt\rrrule {W} {B} {W} {B} {W} {W} {B} {}\hfill}
}
\hfill
\vtop{\leftskip 0pt\parindent 0pt\hsize=\largeouille
\ligne{\hfill\ftt{case $c=2$ } \hfill}
\ligne{\hfill\gzzz\ftt{\the\compteregle \ }  
\llrule {W} {R} {Y} {G} {B} {W} {W} 
\hskip-3.5pt\rrrule {W} {B} {W} {B} {W} {W} {W} {}\hfill}
\vskip-3pt
\ligne{\hfill\gzzz\ftt{\the\compteregle \ }  
\llrule {W} {R} {R} {B} {B} {W} {W} 
\hskip-3.5pt\rrrule {W} {B} {W} {B} {W} {W} {W} {}\hfill}
}
\hfill}
\ligne{\hfill the cases $c = 1$ and $c=0$ \hfill}
\ligne{\hfill
\vtop{\leftskip 0pt\parindent 0pt\hsize=\largeouille
\ligne{\hfill\ftt{case $c = 1$ } \hfill}
\ligne{\hfill\gzzz\ftt{\the\compteregle \ }  
\llrule {W} {R} {Y} {B} {B} {W} {R} 
\hskip-3.5pt\rrrule {W} {B} {W} {B} {W} {W} {G} {}\hfill}
\vskip-3pt
\ligne{\hfill\gzzz\ftt{\the\compteregle \ }  
\llrule {G} {R} {Y} {B} {B} {W} {W} 
\hskip-3.5pt\rrrule {W} {B} {W} {B} {W} {W} {G} {}\hfill}
\vskip-3pt
\ligne{\hfill\gzzz\ftt{\the\compteregle \ }  
\llrule {G} {R} {R} {B} {B} {W} {W} 
\hskip-3.5pt\rrrule {B} {B} {B} {B} {B} {B} {B} {}\hfill}
\vskip-3pt
\ligne{\hfill\gzzz\ftt{\the\compteregle \ }  
\llrule {B} {R} {Y} {B} {B} {W} {W} 
\hskip-3.5pt\rrrule {W} {B} {W} {B} {W} {W} {B} {}\hfill}
\ligne{\hfill\gzzz\ftt{\the\compteregle \ }  
\llrule {W} {W} {W} {G} {W} {W} {W} 
\hskip-3.5pt\rrrule {Y} {W} {W} {Y} {Y} {W} {W} {}\hfill}
\ligne{\hfill\gzzz\ftt{\the\compteregle \ }  
\llrule {Y} {B} {G} {W} {W} {B} {Y} 
\hskip-3.5pt\rrrule {B} {W} {W} {R} {W} {R} {R} {}\hfill}
\vskip-3pt
\ligne{\hfill\gzzz\ftt{\the\compteregle \ }  
\llrule {R} {B} {G} {W} {W} {B} {Y} 
\hskip-3.5pt\rrrule {B} {W} {W} {R} {W} {R} {Y} {}\hfill}
\vskip-3pt
\ligne{\hfill\gzzz\ftt{\the\compteregle \ }  
\llrule {Y} {B} {B} {W} {W} {B} {Y} 
\hskip-3.5pt\rrrule {B} {W} {W} {R} {W} {R} {Y} {}\hfill}
\vskip-3pt
\ligne{\hfill\gzzz\ftt{\the\compteregle \ }  
\llrule {Y} {B} {B} {R} {W} {B} {Y} 
\hskip-3.5pt\rrrule {B} {W} {W} {R} {W} {R} {Y} {}\hfill}
\ligne{\hfill\gzzz\ftt{\the\compteregle \ }  
\llrule {R} {G} {B} {W} {W} {B} {R} 
\hskip-3.5pt\rrrule {B} {W} {W} {B} {W} {W} {R} {}\hfill}
\vskip-3pt
\ligne{\hfill\gzzz\ftt{\the\compteregle \ }  
\llrule {R} {B} {B} {W} {W} {B} {R} 
\hskip-3.5pt\rrrule {B} {W} {W} {B} {W} {W} {R} {}\hfill}
\vskip-3pt
}
\hfill
\vtop{\leftskip 0pt\parindent 0pt\hsize=\largeouille
\ligne{\hfill\ftt{case $c = 0$ } \hfill}
\ligne{\hfill\ftt{\RR$_c$(0) } \hfill}
\ligne{\hfill\gzzz\ftt{\the\compteregle \ }  
\llrule {B} {R} {Y} {B} {B} {W} {G} 
\hskip-3.5pt\rrrule {W} {B} {W} {B} {W} {W} {B} {}\hfill}
\ligne{\hfill\ftt{\RR$_c$(-1) } \hfill}
\ligne{\hfill\gzzz\ftt{\the\compteregle \ }  
\llrule {W} {W} {W} {B} {W} {R} {W} 
\hskip-3.5pt\rrrule {Y} {W} {W} {Y} {Y} {W} {G} {}\hfill}
\vskip-3pt
\ligne{\hfill\gzzz\ftt{\the\compteregle \ }  
\llrule {G} {W} {W} {B} {W} {W} {W} 
\hskip-3.5pt\rrrule {Y} {W} {W} {Y} {Y} {W} {G} {}\hfill}
\vskip-3pt
\ligne{\hfill\gzzz\ftt{\the\compteregle \ }  
\llrule {G} {W} {W} {B} {B} {W} {W} 
\hskip-3.5pt\rrrule {Y} {W} {B} {Y} {Y} {B} {W} {}\hfill}
\vskip-3pt
\ligne{\hfill\gzzz\ftt{\the\compteregle \ }  
\llrule {W} {W} {W} {B} {W} {W} {W} 
\hskip-3.5pt\rrrule {Y} {W} {W} {Y} {Y} {W} {W} {}\hfill}
\ligne{\hfill\ftt{\RR$_c$(-2)$_b$ } \hfill}
\ligne{\hfill\gzzz\ftt{\the\compteregle \ }  
\llrule {W} {W} {W} {W} {W} {G} {W} 
\hskip-3.5pt\rrrule {W} {W} {Y} {Y} {W} {Y} {B} {}\hfill}
\vskip-3pt
\ligne{\hfill\gzzz\ftt{\the\compteregle \ }  
\llrule {B} {W} {W} {W} {W} {G} {W} 
\hskip-3.5pt\rrrule {W} {W} {Y} {Y} {W} {Y} {W} {}\hfill}
\vskip-3pt
\ligne{\hfill\gzzz\ftt{\the\compteregle \ }  
\llrule {W} {W} {W} {W} {W} {B} {W} 
\hskip-3.5pt\rrrule {W} {W} {Y} {Y} {W} {Y} {W} {}\hfill}
}
\hfill
\vtop{\leftskip 0pt\parindent 0pt\hsize=\largeouille
\ligne{\hfill\ftt{case $c = 0$ } \hfill}
\ligne{\hfill\ftt{\RR$_d$(-1) } \hfill}
\ligne{\hfill\gzzz\ftt{\the\compteregle \ }  
\llrule {W} {W} {G} {W} {W} {W} {Y} 
\hskip-3.5pt\rrrule {Y} {W} {W} {Y} {R} {W} {W} {}\hfill}
%
\ligne{\hfill\ftt{\RR$_i$(-1) } \hfill}
\ligne{\hfill\gzzz\ftt{\the\compteregle \ }  
\llrule {W} {G} {W} {W} {W} {W} {R} 
\hskip-3.5pt\rrrule {R} {W} {W} {R} {Y} {W} {W} {}\hfill}
%
\ligne{\hfill\ftt{\RR$_c$(-2) } \hfill}
\ligne{\hfill\gzzz\ftt{\the\compteregle \ }  
\llrule {W} {W} {W} {G} {W} {W} {W} 
\hskip-3.5pt\rrrule {B} {W} {W} {B} {B} {W} {W} {}\hfill}
\ligne{\hfill\ftt{\RR$_c$(-2)$_a$ } \hfill}
\ligne{\hfill\gzzz\ftt{\the\compteregle \ }  
\llrule {W} {W} {W} {G} {W} {W} {W} 
\hskip-3.5pt\rrrule {R} {B} {W} {R} {R} {W} {W} {}\hfill}
}
\hfill}
}
\hfill}
\vskip 10pt
   Table~\ref{tregdec} gives the rules for the decrementation in the general case and 
when $c=2$. In the general case, rules 145 up to~152 deal with \RR$_c$: rule~145
depicts the situation which triggers the appearance of the \sgg-cell at \RR$_c(\nu)$.
Rule~146, 147 indicates that the cells \RR$_c(\nu$$-$$1)$ and \RR$_c(\nu$+$1)$ 
respectively keep their state, \sww{} and \sbb{} respectively. Rule~148 indicates that
\sgg- remains for one more time and rule~148 makes it return to~\sbb{} indicating at the
same times which faces are covered with \sbb-neighbours. Rule~149 concerns 
\RR$_c(\nu$$-$$1)$ which can see the appearance of a red locomotive on~\RR$_d$.
Rule~151 concerns a white cell of~\RR$_c$ seeing the red locomotive on~\RR$_d$. Rule~152
does the same for \RR$_c$(0) as witnessed by its decoration.

Rule 153 up to 158 manage the reaction of the cells of the other strands at the appearance
of \sgg{} and, also of the red locomotive on \RR$_d$. Rules~153 up to~155 manage
the motion of the locomotive on \RR$_d$. Rule~156, 157 say that \RR$_i$, \RR$_s$ 
respectively witness only what happens on \RR$_c$. Rule~158 deals with \RR$_s$(0) seeing
the locomotive about to leave \RR$_d$.

Two additional rules only are needed when $c=2$. As indicated by the \ftt{WBWW } suffix
of the neighbourhoods in the rules, they deal with \RR$_c$(0) which can see the appearance
of \sgg{} at \RR$_c$(1). Rule 160 says that the red locomotive of \RR$_d$ can be seen from
\RR$_c$(0). Rules 161 up to 164 manage \RR$_c$(0) and \RR$_c$(1), while rule~165
manage \RR$_c$(-1), saying that the cell does not react to the occurrence of~\sgg{}
in \RR$_c$(0).

Rules~166 up to~169 deal with \RR$_d$ giving motion rules for \RR$_d$(-1). Rules~170 and
171 say that \RR$_i$(0) remains unchanged at that time.

At last, in the case $c=0$, \sgg{} occurs in \RR$_c$(-1) so that, rule~172 says that
\RR$_c$(0) can witness that event. It is the single cell of the strands of the register 
which is touched by the operation. The fact that the decrementation could not be
performed requires two facts: first, no locomotive is triggered on \RR$_d$, which is
indicated by rule~180. Secondly, a locomotive must be sent on the $Z$-path which 
precisely starts from \RR$_c$(-2)$_c$ to which rules~173 up to~179 are devoted:
rules~173 up to 176 manage the occurrence of~\sgg{} in \RR$_c$(-1). Rule~177 says that
seeing \sgg{} in \RR$_c$(-1), \RR$_c$(-2)$_b$ triggers a blue
locomotive, and rules~178 together with rule~179 launch the locomotive on the Z-path.
Note that, in principle, rule~179 indicates that the cell can see the \sbb-neighbour 
through its face~4, which is the case when a blue locomotive passes through \RR$_c$(-1),
a situation which occurs for an incrementing locomotive. But rule~179 and the expected 
rule $(\rho)$ \hbox{\llrule {W} {W} {W} {B} {W} {W} {W} 
\hskip-3.5pt\rrrule {W} {W} {Y} {Y} {W} {Y} {W} {} } have the same 
minimal form
\hbox{\llrule {W} {W} {W} {W} {W} {W} {W}
\hskip-3.5pt\rrrule {W} {B} {W} {Y} {Y} {Y} {W} {} }. The rotation around 
the vertices 0-1-2 and 9-10-11 we already met transforms rule~179 into that minimal form
while the rotation around faces~4 and~7 transforms~$(\rho)$ into the same minimal form.

The other rules about $\mathcal S$(-1) say that the corresponding cell do not react
to the occurrence of \sgg{} in \RR$_c$(-1). In particular, rule~180, \RR$_d$(-1) remains
\sww{} when it sees \sgg{} on \RR$_c$(-1). Rule~181 deals with \RR$_i$(-1). Rule~182
says that the path arriving to \RR$_c$(-1) from \DDI{} does not react and
Rule~183 says the same for the path arriving to \RR$_c$(-1) from \DDD.

 In the Appendix, we give traces of execution for the decrementation, in the general case,
see Table~\ref{texdecgen}, and in the particular case, when $c=2$, when $c=1$ and 
when $c=0$, see Table~\ref{texdecpart}.

\subsubsection{Incrementation of the register}\label{sbbrreginc}

   Presently, we arrive to the implementation of an incrementation of the register.
The incrementation as well as the decrementation are performed on~\RR$_c$ as far as
its content~$c$ is concerned. We remind the reader that the decrementing path arrives
through \RR$_c$(-2)$_b$ which is seen by \RR$_c$(-1) through its face~4. That makes a
difference for \RR$_c$(-1) so that the blue locomotive remains blue when it enters the
register on the \RR$_c$ strand.

    Here too, we shall follow a general pattern as long as it will be possible. It is 
given by (10) where the same convention as in (9) are followed. We shall see that, in the
case of incrementations, the execution is more faithful to the scheme of (10) that
the execution of decrementations with respect to (9). The reason is the use of~\syy{}
instead of \sgg{} and, mainly, to the reason that the case $c=0$ is not really 
exceptional: the content is actually incremented in that case too.
\vskip 10pt
\ligne{\hfill
\vtop{\leftskip 0pt\parindent 0pt\hsize=95pt
$\vcenter{\vbox{
\ftt{
0 W B W W W B B B\vszz
1 W W B W W B B B\vszz
2 W W W B W B B B\vszz
3 W W W W Y B B B\vszz
4 W W W W Y R B B\vszz
5 W W W W W W B B\vszz
6 W W W W W W B B\vszz
}
}}$
}
\hfill (\numerrel)\hskip 10pt}
\vskip 10pt
    However, the arrival of the locomotive is different: the path coming from the \DDI{} 
devoted to the register arrives at \RR$_c$(-2)$_a$ which can see \RR$_c$(-1) through its
face~2 while it seen from that latter cell from its face~5.

\ligne{\hfill
\vtop{
\begin{tab}\label{treginc}
\leurre
Rules for the incrementation. General case and then particular cases when the content $c$
is in $\{0,1,2\}$.
\end{tab}
\vskip -7pt
\ligne{\hfill general case\hfill}
\ligne{\hfill
\vtop{\leftskip 0pt\parindent 0pt\hsize=\largeouille
\ligne{\hfill\gzzz\ftt{\the\compteregle \ }  
\llrule {W} {W} {W} {W} {W} {W} {B} 
\hskip-3.5pt\rrrule {Y} {W} {W} {Y} {Y} {W} {B} {}\hfill}
\vskip-3pt
\ligne{\hfill\gzzz\ftt{\the\compteregle \ }  
\llrule {B} {W} {W} {W} {W} {W} {W} 
\hskip-3.5pt\rrrule {Y} {W} {W} {Y} {Y} {W} {W} {}\hfill}
\vskip-3pt
\ligne{\hfill\gzzz\ftt{\the\compteregle \ }  
\llrule {W} {B} {W} {W} {W} {W} {R} 
\hskip-3.5pt\rrrule {R} {W} {W} {R} {Y} {W} {W} {}\hfill}
\vskip-3pt
\ligne{\hfill\gzzz\ftt{\the\compteregle \ }  
\llrule {W} {W} {B} {W} {W} {W} {Y} 
\hskip-3.5pt\rrrule {Y} {W} {W} {Y} {R} {W} {W} {}\hfill}
\vskip-3pt
\ligne{\hfill\gzzz\ftt{\the\compteregle \ }  
\llrule {W} {W} {W} {W} {W} {B} {W} 
\hskip-3.5pt\rrrule {W} {W} {Y} {Y} {W} {Y} {W} {}\hfill}
\ligne{\hfill\gzzz\ftt{\the\compteregle \ }  
\llrule {B} {R} {Y} {W} {B} {W} {W} 
\hskip-3.5pt\rrrule {W} {B} {W} {B} {W} {W} {W} {}\hfill}
\vskip-3pt
\ligne{\hfill\gzzz\ftt{\the\compteregle \ }  
\llrule {W} {B} {Y} {W} {B} {W} {W} 
\hskip-3.5pt\rrrule {W} {B} {W} {B} {W} {W} {W} {}\hfill}
\vskip-3pt
\ligne{\hfill\gzzz\ftt{\the\compteregle \ }  
\llrule {W} {R} {Y} {B} {B} {W} {B} 
\hskip-3.5pt\rrrule {W} {B} {W} {W} {W} {W} {Y} {}\hfill}
\vskip-3pt
\ligne{\hfill\gzzz\ftt{\the\compteregle \ }  
\llrule {Y} {R} {Y} {B} {B} {W} {W} 
\hskip-3.5pt\rrrule {W} {B} {W} {W} {W} {W} {Y} {}\hfill}
\vskip-3pt
\ligne{\hfill\gzzz\ftt{\the\compteregle \ }  
\llrule {W} {R} {Y} {Y} {B} {W} {W} 
\hskip-3.5pt\rrrule {W} {B} {W} {W} {W} {W} {W} {}\hfill}
}
\vtop{\leftskip 0pt\parindent 0pt\hsize=\largeouille
\ligne{\hfill\gzzz\ftt{\the\compteregle \ }  
\llrule {B} {R} {Y} {B} {B} {W} {Y} 
\hskip-3.5pt\rrrule {W} {B} {W} {W} {W} {W} {R} {}\hfill}
\vskip-3pt
\ligne{\hfill\gzzz\ftt{\the\compteregle \ }  
\llrule {R} {Y} {B} {R} {W} {B} {R} 
\hskip-3.5pt\rrrule {B} {W} {W} {W} {W} {W} {B} {}\hfill}
\vskip-3pt
\ligne{\hfill\gzzz\ftt{\the\compteregle \ }  
\llrule {R} {R} {Y} {B} {B} {W} {Y}                        
\hskip-3.5pt\rrrule {W} {B} {W} {W} {W} {W} {W} {}\hfill}    
\vskip-3pt
\ligne{\hfill\gzzz\ftt{\the\compteregle \ }  
\llrule {Y} {B} {Y} {R} {B} {W} {W} 
\hskip-3.5pt\rrrule {W} {B} {W} {W} {W} {W} {W} {}\hfill}    
\vskip-3pt
\ligne{\hfill\gzzz\ftt{\the\compteregle \ }  
\llrule {B} {R} {Y} {B} {B} {W} {R}                     
\hskip-3.5pt\rrrule {W} {B} {W} {W} {W} {W} {B} {}\hfill}
%
\ligne{\hfill\gzzz\ftt{\the\compteregle \ }  
\llrule {W} {B} {Y} {W} {B} {W} {W} 
\hskip-3.5pt\rrrule {W} {B} {W} {W} {W} {W} {W} {}\hfill}
\vskip-3pt
\ligne{\hfill\gzzz\ftt{\the\compteregle \ }  
\llrule {Y} {B} {Y} {Y} {W} {B} {Y} 
\hskip-3.5pt\rrrule {B} {W} {W} {W} {W} {W} {Y} {}\hfill}
\vskip-3pt
\ligne{\hfill\gzzz\ftt{\the\compteregle \ }  
\llrule {B} {Y} {B} {B} {B} {W} {B} 
\hskip-3.5pt\rrrule {W} {B} {W} {W} {W} {W} {B} {}\hfill}
\vskip-3pt
\ligne{\hfill\gzzz\ftt{\the\compteregle \ }  
\llrule {B} {Y} {B} {R} {W} {B} {R} 
\hskip-3.5pt\rrrule {B} {W} {W} {W} {W} {W} {R} {}\hfill}
\vskip-3pt
\ligne{\hfill\gzzz\ftt{\the\compteregle \ }  
\llrule {R} {R} {B} {B} {W} {B} {R} 
\hskip-3.5pt\rrrule {B} {W} {W} {W} {W} {W} {R} {}\hfill}
}
\vtop{\leftskip 0pt\parindent 0pt\hsize=\largeouille
\ligne{\hfill\gzzz\ftt{\the\compteregle \ }  
\llrule {B} {Y} {B} {B} {B} {W} {W} 
\hskip-3.5pt\rrrule {W} {B} {W} {B} {W} {W} {B} {}\hfill}
\ligne{\hfill\gzzz\ftt{\the\compteregle \ }  
\llrule {W} {W} {W} {W} {W} {W} {B} 
\hskip-3.5pt\rrrule {R} {W} {W} {R} {Y} {W} {B} {}\hfill}
\vskip-3pt
\ligne{\hfill\gzzz\ftt{\the\compteregle \ }  
\llrule {B} {W} {W} {W} {W} {W} {R} 
\hskip-3.5pt\rrrule {R} {W} {W} {R} {Y} {W} {W} {}\hfill}
\vskip-3pt
\ligne{\hfill\gzzz\ftt{\the\compteregle \ }  
\llrule {W} {W} {W} {B} {W} {W} {R} 
\hskip-3.5pt\rrrule {R} {W} {W} {R} {Y} {W} {W} {}\hfill}
\ligne{\hfill\gzzz\ftt{\the\compteregle \ }  
\llrule {W} {B} {W} {W} {W} {W} {W} 
\hskip-3.5pt\rrrule {Y} {W} {W} {Y} {Y} {W} {W} {}\hfill}
\ligne{\hfill\gzzz\ftt{\the\compteregle \ }  
\llrule {W} {W} {B} {B} {W} {W} {W} 
\hskip-3.5pt\rrrule {R} {B} {W} {R} {R} {W} {W} {}\hfill}
}
\hfill}
\vskip 5pt
\ligne{\hfill cases $c=2$, $c=1$ and $c=0$\hfill}
\ligne{\hfill
\vtop{\leftskip 0pt\parindent 0pt\hsize=\largeouille
\ligne{\hfill\ftt{$c=2$ } \hfill}
\ligne{\hfill\gzzz\ftt{\the\compteregle \ }  
\llrule {W} {R} {Y} {Y} {B} {W} {W}                     
\hskip-3.5pt\rrrule {W} {B} {W} {B} {W} {W} {W} {}\hfill}
\ligne{\hfill\ftt{$c=1$ } \hfill}
\ligne{\hfill\gzzz\ftt{\the\compteregle \ }  
\llrule {W} {R} {Y} {B} {B} {W} {B} 
\hskip-3.5pt\rrrule {W} {B} {W} {B} {W} {W} {Y} {}\hfill}
\vskip-3pt
\ligne{\hfill\gzzz\ftt{\the\compteregle \ }  
\llrule {Y} {R} {Y} {B} {B} {W} {W} 
\hskip-3.5pt\rrrule {W} {B} {W} {B} {W} {W} {Y} {}\hfill}
\vskip-3pt
\ligne{\hfill\gzzz\ftt{\the\compteregle \ }  
\llrule {R} {Y} {B} {W} {W} {B} {R} 
\hskip-3.5pt\rrrule {B} {W} {W} {B} {W} {W} {B} {}\hfill}
\vskip-3pt
\ligne{\hfill\gzzz\ftt{\the\compteregle \ }  
\llrule {B} {Y} {B} {W} {W} {B} {R} 
\hskip-3.5pt\rrrule {B} {W} {W} {B} {W} {W} {R} {}\hfill}
\vskip-3pt
\ligne{\hfill\gzzz\ftt{\the\compteregle \ }  
\llrule {Y} {B} {Y} {R} {B} {W} {W} 
\hskip-3.5pt\rrrule {W} {B} {W} {B} {W} {W} {W} {}\hfill}
\ligne{\hfill\gzzz\ftt{\the\compteregle \ }  
\llrule {W} {W} {W} {Y} {W} {W} {W} 
\hskip-3.5pt\rrrule {Y} {W} {W} {Y} {Y} {W} {W} {}\hfill}
\vskip-3pt
\ligne{\hfill\gzzz\ftt{\the\compteregle \ }  
\llrule {Y} {B} {Y} {W} {W} {B} {Y} 
\hskip-3.5pt\rrrule {B} {W} {W} {R} {W} {R} {Y} {}\hfill}
\vskip-3pt
}
\hfill
\vtop{\leftskip 0pt\parindent 0pt\hsize=\largeouille
\ligne{\hfill\ftt{$c=0$ } \hfill}
\ligne{\hfill\gzzz\ftt{\the\compteregle \ }  
\llrule {W} {W} {W} {B} {W} {W} {B} 
\hskip-3.5pt\rrrule {Y} {W} {W} {Y} {Y} {W} {Y} {}\hfill}
\vskip-3pt
\ligne{\hfill\gzzz\ftt{\the\compteregle \ }  
\llrule {Y} {W} {W} {B} {W} {W} {W} 
\hskip-3.5pt\rrrule {Y} {W} {W} {Y} {Y} {W} {Y} {}\hfill}
\vskip-3pt
\ligne{\hfill\gzzz\ftt{\the\compteregle \ }  
\llrule {B} {R} {Y} {B} {B} {W} {Y} 
\hskip-3.5pt\rrrule {W} {B} {W} {B} {W} {W} {R} {}\hfill}
\vskip-3pt
\ligne{\hfill\gzzz\ftt{\the\compteregle \ }  
\llrule {Y} {B} {W} {R} {W} {W} {W} 
\hskip-3.5pt\rrrule {Y} {W} {W} {Y} {Y} {W} {W} {}\hfill}
\vskip-3pt
\ligne{\hfill\gzzz\ftt{\the\compteregle \ }  
\llrule {R} {R} {Y} {B} {B} {W} {Y} 
\hskip-3.5pt\rrrule {W} {B} {W} {B} {W} {W} {W} {}\hfill}
\vskip-3pt
\ligne{\hfill\gzzz\ftt{\the\compteregle \ }  
\llrule {W} {Y} {W} {W} {W} {W} {R} 
\hskip-3.5pt\rrrule {R} {W} {W} {R} {Y} {W} {B} {}\hfill}
\vskip-3pt
\ligne{\hfill\gzzz\ftt{\the\compteregle \ }  
\llrule {B} {Y} {W} {W} {W} {W} {R} 
\hskip-3.5pt\rrrule {R} {W} {W} {R} {Y} {W} {W} {}\hfill}
\vskip-3pt
\ligne{\hfill\gzzz\ftt{\the\compteregle \ }  
\llrule {W} {W} {Y} {W} {W} {W} {Y} 
\hskip-3.5pt\rrrule {Y} {W} {W} {Y} {R} {W} {W} {}\hfill}
\ligne{\hfill\gzzz\ftt{\the\compteregle \ }  
\llrule {R} {R} {B} {B} {W} {B} {R} 
\hskip-3.5pt\rrrule {B} {W} {W} {B} {W} {W} {R} {}\hfill}
\vskip-3pt
\ligne{\hfill\gzzz\ftt{\the\compteregle \ }  
\llrule {W} {W} {W} {W} {W} {Y} {W} 
\hskip-3.5pt\rrrule {W} {W} {Y} {Y} {W} {Y} {W} {}\hfill}
\vskip-3pt
\ligne{\hfill\gzzz\ftt{\the\compteregle \ }  
\llrule {W} {W} {W} {Y} {W} {W} {W} 
\hskip-3.5pt\rrrule {R} {B} {W} {R} {R} {W} {W} {}\hfill}
\vskip-3pt
}
\hfill}
}
\hfill}
\vskip 10pt
   In the part of Table~\ref{treginc} devoted to the general case, the rules up to~198
deal with the arrival of the locomotive to the register and to the
incrementation itself. Starting from rule~199, they deal with the locomotive which
occurs on \RR$_i$, the return path of the locomotive after performing the incrementation.

   Table~\ref{tregdeb} gives the rule for the motion of a blue locomotive in the \sww-part
of \RR$_c$. But before arriving at \RR$_c$(0), the locomotive must cross \RR$_c$(-1). 
Rules 184, 185 and~132 in Table~\ref{treginc} allow the locomotive to do it. Of 
course, \RR$_i$(-1), \RR$_d$(-1) witness that passage of the locomotive, whence 
rules~186, 187 respectively. Also \RR$_c$(-2)$_b$ witnesses the same passage, which is 
the role of rule~188.

In the white part of~\RR$_c$, rules~189 and~190 complete the rules of 
Table~\ref{tregdeb}, so that we arrive at rule~191 which shows us the blank cell 
of~\RR$_c$ which can see the first \sbb-cell of the register on front of it and, behind 
it, the blue locomotive. At that moment, rule~191 tells us that \sww{} is 
replaced by~\syy, the signal of the incrementation. Rule~192 keeps \syy{} for one more 
step. In the mean while, rule~193 show us that the \sww-cell which is behind \syy{} 
remains \sww{} and rule~194 shows us that the \sbb-cell in front of \syy{} becomes \srr:
it is the preparation of the incrementation. Rule~195 shows us that the locomotive is
triggered on \RR$_i$ when \syy{} is seen. Rules~196 and~197 apply to the pattern 
\syy\srr{} which becomes \sww\sww{} performing the incrementation in that simple way.
Rule~198 tells us that the occurrence of~\srr{} does not affect the \sbb-cell which is in
front of it.

   As already mentioned, starting from rule~199, the rules are devoted to the motion of 
the locomotive
on~\RR$_i$. The locomotive goes towards R$_i$(-1) through rules~86 to~92 of
Table~\ref{tregdeb}, the rules~90 to~92 dealing specifically with \RR$_i$(0). Note that
\syy{} was triggered on the \sww-part of~\RR$_c$ so that the motion of the returning
locomotive is seen by the \sww-part of~\RR$_c$. Rule~199 tells us that in that case,
the \sww-cell is simply a witness. Rule~200 says that \RR$_d$ did not react to the 
\syy-signal on \RR$_c$. Note that \RR$_s$ which cannot see \RR$_c$ can however see 
\RR$_i$, in particular the locomotive triggered on that strand by the signal~\syy{} on
\RR$_c$. Rule 201 says that \RR$_s$ remains unchanged. Rules 202 and 203 indicate the 
reaction of \RR$_i$ to the pattern \syy\srr{} present on \RR$_c$. The first occurrence
of \syy{} raised a locomotive on \RR$_i$, rule 195 as already mentioned, but the
second time of the presence of~\syy, at that time accompanied by \srr{} makes the 
locomotive advance towards \RR$_i$(0) while rule 203 says that \RR$_i$ remains unchanged 
by the occurrence of \srr{} on \RR$_c$. Rule~204 says that \RR$_s$(0) too is not affected
by the presence of the locomotive in \RR$_i$(0).

   Rules 205 up to 207 manage the motion of the leaving locomotive through \RR$_i$(-1).
Rules~208, 209 say that the occurrence of the locomotive in \RR$_i$(-1) does not
affect \RR$_c$(-1) nor \RR$_d$(-1) respectively.

    In the part of Table~\ref{treginc} devoted to the particular case when $c\leq2$,
we can see for $c=2$ the single rule~210 which indicates that \RR$_c$(0) can see
\syy{} in \RR$_c$(1). Note that \RR$_c$(0) cannot see \RR$_c$(2) and as far as nothing
happens on \RR$_c$($\ell$) when \hbox{$0\leq\ell<2$} and as far the rules for the 
locomotive on \RR$_i$ are the same as in the general case, no other rule is needed for
that case.

    The situation is different for~$c=1$ and for $c=0$ as far as \syy{} occurs
in \RR$_c$(0) for $c=1$ and in \RR$_c$(-1) for $c=0$.

    When $c=1$, \syy{} is triggered in \RR$_c$(0): rule~211. Rule~212 keeps \syy{}
for one more step while rule~213 triggers the returning locomotive in \RR$_i$(0). Rule~214
allows \RR$_i$ to recover its idle configuration as far as the locomotive is leaving
\RR$_i$(0). The remaining rules, 216 and 217 tell us that \RR$_c$(-1) and \RR$_d$(0)
respectively are not changed by the occurrence of~\syy{} in \RR$_i$(0).

   The rules when $c=0$ essentially deal with \RR$_c$(-1) and $\mathcal S$(0) for all
strands, \RR$_s$ being excepted as far as it cannot see what happens on \RR$_c$ and
as far as \RR$_i$ and \RR$_d$ are unchanged in the case when $c=0$. Rules~218 and 219
trigger \syy{} in \RR$_c$(-1) and keep it for two steps. In those rules, we notice
that \RR$_c$(-1) can see \sbb{} in \RR$_c$(0), the translation onto \RR$_c$ that $c=0$.
Rule~220 says that \sbb{} in \RR$_c$(0) can see \syy{} behind it so that it becomes~\srr.
Rules~221 and~222 transform the pattern \syy\srr{} into \sww\sww{} which performs the
incrementation: \RR$_c$(0) is \sww{} while \RR$_c$(1) remains \sbb. Rule~223 triggers
the locomotive in \RR$_i$(-1) as it can see the first occurrence of ~\syy{} 
in \RR$_c$(-1). Rule~224 says that the second occurrence makes the locomotive completely
leave \RR$_i$(-1). Rule 225 says that \RR$_d$(-1) is not affected by what happens on
\RR$_i$(-1). The remaining rules, 226 up to 228 say that \RR$_i$(0) and \RR$_c$(-2)$_c$
together with \RR$_c$(-2)$_b$ are no more affected by what they can see: the 
occurrence of~\srr{} in \RR$_c$(0), the occurrence of \syy{} in \RR$_c$(-1) for
\RR$_c$(-2)$_c$ and \RR$_c$(-2)$_b$ respectively.

\subsection{Stopping the computation}\label{srstop}

We arrive to the last Subsection of the present section. It is not the least important
as far as it deals with the end of the computation. We remind the reader that, by
definition, the computation of a cellular automaton has no halting state. In fact, it
is usually considered that when two successive configurations are identical, it should
be considered as the end of the computation. Indeed, nothing new can appear in a 
configuration which is constantly repeated cell by cell. Consequently, stopping the 
computation will consist in realizing a configuration which is endlessly 
identically repeated. Moreover, such a situation is algorithmically detectable.

   We noted that, outside the simulation of the register machine, we decided to construct
the register according a potentially non stopping process. It was the growth of the
register as defined in Sub-subsection~\ref{sbbrgrow}. In that Sub-subsection, we
underlined that the speed of the construction was $\displaystyle{1\over2}$: it needs two
tips of the clock in order to advance the \sgg-cells of the register by one step forward.
It is enough for stopping the computation. When the locomotive arrives to the halting
instruction of the register machine, the implementation of that instruction consists in
sending a locomotive to all~\RR$_s$ of all registers of the simulated register
machine. If we have $n$ registers, $k$ forks will allow us to do that where
\hbox{$2^{k-1}< n\leq 2^k$}. The locomotive arrives at~\RR$_s$ at some time~$t$.
We may assume that $t$ is the same for all registers. At time~$t$, the \sgg-cells are 
at some distance~$L$ from \RR$_s$(0). As far as 
one step forward of the \sgg-cells requires two tips of the clock, if the locomotive 
in~\RR$_s$ advances at speed~1, it reaches the \sgg-cells at time~$t$+$2L$. Taking
for $L$ the biggest one for all registers, we can then compute the time at which
the computation of the automaton halts when we know the time of the halting of the
register machine. If that latter one does not halt, no signal is sent to any \RR$_s$
so that in that case, the registers are endlessly growing.

   Table~\ref{tregstop} gives the rule for stopping the computation. Rule~229 gives
the signal of the process which will lead to the destruction of the \sgg-block of four 
\sgg-cells which allows the register to grow. It is interesting to note that all rules
of Table~\ref{tregstop} have \sww{} as the new state. It is clearly the destruction mark.
The process followed by the rules can be schematised by the figure of~(11).
The idea behind the principle is that \sgg{} can advance thanks to the covering by \sbb{}
and the transformation of \sbb{} into~\sgg{} for those who share a side with the 
line~$\ell$. The process for advancing the strands is disturbed by the erasing of 
one~\sgg{} and much more when other~\sgg-cells are destroyed.

\vskip 10pt
\ligne{\hfill
\vtop{\leftskip 0pt\parindent 0pt\hsize=90pt
$\vcenter{\vbox{
\ftt{
B R B B G W W W W\vszz
B B R B G B W W W\vszz
B B B W B G W W W\vszz
B B B W W G B W W\vszz
B B B W W W G W W\vszz
B B B W W W W B W\vszz
}
}}$
}
\hfill(\numerrel)\hskip 10pt}
\vskip 10pt
\ligne{\hfill
\vtop{
\begin{tab}\label{tregstop}
\leurre
Rules for stopping the computation.
\end{tab}
\vskip -7pt
\ligne{\hfill
\vtop{\leftskip 0pt\parindent 0pt\hsize=\largeouille
\ligne{\hfill\gzzz\ftt{\the\compteregle \ }  
\llrule {B} {Y} {R} {G} {B} {W} {R} 
\hskip-3.5pt\rrrule {W} {B} {W} {W} {W} {W} {W} {}\hfill}
\vskip-3pt
\ligne{\hfill\gzzz\ftt{\the\compteregle \ }  
\llrule {R} {B} {W} {R} {W} {B} {R} 
\hskip-3.5pt\rrrule {B} {W} {W} {W} {W} {W} {W} {}\hfill}
\vskip-3pt
\ligne{\hfill\gzzz\ftt{\the\compteregle \ }  
\llrule {Y} {W} {B} {Y} {W} {B} {Y} 
\hskip-3.5pt\rrrule {B} {W} {W} {W} {W} {W} {W} {}\hfill}
\vskip-3pt
\ligne{\hfill\gzzz\ftt{\the\compteregle \ }  
\llrule {W} {Y} {R} {B} {B} {W} {B} 
\hskip-3.5pt\rrrule {W} {B} {W} {W} {W} {W} {W} {}\hfill}
\vskip-3pt
\ligne{\hfill\gzzz\ftt{\the\compteregle \ }  
\llrule {B} {Y} {R} {G} {B} {W} {W} 
\hskip-3.5pt\rrrule {W} {B} {W} {W} {W} {W} {W} {}\hfill}
\ligne{\hfill\gzzz\ftt{\the\compteregle \ }  
\llrule {R} {B} {W} {W} {W} {B} {G} 
\hskip-3.5pt\rrrule {B} {W} {W} {W} {W} {W} {W} {}\hfill}
\vskip-3pt
\ligne{\hfill\gzzz\ftt{\the\compteregle \ }  
\llrule {Y} {W} {B} {W} {W} {B} {G} 
\hskip-3.5pt\rrrule {B} {W} {W} {W} {W} {W} {W} {}\hfill}
\vskip-3pt
\ligne{\hfill\gzzz\ftt{\the\compteregle \ }  
\llrule {W} {B} {W} {R} {W} {B} {R} 
\hskip-3.5pt\rrrule {B} {W} {W} {W} {W} {W} {W} {}\hfill}
\vskip-3pt
\ligne{\hfill\gzzz\ftt{\the\compteregle \ }  
\llrule {W} {W} {B} {Y} {W} {B} {Y} 
\hskip-3.5pt\rrrule {B} {W} {W} {W} {W} {W} {W} {}\hfill}
\ligne{\hfill\gzzz\ftt{\the\compteregle \ }  
\llrule {B} {W} {W} {B} {B} {W} {B} 
\hskip-3.5pt\rrrule {W} {B} {W} {W} {W} {W} {W} {}\hfill}
\vskip-3pt
\ligne{\hfill\gzzz\ftt{\the\compteregle \ }  
\llrule {W} {W} {W} {W} {W} {B} {W} 
\hskip-3.5pt\rrrule {B} {W} {B} {W} {W} {W} {W} {}\hfill}
\vskip-3pt
\ligne{\hfill\gzzz\ftt{\the\compteregle \ }  
\llrule {W} {Y} {R} {G} {B} {W} {W} 
\hskip-3.5pt\rrrule {W} {B} {W} {W} {W} {W} {W} {}\hfill}
\vskip-3pt
}
\hfill
\vtop{\leftskip 0pt\parindent 0pt\hsize=\largeouille
\ligne{\hfill\gzzz\ftt{\the\compteregle \ }  
\llrule {G} {G} {G} {W} {W} {W} {W} 
\hskip-3.5pt\rrrule {W} {W} {W} {W} {W} {W} {W} {}\hfill}
\vskip-3pt
\ligne{\hfill\gzzz\ftt{\the\compteregle \ }  
\llrule {G} {G} {G} {B} {B} {W} {W} 
\hskip-3.5pt\rrrule {W} {B} {B} {B} {B} {B} {W} {}\hfill}
\vskip-3pt
\ligne{\hfill\gzzz\ftt{\the\compteregle \ }  
\llrule {W} {B} {W} {W} {W} {B} {R} 
\hskip-3.5pt\rrrule {B} {W} {W} {W} {W} {W} {W} {}\hfill}
\vskip-3pt
\ligne{\hfill\gzzz\ftt{\the\compteregle \ }  
\llrule {W} {W} {B} {W} {W} {B} {Y} 
\hskip-3.5pt\rrrule {B} {W} {W} {W} {W} {W} {W} {}\hfill}
\vskip-3pt
\ligne{\hfill\gzzz\ftt{\the\compteregle \ }  
\llrule {W} {W} {W} {R} {W} {B} {W} 
\hskip-3.5pt\rrrule {B} {W} {W} {W} {W} {W} {W} {}\hfill}
\vskip-3pt
\ligne{\hfill\gzzz\ftt{\the\compteregle \ }  
\llrule {W} {W} {W} {Y} {W} {B} {W} 
\hskip-3.5pt\rrrule {B} {W} {W} {W} {W} {W} {W} {}\hfill}
\vskip-3pt
\ligne{\hfill\gzzz\ftt{\the\compteregle \ }  
\llrule {W} {W} {W} {W} {B} {W} {W} 
\hskip-3.5pt\rrrule {W} {B} {W} {W} {W} {W} {W} {}\hfill}
\vskip-3pt
\ligne{\hfill\gzzz\ftt{\the\compteregle \ }  
\llrule {B} {W} {W} {G} {B} {W} {W} 
\hskip-3.5pt\rrrule {W} {B} {W} {W} {W} {W} {W} {}\hfill}
\vskip-3pt
\ligne{\hfill\gzzz\ftt{\the\compteregle \ }  
\llrule {W} {B} {W} {W} {W} {B} {G} 
\hskip-3.5pt\rrrule {B} {W} {W} {W} {W} {W} {W} {}\hfill}
\vskip-3pt
\ligne{\hfill\gzzz\ftt{\the\compteregle \ }  
\llrule {G} {G} {W} {W} {W} {B} {B} 
\hskip-3.5pt\rrrule {B} {W} {B} {B} {B} {B} {W} {}\hfill}
\vskip-3pt
\ligne{\hfill\gzzz\ftt{\the\compteregle \ }  
\llrule {G} {W} {G} {W} {W} {B} {B} 
\hskip-3.5pt\rrrule {B} {W} {B} {B} {B} {B} {W} {}\hfill}
\vskip-3pt
\ligne{\hfill\gzzz\ftt{\the\compteregle \ }  
\llrule {W} {G} {G} {B} {B} {W} {W} 
\hskip-3.5pt\rrrule {W} {B} {B} {B} {B} {B} {W} {}\hfill}
\vskip-3pt
}
\hfill
\vtop{\leftskip 0pt\parindent 0pt\hsize=\largeouille
\ligne{\hfill\gzzz\ftt{\the\compteregle \ }  
\llrule {G} {G} {B} {W} {W} {W} {W} 
\hskip-3.5pt\rrrule {W} {W} {W} {W} {W} {W} {W} {}\hfill}
\vskip-3pt
\ligne{\hfill\gzzz\ftt{\the\compteregle \ }  
\llrule {W} {W} {W} {B} {B} {W} {W} 
\hskip-3.5pt\rrrule {W} {B} {B} {B} {B} {B} {W} {}\hfill}
\vskip-3pt
\ligne{\hfill\gzzz\ftt{\the\compteregle \ }  
\llrule {B} {G} {G} {W} {W} {W} {W} 
\hskip-3.5pt\rrrule {W} {W} {W} {W} {W} {W} {W} {}\hfill}
\vskip-3pt
\ligne{\hfill\gzzz\ftt{\the\compteregle \ }  
\llrule {W} {W} {W} {G} {B} {W} {W} 
\hskip-3.5pt\rrrule {W} {B} {W} {W} {W} {W} {W} {}\hfill}
\vskip-3pt
\ligne{\hfill\gzzz\ftt{\the\compteregle \ }  
\llrule {G} {W} {W} {B} {B} {W} {W} 
\hskip-3.5pt\rrrule {W} {B} {B} {B} {B} {B} {W} {}\hfill}
\vskip-3pt
\ligne{\hfill\gzzz\ftt{\the\compteregle \ }  
\llrule {W} {G} {W} {W} {W} {B} {B} 
\hskip-3.5pt\rrrule {B} {W} {B} {B} {B} {B} {W} {}\hfill}
\vskip-3pt
\ligne{\hfill\gzzz\ftt{\the\compteregle \ }  
\llrule {W} {W} {G} {W} {W} {B} {B} 
\hskip-3.5pt\rrrule {B} {W} {B} {B} {B} {B} {W} {}\hfill}
\vskip-3pt
\ligne{\hfill\gzzz\ftt{\the\compteregle \ }  
\llrule {W} {W} {W} {W} {B} {W} {W} 
\hskip-3.5pt\rrrule {W} {B} {B} {B} {B} {B} {W} {}\hfill}
\vskip-3pt
\ligne{\hfill\gzzz\ftt{\the\compteregle \ }  
\llrule {G} {B} {B} {W} {W} {W} {W} 
\hskip-3.5pt\rrrule {W} {W} {W} {W} {W} {W} {W} {}\hfill}
\vskip-3pt
}
\hfill}
}
\hfill}
\vskip 10pt
When the \sww{} on \RR$_s$ is close to~\sgg, it is seen by a cell of~\RR$_i$ and~\RR$_d$
which both can see \RR$_c$. Accordingly, the \sww{} signal which occurred on \RR$_s$ is
transported onto \RR$_i$ and \RR$_d$ by rules~230 and~231 respectively. Rule~232 confirms
the occurrence of that \sww-cell. In the meanwhile, \sgg{} advanced a bit so that now, 
on \RR$_s$, there is a \sbb-cell in between the created \sww{} and \sgg. Rule~233 says 
that the considered \sbb{} becomes \sww. The \sww-cells go on propagating on
\RR$_i$, rules 234 and 236, and on \RR$_d$, rules 235 and 237. Those cells also propagate
on~\RR$_s$, rule~238. Rules 239 and 240 confirm a created \sww{} as a permanent one.

Rules~241 and~242 are the first rules which change the \sgg-state of a cell into an
\sww-one. Note that in both rules the cell under the \sgg-state can see two neighbours
in the \sgg-state. Rules~115 up to 117 which keep the \sgg-state require the occurrence
of two \sgg-neighbours but also, on the same strand of a non-blank cell of the strand.
Rules~241 and~242 indicate the occurrence of a \sww-neighbour through face~5 which is the
case on \RR$_s$. Before looking further, let us go back to~(11). We can see that \sww{}
appears behind \sgg{} when we have the pattern \sbb\sgg\sww{} on the strand. As shown on
(12), it may also appear by one step sooner and we have the pattern \sbb\sgg\sbb\sww.
\vskip 10pt
\ligne{\hfill
\def\gzgz{\hskip 30pt{}}
\vtop{\leftskip 0pt\parindent 0pt\hsize=220pt
$\vcenter{\vbox{
\ftt{
B R B B G W W W W\gzgz R B B B G W W W W W \vszz
B B R B G B W W W\gzgz B R B B G B W W W W \vszz
B B B W B G W W W\gzgz B B R B B G W W W W \vszz
B B B W W G B W W\gzgz B B B R B G B W W W \vszz
B B B W W W G W W\gzgz B B B B W B G W W W \vszz
B B B W W W W B W\gzgz B B B B W W G B W W \vszz
B B B W W W W W W\gzgz B B B B W W W G W W \vszz
B B B W W W W W W\gzgz B B B B W W W W B W \vszz
B B B W W W W W W\gzgz B B B B W W W W W W \vszz
}
}}$
}
\hfill(\numerrel)\hskip 10pt}
\vskip 10pt
As shown on (12), when it is the case, we have an occurrence of the pattern \sww\sgg\sww{}
which occurs three steps later.

Rules 244 and 245 keep the created \sww-cells on \RR$_i$ and on R$_d$ respectively.
Rule 246 and 247 also keep an \sww-state on \RR$_d$ and on \RR$_s$ respectively.
Rule 249 replaces an \sbb-state by \sww{} on \RR$_s$ where we have the pattern 
\sww\sbb\sgg{} on the strand. Rule~249 again keep the \sww-state on both \RR$_i$ 
and \RR$_d$. 
    
  In rules~250 and 251,we have a \sgg-cell on \RR$_i$ and \RR$_d$ respectively which
can see a single \sgg-cell on \RR$_c$: on \RR$_s$ there is no \sgg-cell at that place.
The rules say that in such a case, \sgg{} becomes \sww. That situation occurs as far as
the propagation of the \sww-cell on whichever strand happens at speed~1. When that 
situation occurs, it remains a \sgg-cell on \RR$_c$ only which will also be soon 
converted into an \sww-cell. When a \sbb-cell at the end of the strand is erased
by reaching \sww, that \sbb-cell can no more contribute to the creation of a \sgg-cell
at the end of the configuration.

   However, as proved by the traces of execution as will soon be seen, several \sbb-cells
are created which are never turned to~\sww. That does not change the conclusion: they 
belong to a configuration which is endlessly repeated step by step. 

   Consider the time~$t$ when we reach the configuration \sww\sgg\sww{} or 
\sww\sgg\sbb\sww{} at the end of the non-blank part of \RR$_s$. At that time rule~240
applies to the \sww-cell before \sgg{} and rule~241 or rule~242 applies to~\sgg. But,
at the same time, rule~118 applies to many neighbours of~\sgg, namely neighbours~2, 3, 
7, 8, 9, 10 and~11. When \sgg{} changes to \sbb, its \sgg-neighbours turn to \sww,
neighbours~3 and~7 being excepted. When \sgg{} changes to \sww, the rule~124 which
allowed some \sbb-neighbours to turn to~\sww{} no more applies so that, in that case,
rule~40 applies which keeps a \sbb-cell when all its neighbours are blank. Accordingly,
the end of the computation is accompanied by a kind of cloud of isolated \sbb-cells.
Those cells remain for ever and their number no more increases when the last \sgg-cell
on a strand is turned to~\sww.

   Rules~252 up to~262 deal with that question and they were found with the help of
the simulating computer program. Table~\ref{texstop} gives the traces of execution
of the rules of Table~\ref{tregstop} in the case where the \sbb-cell before the \sgg-cell
on \RR$_s$ becomes \sww{} at a moment when the end of the computation has the
pattern \sbb\sgg\sww.
    
   That last point completes the proof of Theorem~\ref{letheo}. \hfill$\Box$

\section{Conclusion}\label{conclude}

There are 258 rules which are rotationally pairwise independent. The tables indicate
262 rules which are rotationally pairwise independent but a few rules among rules~1 up 
to~39 are not used in the computation as far as the configuration they imply does not 
occur. As an example, rule~18 is never used as far as the configuration of white cells
neighbouring a \sgg-cell has never three \sgg-neighbours. It has a single \sgg-neighbour
together, sometimes, with another non-blank neighbour, \sbb, \srr{} or \syy.

We said that the rotation invariance is a huge constraint. If we restrict 
the test to a set of generators of the whole group \RR{} of the rotations leaving the  
dodecahedron globally invariant, the minimum reached by the generators is bigger than
that obtained by the whole group for many rules. It was checked by the computer program.
And so, the argument based on the strength of the constraint raised by the requirement
of rotation invariance is sound.

   It should be mentioned that the solution given in the paper is not unique. There could
be a permutation in the role of the colours. Perhaps, the registers could be organized 
somehow differently. The true question is: is it still possible to obtain strong 
universality with less states? I do not know presently how to do. That does not 
mean, of course, that it would be impossible. Even with five states it could be
possible to devise a simpler solution. 
It might also be possible to reduce the number of independent rules. 
However, it is more interesting to reduce the number of states. 
Experience shows that such a reduction requires to change something in the model. Indeed, the present result avoided several 
structures used in planar simulations. So, to change something in the model is the way. 
To change what exactly? That is the question.

\def\fns{\footnotesize}
\newcount\compteregle\compteregle=0
\def\gzzz{\global\advance\compteregle by 1}
\def\qqs{\hskip 1.5pt}
\def\numvoissynth #1 {
\ligne{\hskip #1{\tiny\tt 0\qqs.\qqs.\qqs.\qqs.\qqs.\qqs6\qqs.\qqs.\qqs.\qqs.\qqs. } }
\vskip-2pt
}
\newpage
\vskip 15pt
\ligne{\hfill\bf\large Appendix\hfill}

\ligne{\hfill
\vtop{
\begin{tab}\label{texdecgen}
\leurre Traces of execution of the rules of Table~{\rm\ref{tregdec}}: decrementation of
a register in the general case.
\end{tab}
\vskip -7pt
\ligne{\hfill
\vtop{\leftskip 0pt\parindent 0pt\hsize=120pt
\lalongue=115pt
\traceline {0} {+ - - - - + - - - - +}{}
\traceline {c} {W  W  W  W  W  W  W  W  B  B  B}{}
\traceline {d} {W  W  W  Y  Y  Y  Y  Y  Y  Y  Y}{} 
\traceline {p} {B  W  W  W  W  W  W  W  W  W  W}{}  
\vskip 5pt
\traceline {1} {+ - - - - + - - - - +}{}
\traceline {c} {W  W  W  W  W  W  W  W  B  B  B}{} 
\traceline {d} {W  W  W  Y  Y  Y  Y  Y  Y  Y  Y}{}  
\traceline {p} {W  B  W  W  W  W  W  W  W  W  W}{}  
\vskip 5pt
\traceline {2} {+ - - - - + - - - - +}{}
\traceline {c} {W  W  W  W  W  W  W  W  B  B  B}{} 
\traceline {d} {W  W  W  Y  Y  Y  Y  Y  Y  Y  Y}{}  
\traceline {p} {W  W  B  W  W  W  W  W  W  W  W}{}  
\vskip 5pt
\traceline {3} {+ - - - - + - - - - +}{}
\traceline {c} {W  W  W  W  W  W  W  W  B  B  B}{} 
\traceline {d} {W  W  W  Y  Y  Y  Y  Y  Y  Y  Y}{}  
\traceline {p} {W  W  W  R  W  W  W  W  W  W  W}{}  
\vskip 5pt
\traceline {4} {+ - - - - + - - - - +}{}
\traceline {c} {W  W  R  W  W  W  W  W  B  B  B}{}  
\traceline {d} {W  W  W  Y  Y  Y  Y  Y  Y  Y  Y}{}  
\traceline {p} {W  W  W  W  W  W  W  W  W  W  W}{}  
\vskip 5pt
\traceline {5} {+ - - - - + - - - - +}{}
\traceline {c} {W  W  W  R  W  W  W  W  B  B  B}{}  
\traceline {d} {W  W  W  Y  Y  Y  Y  Y  Y  Y  Y}{}  
\traceline {p} {W  W  W  W  W  W  W  W  W  W  W}{}  
\vskip 5pt
\traceline {6} {+ - - - - + - - - - +}{}
\traceline {c} {W  W  W  W  R  W  W  W  B  B  B}{}  
\traceline {d} {W  W  W  Y  Y  Y  Y  Y  Y  Y  Y}{}  
\traceline {p} {W  W  W  W  W  W  W  W  W  W  W}{}  
\vskip 5pt
\traceline {7} {+ - - - - + - - - - +}{}
\traceline {c} {W  W  W  W  W  R  W  W  B  B  B}{}  
\traceline {d} {W  W  W  Y  Y  Y  Y  Y  Y  Y  Y}{}  
\traceline {p} {W  W  W  W  W  W  W  W  W  W  W}{}  
\vskip 5pt
\traceline {8} {+ - - - - + - - - - +}{}
\traceline {c} {W  W  W  W  W  W  R  W  B  B  B}{}  
\traceline {d} {W  W  W  Y  Y  Y  Y  Y  Y  Y  Y}{}  
\traceline {p} {W  W  W  W  W  W  W  W  W  W  W}{}  
\vskip 5pt
\traceline {9} {+ - - - - + - - - - +}{}
\traceline {c} {W  W  W  W  W  W  R  W  B  B  B}{}  
\traceline {d} {W  W  W  Y  Y  Y  Y  Y  Y  Y  Y}{}  
\traceline {p} {W  W  W  W  W  W  W  W  W  W  W}{}  
}
\hfill
\vtop{\leftskip 0pt\parindent 0pt\hsize=120pt
\lalongue=115pt
\traceline {10} {+ - - - - + - - - - +}{}
\traceline {c}  {W  W  W  W  W  W  W  G  B  B  B}{}  
\traceline {d}  {W  W  W  Y  Y  Y  Y  Y  Y  Y  Y}{}  
\traceline {p}  {W  W  W  W  W  W  W  W  W  W  W}{}  
\vskip 5pt
\traceline {11} {+ - - - - + - - - - +}{}
\traceline {c}  {W  W  W  W  W  W  W  G  B  B  B}{}  
\traceline {d}  {W  W  W  Y  Y  Y  Y  R  Y  Y  Y}{}  
\traceline {p}  {W  W  W  W  W  W  W  W  W  W  W}{}  
\vskip 5pt
\traceline {12} {+ - - - - + - - - - +}{}
\traceline {c}  {W  W  W  W  W  W  W  B  B  B  B}{}  
\traceline {d}  {W  W  W  Y  Y  Y  R  Y  Y  Y  Y}{}  
\traceline {p}  {W  W  W  W  W  W  W  W  W  W  W}{}  
\vskip 5pt
\traceline {13} {+ - - - - + - - - - +}{}
\traceline {c}  {W  W  W  W  W  W  W  B  B  B  B}{}  
\traceline {d}  {W  W  W  Y  Y  R  Y  Y  Y  Y  Y}{}  
\traceline {p}  {W  W  W  W  W  W  W  W  W  W  W}{}  
\vskip 5pt
\traceline {14} {+ - - - - + - - - - +}{}
\traceline {c}  {W  W  W  W  W  W  W  B  B  B  B}{}  
\traceline {d}  {W  W  W  Y  R  Y  Y  Y  Y  Y  Y}{}  
\traceline {p}  {W  W  W  W  W  W  W  W  W  W  W}{}  
\vskip 5pt
\traceline {15} {+ - - - - + - - - - +}{}
\traceline {c}  {W  W  W  W  W  W  W  B  B  B  B}{}  
\traceline {d}  {W  W  W  R  Y  Y  Y  Y  Y  Y  Y}{}  
\traceline {p}  {W  W  W  W  W  W  W  W  W  W  W}{}  
\vskip 5pt
\traceline {16} {+ - - - - + - - - - +}{}
\traceline {c}  {W  W  W  W  W  W  W  B  B  B  B}{}  
\traceline {d}  {W  W  R  Y  Y  Y  Y  Y  Y  Y  Y}{}  
\traceline {p}  {W  W  W  W  W  W  W  W  W  W  W}{}  
\vskip 5pt
\traceline {17} {+ - - - - + - - - - +}{}
\traceline {c}  {W  W  W  W  W  W  W  B  B  B  B}{}  
\traceline {d}  {W  B  W  Y  Y  Y  Y  Y  Y  Y  Y}{}  
\traceline {p}  {W  W  W  W  W  W  W  W  W  W  W}{}  
\vskip 5pt
\traceline {18} {+ - - - - + - - - - +}{}
\traceline {c}  {W  W  W  W  W  W  W  B  B  B  B}{}  
\traceline {d}  {B  W  W  Y  Y  Y  Y  Y  Y  Y  Y}{}  
\traceline {p}  {W  W  W  W  W  W  W  W  W  W  W}{}  
}
\hfill}
}
\hfill}

\ligne{\hfill 
\vtop{
\begin{tab}\label{texdecpart}
\leurre Traces of execution of the rules of Table~{\rm\ref{tregdec}} in the particular
cases of decrementation, when $c\in\{0,1,2\}$ 
\end{tab}
\vskip-7pt
\ligne{\hfill $c=2$\hfill}
\ligne{\hfill
\vtop{\leftskip 0pt\parindent 0pt\hsize=120pt
\lalongue=115pt
\traceline {0}  {+ - - - - + - - -}{}
\traceline {c}  {W  W  W  W  W  B  B  B  B}{}  
\traceline {d}  {W  W  W  Y  Y  Y  Y  Y  Y}{}  
\traceline {p}  {B  W  W  W  W  W  W  W  W}{}  
\vskip 5pt
\traceline {1}  {+ - - - - + - - -}{}
\traceline {c}  {W  W  W  W  W  B  B  B  B}{}  
\traceline {d}  {W  W  W  Y  Y  Y  Y  Y  Y}{}  
\traceline {p}  {W  B  W  W  W  W  W  W  W}{}  
\vskip 5pt
\traceline {2}  {+ - - - - + - - -}{}
\traceline {c}  {W  W  W  W  W  B  B  B  B}{}  
\traceline {d}  {W  W  W  Y  Y  Y  Y  Y  Y}{}  
\traceline {p}  {W  W  B  W  W  W  W  W  W}{}  
\vskip 5pt
\traceline {3}  {+ - - - - + - - -}{}
\traceline {c}  {W  W  W  W  W  B  B  B  B}{}  
\traceline {d}  {W  W  W  Y  Y  Y  Y  Y  Y}{}  
\traceline {p}  {W  W  W  R  W  W  W  W  W}{}  
\vskip 5pt
\traceline {4}  {+ - - - - + - - -}{}
\traceline {c}  {W  W  R  W  W  B  B  B  B}{}  
\traceline {d}  {W  W  W  Y  Y  Y  Y  Y  Y}{}  
\traceline {p}  {W  W  W  W  W  W  W  W  W}{}  
\vskip 5pt
\traceline {5}  {+ - - - - + - - -}{}
\traceline {c}  {W  W  W  R  W  B  B  B  B}{}  
\traceline {d}  {W  W  W  Y  Y  Y  Y  Y  Y}{}  
\traceline {p}  {W  W  W  W  W  W  W  W  W}{}  
}
\hfill
\vtop{\leftskip 0pt\parindent 0pt\hsize=120pt
\lalongue=115pt
\traceline {6}  {+ - - - - + - - -}{}
\traceline {c}  {W  W  W  W  G  B  B  B  B}{}  
\traceline {d}  {W  W  W  Y  Y  Y  Y  Y  Y}{}  
\traceline {p}  {W  W  W  W  W  W  W  W  W}{}  
\vskip 5pt
\traceline {7}  {+ - - - - + - - -}{}
\traceline {c}  {W  W  W  W  G  B  B  B  B}{}  
\traceline {d}  {W  W  W  Y  R  Y  Y  Y  Y}{}  
\traceline {p}  {W  W  W  W  W  W  W  W  W}{}  
\vskip 5pt
\traceline {8}  {+ - - - - + - - -}{}
\traceline {c}  {W  W  W  W  B  B  B  B  B}{}  
\traceline {d}  {W  W  W  R  Y  Y  Y  Y  Y}{}  
\traceline {p}  {W  W  W  W  W  W  W  W  W}{}  
\vskip 5pt
\traceline {9}  {+ - - - - + - - -}{}
\traceline {c}  {W  W  W  W  B  B  B  B  B}{}  
\traceline {d}  {W  W  R  Y  Y  Y  Y  Y  Y}{}  
\traceline {p}  {W  W  W  W  W  W  W  W  W}{}  
\vskip 5pt
\traceline {10} {+ - - - - + - - -}{}
\traceline {c}  {W  W  W  W  B  B  B  B  B}{}  
\traceline {d}  {W  B  W  Y  Y  Y  Y  Y  Y}{}  
\traceline {p}  {W  W  W  W  W  W  W  W  W}{}  
\vskip 5pt
\traceline {11} {+ - - - - + - - -}{}
\traceline {c}  {W  W  W  W  B  B  B  B  B}{}  
\traceline {d}  {B  W  W  Y  Y  Y  Y  Y  Y}{}  
\traceline {p}  {W  W  W  W  W  W  W  W  W}{}  
}
\hfill}
\vskip 10pt
\ligne{\hfill decrementation when $c = 1$ \hfill}
\ligne{\hfill
\vtop{\leftskip 0pt\parindent 0pt\hsize=110pt
\lalongue=105pt
\traceline {0}  {+ - - - - + - -}{}
\traceline {c}  {W  W  W  W  B  B  B  B}{}  
\traceline {d}  {W  W  W  Y  Y  Y  Y  Y}{}  
\traceline {p}  {B  W  W  W  W  W  W  W}{}  
\vskip 5pt
\traceline {1}  {+ - - - - + - -}{}
\traceline {c}  {W  W  W  W  B  B  B  B}{}  
\traceline {d}  {W  W  W  Y  Y  Y  Y  Y}{}  
\traceline {p}  {W  B  W  W  W  W  W  W}{}  
\vskip 5pt
\traceline {2}  {+ - - - - + - -}{}
\traceline {c}  {W  W  W  W  B  B  B  B}{}  
\traceline {d}  {W  W  W  Y  Y  Y  Y  Y}{}  
\traceline {p}  {W  W  B  W  W  W  W  W}{}  
\vskip 5pt
\traceline {3}  {+ - - - - + - -}{}
\traceline {c}  {W  W  W  W  B  B  B  B}{}  
\traceline {d}  {W  W  W  Y  Y  Y  Y  Y}{}  
\traceline {p}  {W  W  W  R  W  W  W  W}{}  
}
\hfill
\vtop{\leftskip 0pt\parindent 0pt\hsize=110pt
\lalongue=105pt
\traceline {4}  {+ - - - - + - -}{}
\traceline {c}  {W  W  R  W  B  B  B  B}{}  
\traceline {d}  {W  W  W  Y  Y  Y  Y  Y}{}  
\traceline {p}  {W  W  W  W  W  W  W  W}{}  
\vskip 5pt
\traceline {5}  {+ - - - - + - -}{}
\traceline {c}  {W  W  W  G  B  B  B  B}{}  
\traceline {d}  {W  W  W  Y  Y  Y  Y  Y}{}  
\traceline {p}  {W  W  W  W  W  W  W  W}{}  
\vskip 5pt
\traceline {6}  {+ - - - - + - -}{}
\traceline {c}  {W  W  W  G  B  B  B  B}{}  
\traceline {d}  {W  W  W  R  Y  Y  Y  Y}{}  
\traceline {p}  {W  W  W  W  W  W  W  W}{}  
}
\hfill
\vtop{\leftskip 0pt\parindent 0pt\hsize=110pt
\lalongue=105pt
\traceline {7}  {+ - - - - + - -}{}
\traceline {c}  {W  W  W  B  B  B  B  B}{}  
\traceline {d}  {W  W  R  Y  Y  Y  Y  Y}{}  
\traceline {p}  {W  W  W  W  W  W  W  W}{}  
\vskip 5pt
\traceline {8}  {+ - - - - + - -}{}
\traceline {c}  {W  W  W  B  B  B  B  B}{}  
\traceline {d}  {W  B  W  Y  Y  Y  Y  Y}{}  
\traceline {p}  {W  W  W  W  W  W  W  W}{}  
\vskip 5pt
\traceline {8}  {+ - - - - + - -}{}
\traceline {c}  {W  W  W  B  B  B  B  B}{}  
\traceline {d}  {B  W  W  Y  Y  Y  Y  Y}{}  
\traceline {p}  {W  W  W  W  W  W  W  W}{}  
}
\hfill}
\vskip 10pt
\ligne{\hfill decrementation when $c = 0$ \hfill}
\ligne{\hfill
\vtop{\leftskip 0pt\parindent 0pt\hsize=105pt
\lalongue=97pt
\traceline {0}  {+ - - - - + -}{}
\traceline {c}  {W  W  W  B  B  B  B}{}  
\traceline {d}  {W  W  W  Y  Y  Y  Y}{}  
\traceline {p}  {B  W  W  W  W  W  W}{}  
\vskip 5pt
\traceline {1}  {+ - - - - + -}{}
\traceline {c}  {W  W  W  B  B  B  B}{}  
\traceline {d}  {W  W  W  Y  Y  Y  Y}{}  
\traceline {p}  {W  B  W  W  W  W  W}{}  
\vskip 5pt
\traceline {2}  {+ - - - - + -}{}
\traceline {c}  {W  W  W  B  B  B  B}{}  
\traceline {d}  {W  W  W  Y  Y  Y  Y}{}  
\traceline {p}  {W  W  B  W  W  W  W}{}  
}
\hfill
\vtop{\leftskip 0pt\parindent 0pt\hsize=105pt
\lalongue=97pt
\traceline {3}  {+ - - - - + -}{}
\traceline {c}  {W  W  W  B  B  B  B}{}  
\traceline {d}  {W  W  W  Y  Y  Y  Y}{}  
\traceline {p}  {W  W  W  R  W  W  W}{}  
\vskip 5pt
\traceline {4}  {+ - - - - + -}{}
\traceline {c}  {W  W  G  B  B  B  B}{}  
\traceline {d}  {W  W  W  Y  Y  Y  Y}{}  
\traceline {p}  {W  W  W  W  W  W  W}{}  
\vskip 5pt
\traceline {5}  {+ - - - - + -}{}
\traceline {c}  {W  W  G  B  B  B  B}{}  
\traceline {d}  {W  W  W  Y  Y  Y  Y}{}  
\traceline {p}  {W  W  W  W  W  W  B}{}  
}
\hfill
\vtop{\leftskip 0pt\parindent 0pt\hsize=105pt
\lalongue=97pt
\traceline {6}  {+ - - - - + -}{}
\traceline {c}  {W  W  W  B  B  B  B}{}  
\traceline {d}  {W  W  W  Y  Y  Y  Y}{}  
\traceline {p}  {W  W  W  W  W  B  W}{}  
\vskip 5pt
\traceline {7}  {+ - - - - + -}{}
\traceline {c}  {W  W  W  B  B  B  B}{}  
\traceline {d}  {W  W  W  Y  Y  Y  Y}{}  
\traceline {p}  {W  W  W  W  B  W  W}{}  
}
\hfill}
}
\hfill}
\vskip 10pt
\ligne{\hfill
\vtop{
\begin{tab}\label{texincgen}
\leurre Traces of execution of the rules of Table~{\rm\ref{treginc}}: incrementation of
a register in the general case.
\end{tab}
\vskip -7pt
\ligne{\hfill
\vtop{\leftskip 0pt\parindent 0pt\hsize=120pt
\lalongue=115pt
\traceline {0} {+ - - - - + - - - - +}{}
\traceline {c} {B  W  W  W  W  W  W  W  B  B  B}{}
\traceline {i} {W  W  W  R  R  R  R  R  R  R  R}{} 
\vskip 5pt
\traceline {1} {+ - - - - + - - - - +}{}
\traceline {c} {W  B  W  W  W  W  W  W  B  B  B}{}
\traceline {i} {W  W  W  R  R  R  R  R  R  R  R}{} 
\vskip 5pt
\traceline {2} {+ - - - - + - - - - +}{}
\traceline {c} {W  W  B  W  W  W  W  W  B  B  B}{}
\traceline {i} {W  W  W  R  R  R  R  R  R  R  R}{} 
\vskip 5pt
\traceline {3} {+ - - - - + - - - - +}{}
\traceline {c} {W  W  W  B  W  W  W  W  B  B  B}{}
\traceline {i} {W  W  W  R  R  R  R  R  R  R  R}{} 
\vskip 5pt
\traceline {4} {+ - - - - + - - - - +}{}
\traceline {c} {W  W  W  W  B  W  W  W  B  B  B}{}
\traceline {i} {W  W  W  R  R  R  R  R  R  R  R}{} 
\vskip 5pt
\traceline {5} {+ - - - - + - - - - +}{}
\traceline {c} {W  W  W  W  W  B  W  W  B  B  B}{}
\traceline {i} {W  W  W  R  R  R  R  R  R  R  R}{} 
\vskip 5pt
\traceline {6} {+ - - - - + - - - - +}{}
\traceline {c} {W  W  W  W  W  W  B  W  B  B  B}{}
\traceline {i} {W  W  W  R  R  R  R  R  R  R  R}{} 
\vskip 5pt
\traceline {7} {+ - - - - + - - - - +}{}
\traceline {c} {W  W  W  W  W  W  W  Y  B  B  B}{}
\traceline {i} {W  W  W  R  R  R  R  R  R  R  R}{} 
}
\hfill
\vtop{\leftskip 0pt\parindent 0pt\hsize=120pt
\lalongue=115pt
\traceline {8} {+ - - - - + - - - - +}{}
\traceline {c} {W  W  W  W  W  W  W  Y  R  B  B}{}
\traceline {i} {W  W  W  R  R  R  R  B  R  R  R}{} 
\vskip 5pt
\traceline {9} {+ - - - - + - - - - +}{}
\traceline {c} {W  W  W  W  W  W  W  W  W  B  B}{}
\traceline {i} {W  W  W  R  R  R  B  R  R  R  R}{} 
\vskip 5pt
\traceline {10} {+ - - - - + - - - - +}{}
\traceline {c}  {W  W  W  W  W  W  W  W  W  B  B}{}
\traceline {i}  {W  W  W  R  R  B  R  R  R  R  R}{} 
\vskip 5pt
\traceline {11} {+ - - - - + - - - - +}{}
\traceline {c}  {W  W  W  W  W  W  W  W  W  B  B}{}
\traceline {i}  {W  W  W  R  B  R  R  R  R  R  R}{} 
\vskip 5pt
\traceline {12} {+ - - - - + - - - - +}{}
\traceline {c}  {W  W  W  W  W  W  W  W  W  B  B}{}
\traceline {i}  {W  W  W  B  R  R  R  R  R  R  R}{} 
\vskip 5pt
\traceline {13} {+ - - - - + - - - - +}{}
\traceline {c}  {W  W  W  W  W  W  W  W  W  B  B}{}
\traceline {i}  {W  W  B  R  R  R  R  R  R  R  R}{} 
\vskip 5pt
\traceline {14} {+ - - - - + - - - - +}{}
\traceline {c}  {W  W  W  W  W  W  W  W  W  B  B}{}
\traceline {i}  {W  B  W  R  R  R  R  R  R  R  R}{} 
\vskip 5pt
\traceline {15} {+ - - - - + - - - - +}{}
\traceline {c}  {W  W  W  W  W  W  W  W  W  B  B}{}
\traceline {i}  {B  W  W  R  R  R  R  R  R  R  R}{} 
}
\hfill}
}
\hfill}

\ligne{\hfill 
\vtop{
\begin{tab}\label{texincpart}
\leurre Traces of execution of the rules of Table~{\rm\ref{treginc}} in the particular
cases of incrementation, when $c\in\{0,1,2\}$ 
\end{tab}
\vskip-7pt
\ligne{\hfill $c=2$\hfill}
\ligne{\hskip-10pt
\vtop{\leftskip 0pt\parindent 0pt\hsize=120pt
\lalongue=115pt
\traceline {0}  {+ - - - - + - - -}{}
\traceline {c}  {B  W  W  W  W  B  B  B  B}{}  
\traceline {d}  {W  W  W  R  R  R  R  R  R}{}  
\vskip 5pt
\traceline {1}  {+ - - - - + - - -}{}
\traceline {c}  {W  B  W  W  W  B  B  B  B}{}  
\traceline {d}  {W  W  W  R  R  R  R  R  R}{}  
\vskip 5pt
\traceline {2}  {+ - - - - + - - -}{}
\traceline {c}  {W  W  B  W  W  B  B  B  B}{}  
\traceline {d}  {W  W  W  R  R  R  R  R  R}{}  
\vskip 5pt
\traceline {3}  {+ - - - - + - - -}{}
\traceline {c}  {W  W  W  B  W  B  B  B  B}{}  
\traceline {d}  {W  W  W  R  R  R  R  R  R}{}  
}
\hfill
\vtop{\leftskip 0pt\parindent 0pt\hsize=120pt
\lalongue=115pt
\traceline {4}  {+ - - - - + - - -}{}
\traceline {c}  {W  W  W  W  Y  B  B  B  B}{}  
\traceline {d}  {W  W  W  R  R  R  R  R  R}{}  
\vskip 5pt
\traceline {5}  {+ - - - - + - - -}{}
\traceline {c}  {W  W  W  W  Y  R  B  B  B}{}  
\traceline {d}  {W  W  W  R  B  R  R  R  R}{}  
\vskip 5pt
\traceline {6}  {+ - - - - + - - -}{}
\traceline {c}  {W  W  W  W  W  W  B  B  B}{}  
\traceline {d}  {W  W  W  B  R  R  R  R  R}{}  
}
\hfill
\vtop{\leftskip 0pt\parindent 0pt\hsize=120pt
\lalongue=115pt
\traceline {7}  {+ - - - - + - - -}{}
\traceline {c}  {W  W  W  W  W  W  B  B  B}{}  
\traceline {d}  {W  W  B  R  R  R  R  R  R}{}  
\vskip 5pt
\traceline {8}  {+ - - - - + - - -}{}
\traceline {c}  {W  W  W  W  W  W  B  B  B}{}  
\traceline {d}  {W  B  W  R  R  R  R  R  R}{}  
\vskip 5pt
\traceline {9}  {+ - - - - + - - -}{}
\traceline {c}  {W  W  W  W  W  W  B  B  B}{}  
\traceline {d}  {B  W  W  R  R  R  R  R  R}{}  
}
\hfill}
\vskip 10pt
\ligne{\hfill
\vtop{\leftskip 0pt\parindent 0pt\hsize=110pt
\lalongue=105pt
\ligne{\hfill $c=1$\hfill}
\traceline {0}  {+ - - - - + - -}{}
\traceline {c}  {B  W  W  W  B  B  B  B}{}  
\traceline {d}  {W  W  W  R  R  R  R  R}{}  
\vskip 5pt
\traceline {1}  {+ - - - - + - -}{}
\traceline {c}  {W  B  W  W  B  B  B  B}{}  
\traceline {d}  {W  W  W  R  R  R  R  R}{}  
\vskip 5pt
\traceline {2}  {+ - - - - + - -}{}
\traceline {c}  {W  W  B  W  B  B  B  B}{}  
\traceline {d}  {W  W  W  R  R  R  R  R}{}  
\vskip 5pt
\traceline {3}  {+ - - - - + - -}{}
\traceline {c}  {W  W  W  Y  B  B  B  B}{}  
\traceline {d}  {W  W  W  R  R  R  R  R}{}  
\vskip 5pt
}
\hfill 
\vtop{\leftskip 0pt\parindent 0pt\hsize=110pt
\lalongue=105pt
\ligne{\hfill $c=1$\hfill}
\traceline {4}  {+ - - - - + - -}{}
\traceline {c}  {W  W  W  Y  R  B  B  B}{}  
\traceline {d}  {W  W  W  B  R  R  R  R}{}  
\vskip 5pt
\traceline {5}  {+ - - - - + - -}{}
\traceline {c}  {W  W  W  W  W  B  B  B}{}  
\traceline {d}  {W  W  B  R  R  R  R  R}{}  
\vskip 5pt
\traceline {6}  {+ - - - - + - -}{}
\traceline {c}  {W  W  W  W  W  B  B  B}{}  
\traceline {d}  {W  B  W  R  R  R  R  R}{}  
\vskip 5pt
\traceline {7}  {+ - - - - + - -}{}
\traceline {c}  {W  W  W  W  W  B  B  B}{}  
\traceline {d}  {B  W  W  R  R  R  R  R}{}  
}
\hfill\hfill
\vtop{\leftskip 0pt\parindent 0pt\hsize=100pt
\lalongue=95pt
\ligne{\hfill $c=0$\hfill}
\traceline {0}  {+ - - - - + -}{}
\traceline {c}  {B  W  W  B  B  B  B}{}  
\traceline {d}  {W  W  W  R  R  R  R}{}  
\vskip 5pt
\traceline {1}  {+ - - - - + -}{}
\traceline {c}  {W  B  W  B  B  B  B}{}  
\traceline {d}  {W  W  W  R  R  R  R}{}  
\vskip 5pt
\traceline {2}  {+ - - - - + -}{}
\traceline {c}  {W  W  Y  B  B  B  B}{}  
\traceline {d}  {W  W  W  R  R  R  R}{}  
\vskip 5pt
\traceline {3}  {+ - - - - + -}{}
\traceline {c}  {W  W  Y  R  B  B  B}{}  
\traceline {d}  {W  W  B  R  R  R  R}{}  
\vskip 5pt
\traceline {4}  {+ - - - - + -}{}
\traceline {c}  {W  W  W  W  B  B  B}{}  
\traceline {d}  {W  B  W  R  R  R  R}{}  
\vskip 5pt
\traceline {5}  {+ - - - - + -}{}
\traceline {c}  {W  W  W  W  B  B  B}{}  
\traceline {d}  {B  W  W  R  R  R  R}{}  
}
\hfill}
}
\hfill}

\ligne{\hfill 
\vtop{
\begin{tab}\label{texstop}
\leurre Traces of execution of the rules of Table~{\rm\ref{tregstop}}.
\end{tab}
\vskip-7pt
\ligne{\hskip-10pt
\vtop{\leftskip 0pt\parindent 0pt\hsize=120pt
\lalongue=115pt
\traceline {0} {+ - - - - + - - - - +}{}
\traceline {c} {B  B  B  B  G  W  W  W  W  W  W}{}
\traceline {i} {R  R  R  R  G  W  W  W  W  W  W}{}
\traceline {d} {Y  Y  Y  Y  G  W  W  W  W  W  W}{}
\traceline {s} {R  B  B  B  G  W  W  W  W  W  W}{}
\vskip 5pt
\traceline {1} {+ - - - - + - - - - +}{}
\traceline {c} {B  B  B  B  G  B  W  W  W  W  W}{}
\traceline {i} {R  R  R  R  G  B  W  W  W  W  W}{}
\traceline {d} {Y  Y  Y  Y  G  B  W  W  W  W  W}{}
\traceline {s} {B  R  B  B  G  B  W  W  W  W  W}{}
\vskip 5pt
\traceline {2} {+ - - - - + - - - - +}{}
\traceline {c} {B  B  B  B  B  G  W  W  W  W  W}{}
\traceline {i} {R  R  R  R  R  G  W  W  W  W  W}{}
\traceline {d} {Y  Y  Y  Y  Y  G  W  W  W  W  W}{}
\traceline {s} {B  B  R  B  B  G  W  W  W  W  W}{}
\vskip 5pt
\traceline {3} {+ - - - - + - - - - +}{}
\traceline {c} {B  B  B  B  B  G  B  W  W  W  W}{}
\traceline {i} {R  R  R  R  R  G  B  W  W  W  W}{}
\traceline {d} {Y  Y  Y  Y  Y  G  B  W  W  W  W}{}
\traceline {s} {B  B  B  R  B  G  B  W  W  W  W}{}
\vskip 5pt
\traceline {4} {+ - - - - + - - - - +}{}
\traceline {c} {B  B  B  B  B  B  G  W  W  W  W}{}
\traceline {i} {R  R  R  R  R  R  G  W  W  W  W}{}
\traceline {d} {Y  Y  Y  Y  Y  Y  G  W  W  W  W}{}
\traceline {s} {B  B  B  B  W  B  G  W  W  W  W}{}
\vskip 5pt
\traceline {5} {+ - - - - + - - - - +}{}
\traceline {c} {B  B  B  B  B  B  G  B  W  W  W}{}
\traceline {i} {R  R  R  R  W  R  G  B  W  W  W}{}
\traceline {d} {Y  Y  Y  Y  W  Y  G  B  W  W  W}{}
\traceline {s} {B  B  B  B  W  W  G  B  W  W  W}{}
\vskip 5pt
\traceline {6} {+ - - - - + - - - - +}{}
\traceline {c} {B  B  B  B  W  B  B  G  W  W  W}{}
\traceline {i} {R  R  R  R  W  W  R  G  W  W  W}{}
\traceline {d} {Y  Y  Y  Y  W  W  Y  G  W  W  W}{}
\traceline {s} {B  B  B  B  W  W  W  G  W  W  W}{}
}
\hskip 20pt
\vtop{\leftskip 0pt\parindent 0pt\hsize=120pt
\lalongue=115pt
\traceline {7} {+ - - - - + - - - - +}{}
\traceline {c} {B  B  B  B  W  W  B  G  B  W  W}{}
\traceline {i} {R  R  R  R  W  W  W  G  B  W  W}{}
\traceline {d} {Y  Y  Y  Y  W  W  W  G  B  W  W}{}
\traceline {s} {B  B  B  B  W  W  W  W  B  W  W}{}
\vskip 5pt
\traceline {8} {+ - - - - + - - - - +}{}
\traceline {c} {B  B  B  B  W  W  W  B  G  W  W}{}
\traceline {i} {R  R  R  R  W  W  W  W  G  W  W}{}
\traceline {d} {Y  Y  Y  Y  W  W  W  W  G  W  W}{}
\traceline {s} {B  B  B  B  W  W  W  W  B  W  W}{}
\vskip 5pt
\traceline {9} {+ - - - - + - - - - +}{}
\traceline {c} {B  B  B  B  W  W  W  W  G  B  W}{}
\traceline {i} {R  R  R  R  W  W  W  W  W  B  W}{}
\traceline {d} {Y  Y  Y  Y  W  W  W  W  W  B  W}{}
\traceline {s} {B  B  B  B  W  W  W  W  W  W  W}{}
\vskip 5pt
\traceline {10} {+ - - - - + - - - - +}{}
\traceline {c}  {B  B  B  B  W  W  W  W  W  G  W}{}
\traceline {i}  {R  R  R  R  W  W  W  W  W  B  W}{}
\traceline {d}  {Y  Y  Y  Y  W  W  W  W  W  B  W}{}
\traceline {s}  {B  B  B  B  W  W  W  W  W  W  W}{}
\vskip 5pt
\traceline {11} {+ - - - - + - - - - +}{}
\traceline {c}  {B  B  B  B  W  W  W  W  W  B  W}{}
\traceline {i}  {R  R  R  R  W  W  W  W  W  W  W}{}
\traceline {d}  {Y  Y  Y  Y  W  W  W  W  W  W  W}{}
\traceline {s}  {B  B  B  B  W  W  W  W  W  W  W}{}
\vskip 5pt
\traceline {12} {+ - - - - + - - - - +}{}
\traceline {c}  {B  B  B  B  W  W  W  W  W  B  W}{}
\traceline {i}  {R  R  R  R  W  W  W  W  W  W  W}{}
\traceline {d}  {Y  Y  Y  Y  W  W  W  W  W  W  W}{}
\traceline {s}  {B  B  B  B  W  W  W  W  W  W  W}{}
\vskip 5pt
}
\hfill}
}
\hfill}
\end{document}